\documentclass[tighten,times,twocolumn]{aastex701}

\graphicspath{{./}{figures/}}

\usepackage{amsmath}
\usepackage{mathtools}

\newcommand{\lp}{\ensuremath{\left(}}        
\newcommand{\rp}{\ensuremath{\right)}}

\def\beq{\begin{equation}}
\def\eeq{\end{equation}}

\renewcommand{\vec}[1]{{\boldsymbol{\mathbf{#1}}}}
\newcommand{\mat}[1]{{\boldsymbol{\mathbf{#1}}}}

\newcommand{\mC}{\mat{C}}

\newcommand{\appropto}{\mathrel{\vcenter{
  \offinterlineskip\halign{\hfil$##$\cr
    \propto\cr\noalign{\kern2pt}\sim\cr\noalign{\kern-2pt}}}}}

\newcommand{\calL}{\mathcal{L}}

\newcommand{\ihMpc}{\;h\:\mathrm{Mpc}^{-1}}

\newcommand\numberthis{\addtocounter{equation}{1}\tag{\theequation}}

\newcommand{\hi}{H{\sc i}}
\usepackage{mathrsfs}
\usepackage{hyperref}
\usepackage{xspace}
\usepackage{graphicx}
\usepackage{hyperref}
\usepackage{booktabs}
\usepackage{savesym}
\savesymbol{tablenum}
\usepackage{siunitx}
\restoresymbol{SIX}{tablenum}
\DeclareSIUnit\jansky{Jy}
\DeclareSIUnit\beam{beam}

\usepackage[capitalize,noabbrev]{cleveref}
\newcommand{\secref}[1]{Section~\ref{#1}}
\newcommand{\appref}[1]{Appendix~\ref{#1}}

\definecolor{darkbg}{rgb}{0.25, 0.78, 0.85}

\newcommand{\dtraw}{\SI{9.9404}{\second}\xspace}

\newcommand{\dnu}{\SI{390.625}{\kilo\hertz}\xspace}
\newcommand{\datestart}{2019 January 1\xspace}
\newcommand{\dateend}{2019 November 5\xspace}
\newcommand{\ns}{NS\xspace}
\newcommand{\ew}{EW\xspace}
\newcommand{\nsample}{\ensuremath{\mathcal{N}}}

\newcommand{\radiometer}{\ensuremath{\varrho}}
\newcommand{\hybrid}{\ensuremath{\breve{V}}}
\newcommand{\hybridnoise}{\ensuremath{\breve{\sigma}}}
\newcommand{\redundancy}{\ensuremath{\mathcal{R}}}
\newcommand{\map}{\ensuremath{M}}
\newcommand{\mask}{\ensuremath{G}}
\newcommand{\HI}{\ensuremath{{\rm HI}}}
\renewcommand{\hi}{\HI}
\newcommand{\sHI}{\ensuremath{{\scriptscriptstyle {\rm HI}}}}

\newcommand{\Tbar}{\ensuremath{\bar{T}_{\rm b}}}

\newcommand{\OmegaHI}{\ensuremath{\Omega_\sHI}}

\newcommand{\tcm}{21$\,$cm\xspace}  

\newcommand{\bHI}{\ensuremath{b_\sHI}}

\newcommand{\zeff}{\ensuremath{z_\text{eff}}}
\newcommand{\AHI}{{\ensuremath{\mathcal{A}_\sHI}}}
\newcommand{\AHIps}{{\ensuremath{\mathcal{A}_\sHI^2}}}

\newcommand{\alphaNL}{\ensuremath{\alpha_{\rm NL}}}
\newcommand{\PL}{\ensuremath{P_{\rm L}}}
\newcommand{\PNL}{\ensuremath{P_{\rm NL}}}
\newcommand{\alphaFoG}{\ensuremath{\alpha_{\rm FoG}}}
\newcommand{\DFoG}[1]{\ensuremath{D_{#1}^{\rm FoG}}}

\newcommand{\fmu}{\langle f \mu^2\rangle}

\newcommand{\CNoise}{\ensuremath{\mC_{\rm N}}}
\newcommand{\AHIpsConstraintFullBand}{\AHIps = 3.55^{+0.96}_{-1.32}\text{(stat.)}\pm0.61\text{(sys.)}}
\newcommand{\AHIpsConstraintUpperBand}{\AHIps = 2.47^{+1.59}_{-1.08}\text{(stat.)}\pm 0.43\text{(sys.)}}
\newcommand{\AHIpsConstraintLowerBand}{\AHIps = 3.24^{+6.79}_{-2.11}\text{(stat.)}\pm 0.56\text{(sys.)}}
\shorttitle{Detection of the 21 cm Auto-power Spectrum with CHIME}

\begin{document}

\title{Detection of the Cosmological 21 cm Signal in Auto-correlation at $z \sim 1$ with the Canadian Hydrogen Intensity Mapping Experiment}



\newcommand{\UBC}{Department of Physics and Astronomy, University of British Columbia, Vancouver, BC, Canada}
\newcommand{\MITP} {Department of Physics, Massachusetts Institute of Technology, Cambridge, MA, USA}
\newcommand{\MITK} {MIT Kavli Institute for Astrophysics and Space Research, Massachusetts Institute of Technology, Cambridge, MA, USA}
\newcommand{\TRU}{Department of Physical Sciences, Thompson Rivers University, Kamloops, BC, Canada}
\newcommand{\PI}{Perimeter Institute for Theoretical Physics, Waterloo, ON, Canada}
\newcommand{\DRAO}{Dominion Radio Astrophysical Observatory, Herzberg Astronomy \& Astrophysics Research Centre, National Research Council Canada, Penticton, BC, Canada}
\newcommand{\UBCO}{Department of Computer Science, Math, Physics, and Statistics, University of British Columbia-Okanagan, Kelowna, BC, Canada}
\newcommand{\McGill}{Department of Physics, McGill University, Montréal, QC, Canada}
\newcommand{\MU}{\McGill}
\newcommand{\TSI}{Trottier Space Institute, McGill University, 3550 rue University, Montréal, QC H3A 2A7, Canada}
\newcommand{\UofTastro}{David A.\ Dunlap Department of Astronomy \& Astrophysics, University of Toronto, Toronto, ON, Canada}
\newcommand{\UTA}{\UofTastro}
\newcommand{\UofTphys}{Department of Physics, University of Toronto, Toronto, ON, Canada}
\newcommand{\WVU} {Department of Computer Science and Electrical Engineering, West Virginia University, Morgantown WV, USA}
\newcommand{\WVUA} {Department of Physics and Astronomy, West Virginia University, Morgantown, WV, USA}
\newcommand{\WVUGWAC} {Center for Gravitational Waves and Cosmology, West Virginia University, Morgantown, WV, USA}
\newcommand{\Yale}{Department of Physics, Yale University, New Haven, CT, USA}
\newcommand{\YUP}{\Yale}
\newcommand{\YaleA}{Department of Astronomy, Yale University, New Haven, CT, USA}
\newcommand{\Dunlap}{Dunlap Institute for Astronomy and Astrophysics, University of Toronto, Toronto, ON, Canada}
\newcommand{\DIAA}{\Dunlap}
\newcommand{\RRI}{Raman Research Institute, Sadashivanagar,   Bengaluru, India}
\newcommand{\ASIAA}{Institute of Astronomy and Astrophysics, Academia Sinica, Taipei, Taiwan}
\newcommand{\CITA}{Canadian Institute for Theoretical Astrophysics, Toronto, ON, Canada}
\newcommand{\CIFAR}{Canadian Institute for Advanced Research,  Toronto, ON, Canada}
\newcommand{\WVUphysastro}{Department of Physics and Astronomy, West Virginia University, Morgantown, WV, USA}
\newcommand{\KIPAC}{Kavli Institute for Particle Astrophysics and Cosmology, Stanford, CA 94305, USA}
\newcommand{\SLAC}{SLAC National Accelerator Laboratory; Menlo Park, CA 94025; USA}
\newcommand{\ASU}{Department of Physics, Arizona State University, Tempe, AZ 85287, USA}
\newcommand{\RUG}{Kapteyn Astronomical Institute, University of Groningen, PO Box 800, 9700 AV Groningen, The Netherlands}
\newcommand{\ASTRON}{ASTRON, The Netherlands Institute for Radio Astronomy, Oude Hoogeveensedijk 4, Dwingeloo, 7991 PD, The Netherlands}
\newcommand{\UWO}{Department of Physics \& Astronomy, University of Western Ontario, 1151 Richmond Street, London, ON, N6A 3K7, Canada}

\shortauthors{CHIME Collaboration}
\collaboration{100}{The CHIME Collaboration:}

\author[0000-0001-6523-9029]{Mandana Amiri}
    \email{mandana@phas.ubc.ca}
    \affiliation{\UBC}

\author[0000-0003-3772-2798]{Kevin Bandura}
    \email{kevin.bandura@mail.wvu.edu}
    \affiliation{\WVU}
    \affiliation{\WVUGWAC}

\author[0000-0002-7758-9859]{Arnab Chakraborty}
    \email{arnab.chakraborty2@mail.mcgill.ca}
    \affiliation{\MU}
    \affiliation{\TSI}

\author[0000-0001-6509-8430]{Jean-François Cliche}
    \email{jfcliche@jfcliche.com}
    \affiliation{\MU}
    \affiliation{\TSI}

\author[0000-0001-7166-6422]{Matt Dobbs}
    \email{matt.dobbs@mcgill.ca}
    \affiliation{\MU}
    \affiliation{\TSI}

\author[0000-0002-0190-2271]{Simon Foreman}
    \email{simon.foreman@asu.edu}
    \affiliation{\ASU}

\author[0000-0003-3986-954X]{Liam Gray}
    \email{liam.gray@ubc.ca}
    \affiliation{\UBC}

\author[0000-0002-1760-0868]{Mark Halpern}
    \email{halpern@physics.ubc.ca}
    \affiliation{\UBC}

\author[0000-0001-7301-5666]{Alex S Hill}
    \email{alex.hill@ubc.ca}
    \affiliation{\UBCO}
    \affiliation{\DRAO}

\author[0000-0002-4241-8320]{Gary Hinshaw}
    \email{hinshaw@phas.ubc.ca}
    \affiliation{\UBC}

\author[0000-0003-4887-8114]{Carolin Höfer}
    \email{hofer@astro.rug.nl}
    \affiliation{\RUG}
    \affiliation{\UBC}

\author[0000-0003-4179-4073]{Albin Joseph}
    \email{ajosep52@asu.edu}
    \affiliation{\ASU}

\author[0009-0004-2241-0550]{Nolan Kruger}
    \email{nckruger@asu.edu}
    \affiliation{\ASU}

\author[0000-0003-1455-2546]{T.L. Landecker}
    \email{tom.landecker.drao@gmail.com}
    \affiliation{\DRAO}

\author[0000-0003-1455-2546]{Rik van Lieshout}
    \email{lieshout@astron.nl}
    \affiliation{\CITA}
    \affiliation{\ASTRON}
    
\author[0000-0001-8064-6116]{Joshua MacEachern}
    \email{joshua.maceachern@nrc-cnrc.gc.ca}
    \affiliation{\DRAO}

\author[0000-0002-4279-6946]{Kiyoshi W. Masui}
    \email{kmasui@mit.edu}
    \affiliation{\MITK}
    \affiliation{\MITP}

\author[0000-0002-0772-9326]{Juan Mena-Parra}
    \email{juan.menaparra@utoronto.ca}
    \affiliation{\DIAA}
    \affiliation{\UTA}

\author[0009-0005-2291-510X]{Kyle Miller}
    \email{ckmiller@berkeley.edu}
    \affiliation{\MU}

\author[0000-0001-8292-0051]{Nikola Milutinovic}
    \email{milni@astro.ubc.ca}
    \affiliation{\UBC}

\author[0000-0002-2626-5985]{Arash Mirhosseini}
    \email{arashmirhosseini@phas.ubc.ca}
    \affiliation{\UBC}

\author[0000-0002-7333-5552]{Laura Newburgh}
    \email{laura.newburgh@yale.edu}
    \affiliation{\YUP}

\author[0000-0002-2465-8937]{Anna Ordog}
    \email{aordog@uwo.ca}
    \affiliation{\UBCO}
    \affiliation{\DRAO}
    \affiliation{\UWO}

\author[0000-0003-2155-9578]{Ue-Li Pen}
    \email{pen@cita.utoronto.ca}
    \affiliation{\DIAA}
    \affiliation{\ASIAA}
    \affiliation{\CITA}
    \affiliation{\CIFAR}
    \affiliation{\PI}

\author[0000-0002-9516-3245]{Tristan Pinsonneault-Marotte}
    \email{tristpm@stanford.edu}
    \affiliation{\KIPAC}
    \affiliation{\SLAC}

\author[0000-0001-6967-7253]{Alex Reda}
    \email{alex.reda@yale.edu}
    \affiliation{\YUP}

\author[0000-0003-3463-7918]{Andre Renard}
    \email{andre@renard.io}
    \affiliation{\DIAA}

\author[0000-0001-5667-8118]{Kana Sakaguri}
    \email{kana.sakaguri@yale.edu}
    \affiliation{\YUP}

\author[orcid=0000-0002-4543-4588]{J. Richard Shaw}
      \email{jrichardshaw@gmail.com}
     \affiliation{\UBC}

\author[0000-0001-6731-0351]{Shabbir Shaikh}
    \email{sshaik14@asu.edu}
    \affiliation{\ASU}
    
\author[0000-0003-2631-6217]{Seth R. Siegel}
    \email{ssiegel@perimeterinstitute.ca}
    \affiliation{\PI}
    \affiliation{\MU}
    \affiliation{\TSI}

\author[0000-0001-7755-902X]{Saurabh Singh}
    \email{saurabhs@rri.res.in}
    \affiliation{\RRI}

\author[0009-0003-6054-8035]{David Spear}
    \email{david.spear@ubc.ca}
    \affiliation{\UBC}
    \affiliation{\DRAO}
    
\author[0009-0003-4114-1301]{Yukari Uchibori}
    \email{yukariu@phas.ubc.ca}
    \affiliation{\UBC}

\author[0000-0003-4535-9378]{Keith Vanderlinde}
    \email{keith.vanderlinde@utoronto.ca}
    \affiliation{\UTA}
    \affiliation{\DIAA}

\author[0000-0002-1491-3738]{Haochen Wang}
    \email{hcwang96@mit.edu}
    \affiliation{\MITP}
    \affiliation{\MITK}

\author[0000-0002-6669-3159]{Donald V. Wiebe}
    \email{dvw@phas.ubc.ca}
    \affiliation{\UBC}

\author[0000-0001-7314-9496]{Dallas Wulf}
    \email{dallas.wulf@mcgill.ca}
    \affiliation{\MU}
    \affiliation{\TSI}

\correspondingauthor{Arnab Chakraborty,  Seth R.~Siegel}
\email{arnab.chakraborty2@mail.mcgill.ca \quad  sethrsiegel@gmail.com}

\begin{abstract}

We present the first detection of the cosmological \tcm intensity mapping signal in auto-correlation at $z \sim 1$ with the Canadian Hydrogen Intensity Mapping Experiment (CHIME).
Using 94 nights of observation, we have measured the \tcm auto-power spectrum over a frequency range from \SIrange{608.2}{707.8}{\mega\hertz} ($z=1.34$ to $1.01$) at $0.4\ihMpc \lesssim k \lesssim 1.5\ihMpc$, with {a detection significance of $12.4\sigma$}. Our analysis employs significant improvements to the CHIME data processing pipeline compared to previous work, including novel radio frequency interference (RFI) detection and masking algorithms, achromatic beamforming techniques, and foreground filtering before time averaging to minimize spectral leakage. We establish the robustness and reliability of our detection through a comprehensive suite of validation tests. We also measure the \tcm signal in two independent sub-bands centered at $z \sim 1.08$ and $z \sim 1.24$ with {detection significance of $8.6\sigma$ and $9.1\sigma$, respectively.} We briefly discuss the theoretical interpretation of these measurements in terms of a power spectrum model, deferring the details to a companion paper.
This auto-power spectrum detection demonstrates CHIME's capability to probe large-scale structure through \tcm intensity mapping without reliance on external galaxy surveys.

\end{abstract}


\section{Introduction} 

The technique of \textit{\tcm intensity mapping} seeks to measure the clustering of atomic hydrogen (\hi) over a wide span of redshift and cosmic volume. Hyperfine transitions in hydrogen atoms are associated with photons having a rest wavelength of $21.106\,{\rm cm}$ (corresponding to $1420.406\,{\rm MHz}$ in frequency), and the cosmic abundance of {\hi} is sufficiently high that radio telescopes can make large-scale maps of the emission or absorption of these photons. The fluctuations in these maps can provide a wealth of information about the behavior of {\hi} in galaxies and the intergalactic medium, along with the background cosmological model.

\tcm intensity mapping can probe the cosmic ``Dark Ages" prior to the formation of the first luminous objects, and instrument concepts for making observations at these high redshifts are currently being explored \citep[e.g.][]{bale2023-lusee,chen2024-laraf,polidan2024-farview,brinkerink2025-dex}. Intensity mapping observations during the cosmic dawn and epoch of reionization can provide insights into the nature of the ionizing sources and the details of how reionization occurred, and steady progress toward a detection is being made by several teams \citep[e.g.][]{abdurashidova2022-hera,mertens2025-lofar,ceccotti2025-lofar,trott2025-mwa}.

In the post-reionization era, most of the {\hi} mass is associated with galaxies, and therefore \tcm intensity maps in this epoch primarily probe the distribution of these galaxies. 
As a result, intensity mapping surveys at these redshifts can provide large-scale structure measurements that are complementary to those from optical galaxy surveys, due to the significant differences in observing technologies and observational systematics at optical and radio frequencies. Furthermore, because intensity mapping directly measures the distribution of {\hi} that fuels star formation in galaxies, it can shed light on how galaxies and their environments evolve through cosmic time.

On the other hand, \tcm intensity mapping observations have proven to be very challenging in practice. At all relevant observing frequencies, Galactic and extragalactic radio foregrounds are several orders of magnitude larger than the cosmological signal. This large dynamic range sets stringent requirements on instrument characterization, removal of radio frequency interference (RFI), mitigation of systematics, removal of foregrounds, and overall robustness of data processing methods (for recent reviews, see e.g.\ \citealt{ansari2018}, \citealt{ahmed2019-RDreview}, and \citealt{Liu&Shaw2020}).

Most \tcm intensity mapping analyses thus far have taken the form of cross-correlations with previous galaxy survey data. After the initial detections using HIPASS \citep{pen2009} and the Green Bank Telescope \citep{chang2010}, several others have followed, at redshifts between $0.05$ and~$2.3$ \citep{masui2013,anderson2018,tramonte2020,li2021-parkesxwigglez,wolz2021,chimestacking,cunnington2023-meerkat,chime-lymanalpha,carucci2024-meerkat,chen2025-meerkat}. 
In addition, \citet{paul2023-21cmauto} have claimed a detection of the \tcm auto power spectrum at $z=0.32$ and $0.44$ at $8.0\sigma$ and $11.5\sigma$ respectively, over the wavenumber range $0.3\,{\rm Mpc^{-1}} \lesssim k < 8\,{\rm Mpc^{-1}}$, using data collected by the MeerKAT telescope.

In this paper, we present a measurement of the \tcm auto power spectrum at a mean redshift of $1.16$, using data from the Canadian Hydrogen Intensity Mapping Experiment (CHIME). CHIME is an interferometric radio telescope located at the Dominion Radio Astrophysical Observatory in British Columbia, Canada. CHIME uses four cylindrical reflectors, each $20\,{\rm m}$ wide and $100\,{\rm m}$ long, to focus incoming radiation onto 256 dual-polarized feeds located along the focal lines of each cylinder. CHIME observes in the $400$-$800\,{\rm MHz}$ band and is pursuing several science goals, including intensity mapping, fast radio bursts \citep{FRB2018}, pulsars \citep{CHIMEPulsar:2021}, Galactic magnetism \citep{chime-tadpole}, and \tcm absorption systems \citep{chime-absorber}. \citet{chimeoverview} describes the telescope design and subsystems in more detail.

In \citet{chimestacking}, 102 nights of CHIME data were used to perform spectral stacking of radio sky maps on luminous red galaxies, emission-line galaxies, and quasars from the extended Baryon Oscillation Spectroscopic Survey (eBOSS), yielding detections with signal-to-noise ratios of $7.1$, $5.7$, and $11.1$. In \citet{chime-lymanalpha}, a reprocessed version of the same dataset was used to measure the cross-correlation with Lyman-$\alpha$ forest measurements from eBOSS, at a mean redshift of $2.3$ and with $9\sigma$ significance. Since those publications, we have made numerous improvements in our data-processing pipeline, including the identification of RFI and the mitigation of other sources of spurious spectral structure.

These improvements have enabled CHIME's first measurement of the \tcm intensity mapping signal that does not rely on external large-scale structure data. Using 94 nights of data, we have measured the \tcm auto power spectrum over a band from \SIrange{608.2}{707.8}{\mega\hertz} ($z=1.34$ to $1.01$) at $0.4\ihMpc \lesssim k \lesssim 1.5\ihMpc$, with {signal-to-noise ratio $12.4$.} 

We have verified that the measurements are consistent across several splits of the data and variations of our analysis choices.
We have also separately measured the power spectrum in each of two halves of this frequency band, {with ${\rm S/N}=9.1$ for \SIrange{608.2}{658.2}{\mega\hertz} and ${\rm S/N}=8.6$ for \SIrange{658.2}{707.8}{\mega\hertz}.}

Our companion paper \citep{chime-auto-interpretation-paper} describes our interpretation of these measurements in terms of the large-scale clustering of {\hi}, but we briefly state our approach here. We use a power spectrum model that depends on the mean density of {\hi}, $\OmegaHI(z)$, and the linear bias of {\hi}, $\bHI(z)$, along with parameters that describe the shape of the power spectrum at the (nonlinear) scales of our measurement. We constrain a degenerate combination of $\OmegaHI$ and $\bHI$, defined as
$\AHIps = 10^6 \OmegaHI^2(\bHI^2+\fmu)^2$,
at the mean redshift of our measurement, where $\fmu$ is a factor capturing the angular average of the Kaiser redshift-space distortion term over the wavenumbers we have measured.
In addition, we compare the measured \tcm power spectrum with that predicted by the distribution of {\hi} in the IllustrisTNG simulations \citep{TNGa,TNGb,TNGc,TNGd,TNGe}.

This paper is organized as follows:
\begin{itemize}
\item In \secref{sec:instrument}, we introduce our conventions for coordinates and visibilities, and describe the dataset used in this work.
\item In \secref{sec:pipeline}, we discuss the data processing pipeline used in this work, focusing on improvements over the pipelines used in previous CHIME cross-correlation studies \citep{chimestacking,chime-lymanalpha}. In particular, we highlight our methods for 
RFI flagging (\secref{sec:rfi} and \secref{sec:residualrfi}), 
foreground removal (\secref{sec:achromatic_pipeline}), 
correcting for residual bandpass-calibration errors (\secref{sec:hyfores}), 
time averaging (\secref{sec:time_averaging}),
mapmaking (\secref{sec:mapmaking}),
and forming pseudo-Stokes maps (\secref{sec:stokesIQ}).
We also describe the identification of several new candidate \hi\ \tcm absorption systems, and how they are masked in this analysis (\secref{sec:absorbers}). 
\item In \secref{sec:stacking}, we repeat the spectral stacking analysis of \citet{chimestacking} using eBOSS quasars and CHIME data processed as in \secref{sec:pipeline}, in order to validate our pipeline.
\item In \secref{sec:power_spectrum_estimation}, we describe technical aspects of our power spectrum estimation methodology: delay spectrum estimation (\secref{subsec:delayspectrum}), spatial masking (\secref{sec:spatial_mask}), binning and normalization (\secref{subsec:ps_estimation}), and noise covariance estimation (\secref{sec:power_spectrum_estimation:noise_covariance_estimation}).
\item In \secref{sec:results}, we present our main results: the power spectrum measurement in our chosen frequency band and two sub-bands (\secref{sec:power_spectrum}). In \secref{sec:results:theoretical_interpretation}, we briefly describe the theoretical model we use for interpreting the measurement, and in \secref{sec:results:detection_significance} we assess the detection significance. 
\item In \secref{sec:validation}, we describe our suite of validation tests, which demonstrate that our measurement is free of significant contamination from various systematics.
\item In \secref{sec:conclusions}, we conclude and discuss the context of our measurements.
\end{itemize}
Several appendices provide further details about 
our algorithms for RFI detection (Appendix~\ref{app:rfi}), 
masking (Appendix~\ref{app:masking}),
delay spectrum estimation (Appendix~\ref{app:varying_prior}), and
estimating detection significance (Appendix~\ref{app:detection_significance}).

For computations requiring a cosmological model, we use cosmological parameters from the final {\em Planck} data release (specifically, the ``TT,TE,EE+lowE+lensing+BAO" parameters from Table 2 of \citealt{planck2020}).


\section{Instrument and Observations}
\label{sec:instrument}

A full description of the CHIME instrument and its real-time pipeline is given in \citet{chimeoverview}, and the offline processing used in earlier cosmology work is described in \citet{chimestacking}. Here we introduce the array geometry, coordinate conventions, and visibility notation that will be used in this work, along with the subset of CHIME observations included in this analysis.

\subsection{Array Layout}
\label{sec:array-layout}

CHIME consists of four cylindrical reflectors separated in the east–west (\ew) direction by $\Delta_{\mathrm{EW}} = \SI{22}{\meter}$, each equipped with 256 dual-polarization feeds along its north–south (\ns) focal line with spacing $\Delta_{\mathrm{NS}} = \SI{0.3048}{\meter}$.  The center of the array is located at geodetic latitude and longitude
\[
(\Lambda,\Phi) = (\SI{49.320709}{\degree},\, \SI{-119.623677}{\degree}) .
\]
To describe baselines and sky directions we work on the plane tangent to the Earth at this location and use a local Cartesian basis $(\mathbf{\hat{x}}, \mathbf{\hat{y}},\mathbf{\hat{z}})$, with $\mathbf{\hat{x}}$ pointing due east on the tangent plane, $\mathbf{\hat{y}}$ pointing due north, and $\mathbf{\hat{z}}$ pointing to the local zenith.

Indexing the designed CHIME baseline grid by an integer pair $(e,n)$, we write
\begin{equation}
    \vec{b}^{(0)}_{en}
    = e\,\Delta_{\mathrm{EW}}\,\mathbf{\hat{x}} + n\,\Delta_{\mathrm{NS}}\,\mathbf{\hat{y}},
    \label{eq:designed-baseline}
\end{equation}
with $e \in [-3,3]$ and $n \in [-255,255]$ for the nominal array.\footnote{Note that in \cite{chimestacking}, we used $(x,y)$ instead of $(e,n)$ as baseline indices, while in this work, we instead use $(x,y)$ for the telescope coordinates described in \secref{sec:coordinate-conventions}. In \cite{chimestacking}, we also used $(d_x,d_y)$ instead of $(\Delta_{\mathrm{EW}}, \Delta_{\mathrm{NS}})$ for cylinder and feed spacings.}
As discussed in \cite{chimestacking}, early analyses modeled small departures from this designed layout as a rotation of the cylinder frame with respect to true north. Subsequent solar \citep{dallassolar} and holographic \citep{alexholo} calibration show that these departures are more simply described as a systematic shift in the \ns position of the feeds with cylinder. We therefore use an effective baseline
\begin{equation}
    \vec{b}_{en}
    = e\,\Delta_{\mathrm{EW}}\,\mathbf{\hat{x}}
      + \bigl(n\,\Delta_{\mathrm{NS}} + \kappa\, e\,\Delta_{\mathrm{EW}}\bigr)\mathbf{\hat{y}},
    \label{eq:shifted-baseline}
\end{equation}
where $\kappa \simeq -0.00124$ is a small, empirically determined slope giving the \ns shift per unit \ew separation.

\subsection{Coordinate Conventions}
\label{sec:coordinate-conventions}

We represent sky positions by the direction cosines of the unit sky vector $\mathbf{\hat{n}}$ in the telescope frame $(\mathbf{\hat{x}},\mathbf{\hat{y}},\mathbf{\hat{z}})$. These are conventionally written $(l,m,n)$, but to prevent confusion with the harmonic index $m$ used later, we write
\[
(x,y,z) \equiv \mathbf{\hat{n}}\cdot(\mathbf{\hat{x}},\mathbf{\hat{y}},\mathbf{\hat{z}}),
\]
and refer to $(x,y,z)$ as the telescope coordinates.

We denote the local Earth Rotation Angle (ERA) by $\phi$. This quantity replaces local sidereal time and is equivalent to the current right ascension of the local meridian in the Celestial Intermediate Reference System (CIRS).

For a point source at right ascension $\alpha$ and declination $\theta$, the hour angle is given by $\mathrm{HA} = \phi - \alpha$, and the telescope-frame coordinates are
\begin{align}
    x &= -\cos\theta \,\sin\mathrm{HA}, \label{eq:tel-x} \\
    y &= \cos\Lambda \,\sin\theta - \sin\Lambda \,\cos\theta \,\cos\mathrm{HA}, \label{eq:tel-y} \\
    z &= \sin\Lambda \,\sin\theta + \cos\Lambda \,\cos\theta \,\cos\mathrm{HA}. \label{eq:tel-z}
\end{align}
which satisfy $x^2 + y^2 + z^2 = 1$.

\subsection{Visibility Data Product}

The CHIME correlator computes the $N^2$ visibilities at a cadence of $\dtraw$ for $N_{\nu}=1024$ frequency channels spanning \SIrange{400}{800}{\mega\hertz}. We apply complex gain calibration and average all redundant copies of a baseline to reduce the data rate by a factor of $\sim 300$, yielding the archived visibilities, which we denote $V_{p f e n t}$.  Here $p \in \{\mathrm{XX}, \mathrm{XY}, \mathrm{YX}, \mathrm{YY}\}$ is the polarization, $f$ indexes frequency channels (with center frequency $\nu_f$), $(e,n)$ label the \ew and \ns baseline indices on the CHIME grid, and $t$ indexes the time.  In this work we consider only the co-polar visibilities $p \in \{\mathrm{XX}, \mathrm{YY}\}$. Conjugate symmetry then implies
\[
V_{p f, -e, -n, t} = \bigl(V_{p f e n t}\bigr)^{*}.
\]

The measured visibilities are the sum of contributions from the sky and from noise. The Stokes-$I$  contribution from the sky to observed polarization $p$ is
\begin{align}
    \label{eq:meas_eq}
\mathcal{V}_{p f e n t}
    = & \int d\phi' \int d\theta'\, \cos\theta' \ I(\nu_f, \theta', \phi') \nonumber \\
 & \times \bigl|A_{p}(\nu_f, \theta', \phi_t - \phi')\bigr|^{2} \nonumber \\
 & \times \exp\bigl\{ j 2 \pi \nu_{f} \ \vec{b}_{en} \cdot \mathbf{\hat{n}}(\theta', \phi_t - \phi') / c \bigr\},
\end{align}
where $I$ is the sky intensity, $A_p$ is the effective voltage primary beam pattern (assumed common to feeds of the same polarization), $c$ is the speed of light, and $\phi_t$ is the local ERA at time $t$.

\subsection{Observations}
\label{subsec:data_selection}

In this work we analyze 102 sidereal days drawn from the period between \datestart and \dateend. These 102 days are the same subset identified and used in \citet{chimestacking}, where they were selected on the basis of the data-quality metrics described in that work.

Because auto power spectrum estimation is more sensitive to residual systematics than the cross-correlation analysis presented in \citet{chimestacking}, we revalidate each of the 102 sidereal days using several independent diagnostics, including the delay–domain power spectrum of the visibilities, $\chi^{2}$ statistics after RFI masking (see Appendix~\ref{app:rfi_spec_filt}), the total fraction of data masked, and sky maps from individual days at a few representative frequencies. This screening identifies eight days that fail one or more of these checks, and we exclude them. Our final dataset therefore consists of 94 well-characterized sidereal days from 2019.  Note that we use only the portions of each sidereal day when the Sun is below the horizon to avoid contamination from solar emission.

CHIME has been operating continuously since 2019 and now has nearly seven years of archival data. Here we focus on this 2019 subset in order to establish and validate the methodology.  Analysis of the full archival dataset will be presented in future work.

\section{Data Processing Pipeline} 
\label{sec:pipeline}

In this section we describe the modifications to the offline data-processing pipeline of \citet{chimestacking} that enable a detection of the \tcm auto power spectrum.

\subsection{Radio Frequency Interference Detection and Masking}
\label{sec:rfi}

We have made substantial improvements to the detection and masking of human-made RFI. The offline algorithms described in \cite{chimestacking} are no longer used; instead, we have developed three new methods, which differ primarily in how they subtract the background sky signal in order to isolate RFI, and in how they combine information across baselines. By averaging over baselines with appropriate weighting, the algorithms enhance sensitivity to weak interference that would not be detectable in individual visibilities.  A detailed description of each algorithm is presented in Appendix~\ref{app:rfi}; here we summarize the main features.

\subsubsection{Radiometer Test Method}
\label{sec:radiometer_test}
The first method flags frequencies and times where the measured variance of the visibilities deviates from that expected for Gaussian, uncorrelated noise, and excludes them from further analysis.

For each frequency channel $f$, coarse-resolution time~$t$, and baseline formed from feeds $a$ and $b$, the real-time software measures the variance of the visibility, $\sigma^{2}_{fabt}$, from the sum of squared differences between even and odd \SI{30}{\milli\second} cadence samples within each \dtraw window. This timescale is chosen so that the sky contribution is approximately constant and cancels in the even–odd difference, leaving only noise and RFI. 

The expected variance is computed from the autocorrelations of the two feeds and the number of integrated samples under the assumption of Gaussian, uncorrelated noise. Both the measured and expected variances are then averaged over all baselines within a given polarization product $p$ to improve sensitivity to low-level deviations. The square root of the ratio of these baseline-averaged values defines the radiometer test statistic $\radiometer_{pft}$ (see Appendix~\ref{app:rfi_radiometer} for details). Under the Gaussian, uncorrelated-noise assumption, $\radiometer_{pft}$ has expectation value equal to unity; deviations from this expectation indicate inconsistency with the noise model and are used to flag contaminated data.

Outliers in $\radiometer_{pft}$ are identified iteratively as a function of frequency and time for each polarization. A slowly varying background level of $\radiometer_{pft}$ is estimated using a robust two-dimensional median filter and subtracted to remove small, systematic variations with frequency and time. The local scatter of the residuals is estimated with a median absolute deviation (MAD) filter.  RFI events are identified using a combination of MAD thresholding for isolated outliers and a custom implementation of the \texttt{SumThreshold} algorithm \citep{rfisumthresh}, which is effective at masking events that are extended in time or frequency.  The mask, background, and scatter estimates are updated over a fixed number of iterations with progressively lower thresholds to improve sensitivity to low-level contamination. Finally, the polarization-dependent masks are combined with a logical OR across polarizations to produce a single frequency--time mask, $Q^{\mathrm{rad}}_{ft}$, where $Q^{\mathrm{rad}}_{ft}=0$ indicates good data and $Q^{\mathrm{rad}}_{ft}=1$ indicates masked data.

We adopt an aggressive masking strategy and prefer to discard marginally contaminated data rather than risk residual RFI. However, we monitor the mask as a function of local ERA to ensure that flagging is not systematically correlated with the sky signal, thereby introducing holes in the sky coverage when averaging over many sidereal days. A more detailed optimization of thresholds is left for future work.

\subsubsection{Fringe-rate Filtering Method}
\label{sec:fringe_rate_filtering}
The second method targets transient RFI by exploiting the fact that emission from steady celestial sources fringes at a well-defined rate set by the Earth’s rotation and the array geometry. For a given frequency, there is a \emph{maximum fringe rate} set by the longest east--west (\ew) baseline and a source at the celestial equator. Any power at higher fringe rates cannot arise from steady signals from the sky and must correspond to transient signals, whether astrophysical or human-made \citep{PArsons_frf_2009,Offringa_filt_2012,shaw2014}.  We apply a high-pass filter in time with a cutoff set by this maximum fringe rate to suppress the celestial contribution. The residual contains only transient signals which we interpret as RFI. We mask data with high-pass residuals above a threshold.

Importantly, this method is applied independently of the radiometer mask described in the previous section; no prior masking is applied at this stage.  To reduce data volume, the $\mathrm{XX}$ and $\mathrm{YY}$ polarization pairs are summed for each baseline. A high-pass filter is then applied along the time axis by subtracting a flattop-convolved copy of the data. The filtered data are beamformed to enhance sensitivity to localized RFI, producing a (frequency, beam, time) cube. For each frequency and beam, the local scatter is estimated over time using a MAD filter and the data are normalized by this scatter. Hysteresis thresholding is then applied in (beam, time) space: a pixel is flagged if it exceeds a high threshold, or if it exceeds a lower threshold and is connected to higher-valued pixels. This approach captures the typical triangular time profiles of transient RFI, where samples at either edge of an event have lower power than the peak, while limiting false positives. A sample is flagged if the fraction of beams identified as outliers exceeds 1\% for that frequency and time, yielding the final mask.

Among other types of transient RFI, this method is particularly effective at identifying distant broadcast television stations briefly reflecting off objects in the sky, such as airplanes, satellites, or meteor ionization trails. These signals often appear approximately Gaussian in time and can evade the radiometer test described in the previous section.

A minimal secondary stage targets persistent narrowband RFI by exploiting a complementary principle. On long \ew baselines, sources fringe rapidly unless they are near the north celestial pole. Applying a low-pass filter in time therefore suppresses the sky contribution in these long \ew baselines while isolating persistent sources of RFI. In practice, we apply this stage to baselines formed from feeds separated by two or three cylinders in the \ew direction. A 30-minute low-pass filter is applied in time, and the visibilities are then averaged over these baselines to improve sensitivity to faint signals. A median-filtered version of the data along frequency is subtracted to isolate narrowband structure. The \texttt{SumThreshold} algorithm is applied in time to detect outliers, and the resulting mask is expanded using a scale-invariant rank operator \citep{rfisir}. While this stage typically flags a small fraction of data, it is effective at capturing weak, narrowband, persistent RFI that is not detected by other methods.  The masks from the transient and persistent stages are combined to form a single frequency--time mask, $Q^{\mathrm{t\text{-}filt}}_{ft}$.

\subsubsection{Spectral Filtering Method}

The delay domain, obtained by Fourier transforming the visibility data along the frequency axis, provides a natural basis for separating the sky signal from many types of RFI \citep{Thompson_2001, PArsons_frf_2009}.  We use this to generate our third RFI filter.  The sky signal is dominated by astrophysical foreground emission, which is spectrally smooth and therefore confined to low-delay modes, whereas RFI typically occupies narrow spectral bands with sharp band edges and thus has power at high delay.  Applying an aggressive high-pass filter in the delay domain removes the spectrally smooth foreground emission and can be used to isolate residual contamination.

For each time and polarization, a high-pass delay filter is constructed using the DAYENU method \citep{ewall-wice2021}, which accounts for frequencies masked by the previous two methods and suppresses delays below the geometric horizon limit plus a buffer. The filter is applied to the visibilities, while the noise variances are propagated through the filter to account for correlations introduced by filtering. A $\chi^2$ per degree of freedom test statistic, $\tilde{\chi}^2_{ft}$, is formed for each frequency and time by taking the sum across baselines of the squared magnitude of the filtered visibilities, weighted by the inverse noise variance, and then dividing by the number of baselines. Under the assumption of thermal noise, this statistic has unit mean and a known variance that decreases with the number of baselines, making it a sensitive metric for detecting residual contamination.

Outliers are flagged in two stages.  First, for each frequency channel we take the median $\tilde{\chi}^2$ over the sidereal day and, if it is an outlier, we mask that channel at all times. Second, for each frequency channel we form the time series of $\tilde{\chi}^2$, compute a slowly varying background in time, and mask only those time samples that depart significantly from this background. This procedure is not applied during daytime, bright source transits, nor around the transit of the pulsar B0329+54, where the statistic is biased by leakage of signal from the astrophysical source. The resulting mask, $Q^{\mathrm{f\text{-}filt}}_{ft}$, identifies residual high-delay RFI not captured by the other methods.

\subsubsection{Final Mask Assembly}
\label{sec:final_masking}
\begin{deluxetable}{l c c}
\tablecolumns{3}
\tablewidth{\columnwidth}
\setlength{\tabcolsep}{6pt}
\tabletypesize{\footnotesize}
\tablecaption{Percentage of Night-time Data Masked \label{table:rfi_masked}}
\tablehead{\colhead{Method} & \colhead{400--800 MHz} & \colhead{608.2--707.8 MHz}}
\startdata
Static & 31.6 & 3.5 \\
Radiometer & 13.1 & 11.9 \\
Fringe-rate & 3.7 & 3.8 \\
Radiometer $\wedge$ Fringe-rate\tablenotemark{a} & 14.4 & 17.3 \\
Spectral & 2.8 & 2.2 \\
\textbf{Total} & \textbf{65.6} & \textbf{38.7} \\
\enddata
\tablenotetext{a}{The radiometer test and fringe-rate filtering masks are constructed independently; this row gives the percentage of data masked by both methods. All other rows give the percentage masked exclusively by that method.}
\end{deluxetable}

The masks obtained from the three methods discussed above are combined using a logical OR to produce the global frequency--time mask,
\begin{equation}
    Q_{ft} = Q^{\mathrm{rad}}_{ft} \lor Q^{\mathrm{t\text{-}filt}}_{ft} \lor Q^{\mathrm{f\text{-}filt}}_{ft},
\end{equation}
where $Q_{ft}=0$ for unmasked (good) data and $Q_{ft}=1$ for masked or missing data.  We also define the complementary ``good data'' flag
\begin{equation}
    \label{eq:flags}
    G^{(0)}_{ft} \equiv 1 - Q_{ft},
\end{equation}
so that $G^{(0)}_{ft}=1$ for unmasked (good) data and $G^{(0)}_{ft}=0$ for masked or missing data. The superscript ``(0)'' indicates that this is the initial version of the good-data flag. This serves as the starting point for downstream stages, which may incorporate additional criteria and define updated versions used in subsequent analysis.

\Cref{table:rfi_masked} summarizes the percentage of night-time data excised by each method and by their combination, for both the full CHIME band (\SIrange{400}{800}{\mega\hertz}) and the analysis sub-band (\SIrange{608.2}{707.8}{\mega\hertz}), over the 94 sidereal days described in Section~\ref{subsec:data_selection}. In this table, \emph{Static} denotes a fixed mask that removes persistently contaminated frequency channels \citep{chimeoverview}. The radiometer and fringe-rate filtering masks are constructed independently, and there is substantial overlap in the samples that both methods identify as contaminated. Among the data excised exclusively by one method or the other, the fringe-rate filtering method is particularly effective at the transient, localized RFI described above (e.g., scattered television channels), whereas the radiometer test method captures many short-duration, broadband events and other miscellaneous RFI. The spectral filtering method is applied after these stages and typically excises only a small additional fraction of data.

\subsection{Foreground Removal}
\label{sec:achromatic_pipeline}

We adopt a foreground removal strategy that exploits the distinct spectral behavior of the cosmological \tcm signal and the astrophysical foregrounds. At the radio frequencies considered, foreground emission is primarily Galactic and extragalactic synchrotron radiation, which in total intensity (Stokes $I$) varies smoothly with frequency \citep{diMatteo,oh2003,Costa_2008}. In contrast, the \tcm signal exhibits structure across a broad range of spectral scales. This is because the observed frequency of the \tcm line corresponds to the redshift of the emitting or absorbing hydrogen gas, which in turn sets its approximate comoving distance along the line of sight. The three-dimensional spatial distribution of neutral hydrogen across cosmological volumes therefore manifests as frequency-dependent fluctuations in the observed \tcm signal, with the scale and amplitude of these fluctuations governed by the line-of-sight power spectrum of HI.  By applying a high-pass filter in frequency, we suppress the spectrally smooth foregrounds while retaining the higher line-of-sight ($k_\parallel$) modes of the \tcm signal. While this comes at the cost of discarding the low-$k_\parallel$ modes that are most sensitive to cosmological parameters, it provides a simple framework for isolating the relatively uncontaminated regions of Fourier space where the \tcm signal is expected to dominate.

In practice, the success of this strategy depends critically on how much additional frequency structure is imprinted on otherwise smooth foregrounds by the telescope and the data processing pipeline. Since the foregrounds are several orders of magnitude brighter than the expected \tcm signal, even weak spectral dependence in the instrument or analysis can cause significant leakage, overwhelming the faint cosmological signal.

Instrumental chromaticity in an interferometer arises from several well-understood physical mechanisms. First, the signal from a source in a given sky direction $\hat{\mathbf{n}}$ reaches the two antennas of a baseline $\mathbf{b}$ with a geometric delay $\tau = \mathbf{b} \cdot \hat{\mathbf{n}} \ / \ c$. When these signals are correlated, the contribution of that source to the visibility exhibits a sinusoidal fringe pattern as a function of frequency, with a period inversely proportional to this delay (see \cref{eq:meas_eq}). Longer baselines produce higher-delay (i.e., more rapidly varying) fringes, and the maximum delay is set by a source on the horizon aligned with the baseline vector. This geometric effect confines smooth-spectrum foreground emission to a characteristic region in Fourier space known as the \emph{foreground wedge} \citep{morales2012-wedge,parsons2012-wedge,liu2014-wedge}, whose extent is determined by the horizon delay limit for each baseline.

Additional sources of chromaticity include mutual coupling between nearby antenna elements, which modifies the effective primary beam pattern in a frequency-dependent way \citep{josaitis:2022}, and reflections in the analog signal chain, which introduce sinusoidal ripples in the bandpass \citep{Josaitis_2022,Kern2019}.  While these effects are not negligible, they are all bounded by the physical layout and electrical length scales of the telescope. As such, contamination from instrumental chromaticity is confined to delay space, typically within the wedge and a modest buffer region.

By contrast, spectral artifacts imprinted on the foregrounds during data processing are not so constrained and can easily contaminate all Fourier modes along the line of sight. Minimizing analysis-induced spectral structure is therefore critical for isolating the cosmological signal. In the remainder of this section, we describe the design choices in our pipeline aimed at suppressing spectral features and ensuring that the high-delay power is dominated by \tcm fluctuations rather than analysis artifacts.

\subsubsection{Sub-band Isolation}
\label{sec:sub-band isolation}
We restrict attention to a relatively RFI-free narrow band of data between \SI{608.2}{\mega\hertz} and \SI{707.8}{\mega\hertz}, which we process separately from the rest of the data. This choice serves several purposes. First, it prevents artifacts from possibly contaminated regions of the spectrum from coupling into the frequencies of interest via the foreground filter. Second, when beamforming we intentionally degrade our maps to a common spatial resolution to enforce spectral smoothness (see \secref{sec:beamforming}), so we prefer to limit the loss of resolution to the portion of the band directly relevant for this analysis. Third, the portion of the sky that is free from geometric aliasing in the north--south (\ns) direction decreases with increasing frequency, so treating the band in contiguous, limited segments ensures that each segment is analyzed with a consistent spatial footprint. The remaining portions of the \SIrange{400}{800}{\mega\hertz} band will be addressed in future work.

\subsubsection{Beamforming in the North--South Direction}  
\label{sec:beamforming}

We beamform over \ns baselines to a uniform grid in the telescope $y$ coordinate, indexed by $d \in \left[0, \ \ldots, \ N_y - 1\right]$, with $N_y = 1217$ points equally spaced between $-1$ and +1:
\begin{align}
    \label{eq:beamform_ns}
    \hybrid_{p f e d t} = \sum_{n} W^{\rm NS}_{fn} \ V_{p f e n t} \ \exp\!\bigl\{-j2 \pi \nu_{f} y_{d} \vec{b}_{en} \cdot \mathbf{\hat{y}} / c \bigr\} \ .
\end{align}
We refer to $\hybrid_{p f e d t}$ as the hybrid beamformed visibilities, since they have been beamformed in the \ns direction but not in the \ew direction.  $W^{\rm NS}_{fn}$ is a window function, taken to be the same for all polarizations, \ew baseline separations, and times.  The window function is normalized separately at each frequency to ensure that the synthesized beam will have a unit peak response:
\begin{align}
    \label{eq:normalized_beamforming_weights}
    W^{\rm NS}_{fn} \equiv \frac{w^{\rm NS}_{fn}}{\sum_{n} w^{\rm NS}_{fn}} \ ,
\end{align}
where $w^{\rm NS}_{fn}$ is the unnormalized window function. We propagate the noise variance through the beamforming operation 
\begin{align}
    \label{eq:beamform_ns_weights}
    \hybridnoise^{2}_{p f e t} =  \sum_{n} \left(W^{\rm NS}_{fn}\right)^2 \sigma_{p f e n t}^{2},
\end{align}
where $\sigma_{p f e n t}^{2}$ is the variance of the visibility estimated by differencing even and odd time samples at \SI{30}{\milli\second} cadence. Here we assume noise that is uncorrelated across baselines, such that the $d$–dependent, geometric phase does not change the beamformed variance, and the same $\hybridnoise^{2}_{p f e t}$ is assigned to all grid points $d$.  We will return to this assumption in \secref{sec:power_spectrum_estimation:noise_covariance_estimation}.

Because sightlines near zenith are well sampled by the array in the \ns direction, emission in the $|y|\lesssim 0.4$ region relevant to this work does not suffer from \ns aliasing. By applying a fixed taper $W^{\rm NS}_{fn}$ across the \ns baselines, we form synthesized beams in $y$ with controlled sidelobes, and the long \ns extent of the array gives these beams narrow main lobes. In effect, the beamforming step projects the visibilities onto sky directions labeled by telescope $y_d$: signals arriving from $y_d$ add coherently (their \ns geometric delay is compensated), while signals from other $y$ are significantly attenuated by the synthesized beam. In the original visibility domain the foreground delay extent grows with \ns baseline length; after projection onto the $y_d$ beams, this dependence is removed, so a single, $y$-independent foreground filter is adequate. Directions with $|y|\gtrsim 0.4$ are not well localized in this band and can exhibit frequency-dependent aliasing; we exclude these directions from power-spectrum estimation for this initial analysis.

This localization will later enable targeted masking of bright sources. Because foreground leakage is primarily multiplicative in sky intensity, the dominant contamination comes from a small number of very bright sources as they transit. By masking the corresponding $y_d$ beams at the relevant times we reduce leakage to high delays and limit contamination of the \tcm signal.

Because the \ew baselines provide only sparse coverage of the corresponding spatial Fourier modes, we cannot form \ew beams with localization comparable to the \ns case, so sky emission at large telescope $x$ is observed with a large \ew geometric delay that we do not correct at this stage. However, sky emission at large telescope $x$ is already significantly attenuated by the primary beam response (e.g.\ \citealt{dallassolar,alexholo}). Additional suppression can be obtained with a fringe–rate filter and by masking the very brightest sources in the sidelobes, as discussed below.

\paragraph{Excluding the Longest North--South Baselines}
Although CHIME’s maximum \ns baseline is \SI{77.7}{\meter}, we exclude baselines longer than \SI{73}{\meter} from the previously discussed beamforming stage. These baselines are few in number and primarily connect feeds at the cylinder ends, which have distinct primary beam patterns relative to the interior feeds. Including them would risk distorting the synthesized beam and compromising localization. Moreover, because of their low redundancy, these baselines are vulnerable to disappearing entirely when feeds are masked in the real-time pipeline, which would induce time-dependent variations in the synthesized beam. While our foreground filtering procedure mitigates sensitivity to such variations, we prefer to maintain a synthesized beam that is stable and straightforward to model. Excluding the longest baselines achieves this with negligible loss of sensitivity.

\paragraph{Applying Achromatic Beamforming Weights}

\begin{figure*}
   \centering \includegraphics[width=0.98\linewidth, keepaspectratio, ]{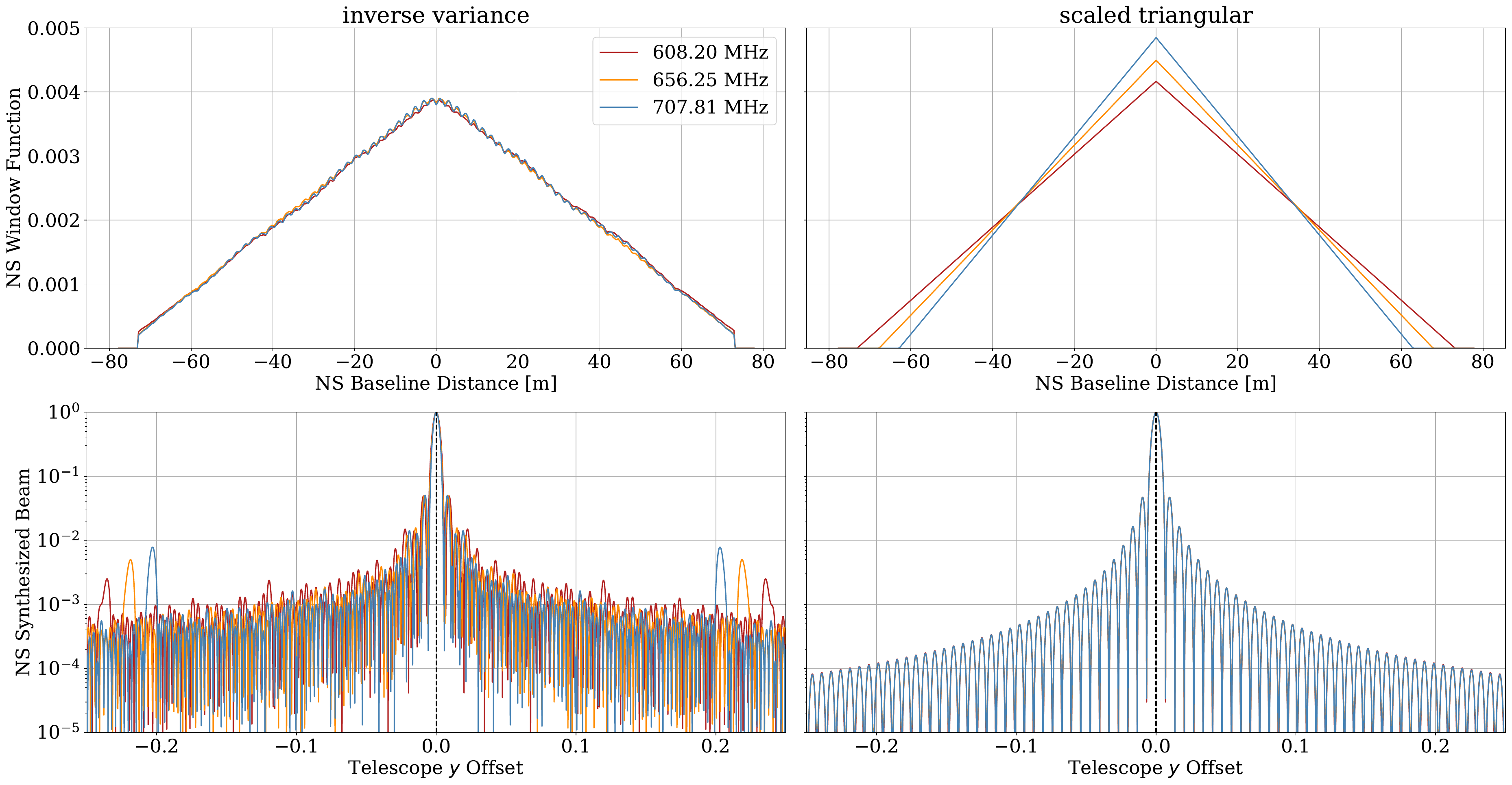}
    \caption{\emph{Top:} Normalized window function, $W^{\rm NS}$, versus \ns baseline distance, shown at three representative frequencies near the bottom, middle, and top of the band. The left column shows the window function used in \cite{chimestacking} which is derived from the inverse-variance weights, while the right column shows the scaled triangular window function adopted in this work.  \emph{Bottom:} Fourier transforms of the normalized window functions, illustrating the expected synthesized beams as a function of offset in the telescope $y$ coordinate. The inverse-variance weights produce a beam that narrows with increasing frequency and exhibits large frequency-dependent sidelobes, in part due to ripple-like structure from noise cross-talk. By contrast, the triangular window is scaled with frequency to maintain an identical synthesized beam across the band.}
    \label{fig:beamforming_weights}
\end{figure*}

We depart from the inverse-variance weighting used in \cite{chimestacking} and instead use a frequency-dependent triangular window versus \ns baseline length that yields an achromatic synthesized beam. These two options are compared in \cref{fig:beamforming_weights}. With inverse-variance weights, the synthesized beam naturally narrows with increasing frequency due to the wavelength dependence of the Fourier transform. The weights can also inherit fine-scale spectral structure from noise cross-talk and residual RFI, imprinting additional artificial chromaticity on the synthesized beam. Our triangular window function is defined as
\begin{equation}
\label{eq:beamforming_weights}
w^{\rm NS}_{fn} =
\begin{cases}
1 - |\rho_{fn}|, & |\rho_{fn}| \le 1 \\
0,               & |\rho_{fn}| > 1
\end{cases}
\end{equation}
with
\begin{equation}
\rho_{fn} \equiv \frac{n}{n_{\max}} \frac{\nu_f}{\nu_{\min}} \, .
\end{equation}
Here $n_{\max} = 239$ is the maximum \ns\ baseline index and $\nu_{\rm min} = \SI{608.2}{\mega\hertz}$ is the minimum frequency considered.  \Cref{eq:beamforming_weights} produces a frequency-independent synthesized beam that approximately matches the inverse-variance scheme at the lowest frequency, but progressively upweights short baselines relative to long baselines at higher frequency to maintain achromaticity. The resulting synthesized response in the $y$ direction (see, e.g., \citealt{masui2019}) is
\begin{align}
    \label{eq:bsynthns}
B^{\rm NS}_{\rm synth}(y - y_{d})
  & \approx \frac{\sin^2\!\left( \pi \,\nu_{\rm min}\, n_{\rm max}\, \Delta_{\rm NS}\, (y - y_{d}) / c \right)}{n_{\rm max}^{2} \sin^2\!\left( \pi \,\nu_{\rm min}\, \Delta_{\rm NS}\, (y - y_{d}) / c \right)}, \notag \\
\end{align}
which has a FWHM of 0.006 in telescope y coordinates, corresponding to roughly $20^\prime$ at zenith.  Crucially, the synthesized response described by \cref{eq:bsynthns} does not vary with frequency.   This approach sacrifices a small amount of point-source sensitivity in favor of enforcing an achromatic synthesized beam and avoiding the introduction of artificial spectral structure into the data prior to foreground filtering.

\subsubsection{Foreground Filtering Before Time Averaging}
\label{sec:fg_filter_timeavg}

\begin{figure}
   \centering \includegraphics[width=0.98\linewidth, keepaspectratio]{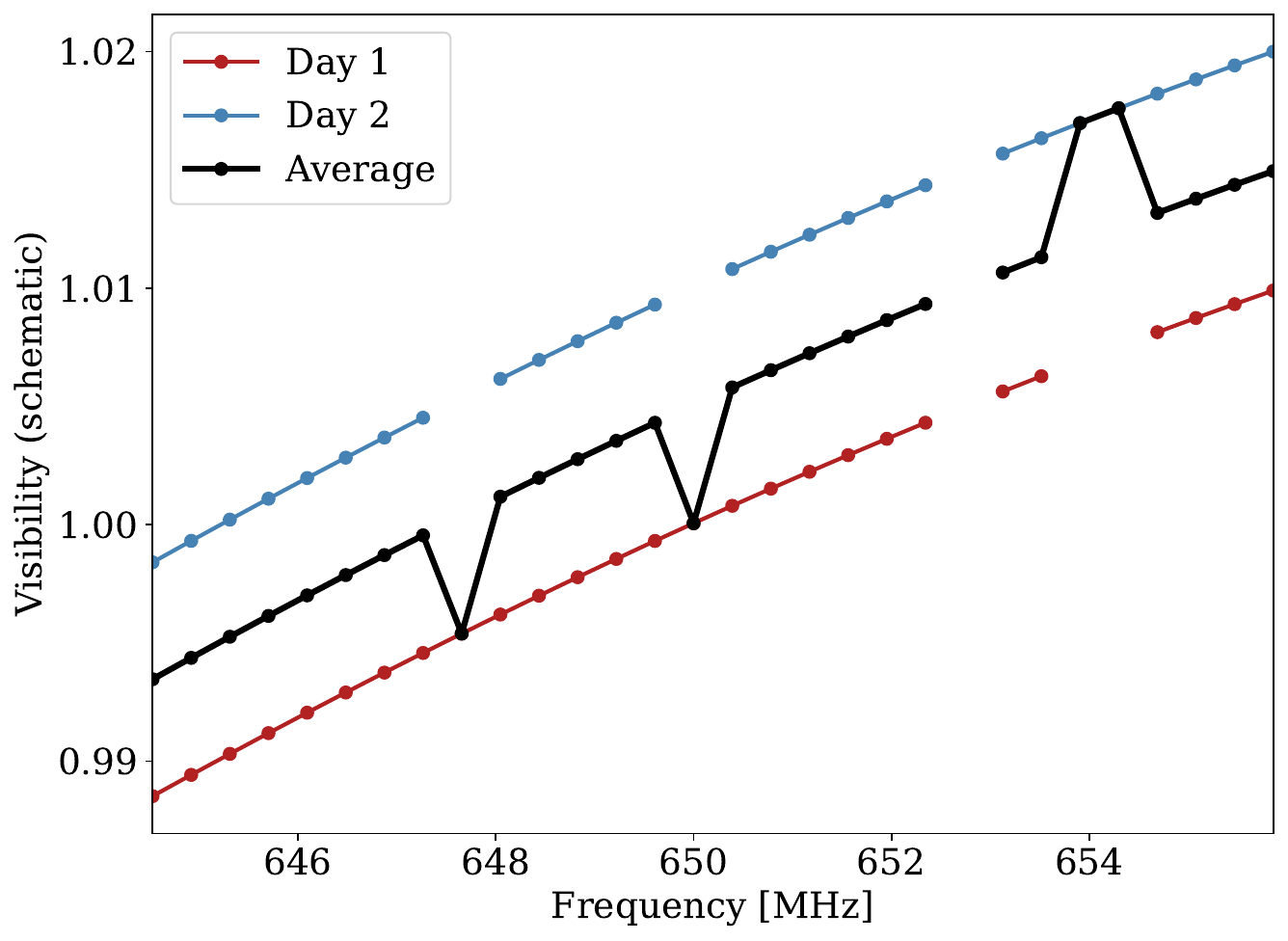}
\caption{Schematic illustration of how spectrally-smooth gain variations can couple to the RFI mask, resulting in leakage of foreground power into high-delay modes. The red and blue curves show the real component of a visibility as a function of frequency over a \SI{10}{\mega\hertz} sub-band for the same local ERA bin on two different days. They differ by an overall $1\%$ multiplicative factor that is approximately constant across this sub-band, representative of gain variations expected in CHIME. Gaps correspond to channels flagged by the RFI mask. The black curve shows the average of the two days, which is no longer spectrally smooth despite each day being smooth individually. This motivates our choice to apply foreground filtering on individual days prior to averaging.}
    \label{fig:demo_gain_mask}
\end{figure}

In addition to the improved RFI masking described in \secref{sec:rfi}, our second major processing change is to apply a foreground filter directly to the \dtraw time series visibilities output by the correlator, prior to any additional time averaging.  Frequency-dependent RFI masks used during time averaging couple to temporal variation in the bright, spectrally smooth foregrounds, imprinting spurious high-delay power (see \cref{fig:demo_gain_mask} for an illustration of this effect). Because these foregrounds are orders of magnitude brighter than the \tcm signal, even small changes over the averaging interval can bias \tcm power-spectrum estimates.  This temporal variation in the foregrounds is caused by Earth’s rotation when rebinning the data to a fixed grid in local ERA, and by day-to-day variability in the instrument response during sidereal-day averaging. To control this effect, we first discard any \dtraw integration for which the real-time system reports more than $2$\% of samples missing or masked, and then apply a per-integration foreground filter along the frequency axis, constructed using the contemporaneous RFI mask.  After filtering, the data are integrated in time further, first rebinning to a common local ERA grid and then averaging over sidereal days. Throughout, we also accumulate the sidereal-day–averaged filter and frequency–frequency noise covariance, which are used to propagate the filtering to 21~cm simulations and to estimate the noise, respectively.

To filter foregrounds we use the DAYENU algorithm \citep{ewall-wice2021} with an implementation based on that described in \citet{chimestacking}, applied to the \ns beamformed visibilities $\hybrid_{p f e d t}$ introduced in \secref{sec:beamforming}. In this space, where the \ns interferometric delays have been removed, and for the region of the sky considered here, the foregrounds should be confined to delays below $\tau \approx 200\ \mathrm{ns}$, beyond which the delay spectrum becomes noise-like (see Figure~10 in \citealt{chimestacking}). We thus adopt a simple model for the foreground plus noise covariance designed to deweight the foreground contaminated delays:
\begin{align}
\tilde{C}(\tau,\tau') &=
\begin{cases}
\epsilon^{-1} \frac{1}{2 \tau_{\rm cut}} \ \delta^D(\tau - \tau'), & |\tau| \leq \tau_{\mathrm{cut}}, \\
\Delta_{\nu} \ \delta^D(\tau - \tau'), & |\tau| > \tau_{\mathrm{cut}},
\end{cases}
\end{align}
where $\delta^D(\tau - \tau')$ is the Dirac delta function, $\Delta_{\nu} = \dnu$ is the channel width, $\tau_{\mathrm{cut}} = 200\ \mathrm{ns}$ is the delay cutoff, and $\epsilon = 10^{-12}$ is the assumed ratio of noise to foreground variance.  Fourier transforming to frequency space yields an analytic expression for the frequency–frequency covariance,
\begin{align}
    \label{eq:dayenu_cov}
\left[ \mathbf{C} \right]_{f f'} &= \epsilon^{-1} \ \mathrm{sinc}\left[ 2\pi \tau_{\mathrm{cut}}(\nu_f - \nu_{f'}) \right] + \delta_{f f'} \,
\end{align}
where here $\delta_{f f'}$ is the Kronecker delta function.  At each time index $t$, we apply the frequency-dependent RFI mask by forming the diagonal masking operator $\left[\mathbf{\mask}_{t}\right]_{ff'} = \delta_{f f'} G^{(0)}_{ft}$, which zeroes flagged frequencies.  We then compute the Moore–Penrose pseudoinverse to obtain the inverse-covariance filter
\begin{align}
\mathbf{H}_{t} &= \left(\mathbf{\mask}_{t} \mathbf{C} \mathbf{\mask}_{t}\right)^{+} \ ,
\label{eq:delay_filter}
\end{align}
where ${}^{+}$ denotes the pseudoinverse.  Foreground removal is achieved by weighting the visibilities with this inverse covariance, i.e.,
\begin{align}
\hybrid^{\mathrm{hpf}}_{p f e d t} &= \sum_{f'} H_{t f f'} \, \hybrid_{p f' e d t}.
\end{align}
Thus the suppression of smooth foregrounds is cast as an inverse covariance weighting problem under this simple foreground covariance model.

The filter $\mathbf{H}_{t}$ depends only on the RFI mask and the assumed covariance, and is identical for all polarisations, \ew baselines, and telescope $y$ coordinates at a given time.  Unlike the implementation in \citet{chimestacking}, we compute a distinct filter for each \dtraw integration and apply it before rebinning and averaging over sidereal days, mitigating spectral leakage from time variability in the visibilities coupling to the RFI mask.

The frequency–frequency noise covariance at time index $t$ is propagated through the same operator:
\begin{align}
\Sigma_{p e t f f'} &= \sum_{f''} H_{t f f''} \ \breve{\sigma}^2_{p f'' e t} \ H_{t f' f''} \ .
\end{align}
where $\breve{\sigma}^2_{p f'' e t}$ is the beamformed noise variance given by \cref{eq:beamform_ns_weights}. Both the filter and the frequency–frequency noise covariance are retained and propagated through the pipeline. These are later used to apply consistent filtering to simulated \tcm signals and to generate noise realizations, both of which are essential for interpreting the measured power spectrum.

\subsection{Estimating and Correcting Residual Bandpass Errors with HyFoReS}
\label{sec:hyfores}

After applying the foreground filter, we observe residual leakage of bright point sources in the hybrid beamformed visibilities. This leakage is spatially coincident with the brightest sources on the sky and is consistent with structure in the instrumental bandpass that couples with foreground emission to evade the delay filter.  To quantify this effect, we construct a model for how a point source $j$ would appear in the \emph{unfiltered} hybrid beamformed visibilities, evaluated at the declination grid point $d_j$ nearest to the source.  For polarization $p$, frequency channel $f$, and \ew baseline index $e$, the template is
\begin{align}
    \Psi_{p f e}(\theta_{j}, \phi - \alpha_{j}) &= S(\nu_{f})\, B_{p f e}(\theta_{j}, \phi - \alpha_{j}),
\end{align}
where $(\alpha_{j}, \theta_{j})$ are the right ascension and declination of the source in CIRS coordinates, $S$ is a model of the source flux density taken from an external catalog of extragalactic point sources, \emph{specfind v3} \citep{Stein2021}, interpolated into the CHIME band, and $B$ is the beam transfer function,
\begin{align}
    \label{eq:beam_transfer}
    B_{p f e}(\theta_{j}, \phi - \alpha_{j}) &= |A_{pf}(\theta_{j}, \phi - \alpha_{j})|^{2} \nonumber \\
    &\quad \times e^{-j 2 \pi \nu_{f} e \Delta_{\rm EW} x(\theta_{j}, \phi - \alpha_{j}) / c},
\end{align}
where $A_{pf}$ is our model for the average primary beam response evaluated at frequency channel $f$ and for polarization $p$, and the exponential factor accounts for the phase introduced by the geometric delay along the projected \ew baseline.

\begin{figure}
   \centering \includegraphics[width=0.98\linewidth, keepaspectratio, ]{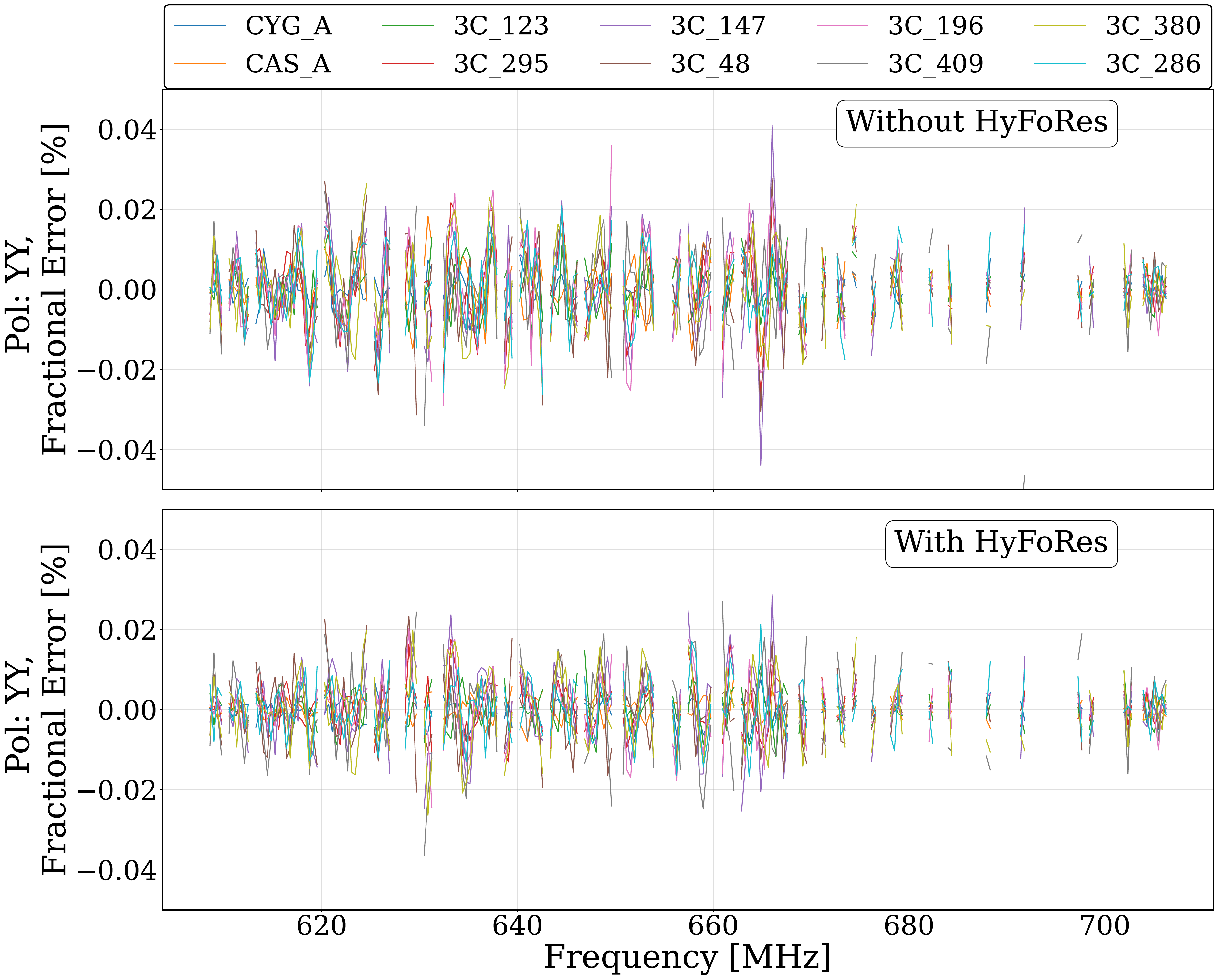}
    \caption{
       Bandpass leakage coefficients (fractional units) for the  YY-polarization, \SI{22}{\meter} \ew baseline as a function of frequency, measured from 10 bright point sources.
        \emph{Top panel:} Leakage coefficients estimated from the sidereal-day stack of foreground-filtered hybrid visibilities using the point-source fitting procedure of \cref{eq:dgcoeff}.
        \emph{Bottom panel:} Leakage coefficients estimated from the corresponding stack when HyFoReS is applied to the per-day foreground-filtered hybrid visibilities before sidereal-day averaging. HyFoReS reduces the amplitude of the bandpass leakage coefficients across most of the band.}
    
    \label{fig:hyfores_fig}
\end{figure}

We fit this template to the \emph{filtered} hybrid visibilities $\hybrid^{\mathrm{hpf}}$ to solve for a multiplicative leakage coefficient $\delta g^{(j)}$ for each polarization, frequency, and \ew baseline:
\begin{equation}
    \label{eq:dgcoeff}
    \widehat{\delta g}^{(j)}_{p f e} =
    \frac{\sum_{t} G^{(0)}_{f t}\, \Psi^{*}_{p f e}(\theta_{j}, \phi_{t} - \alpha_{j})\, \hybrid^{\mathrm{hpf}}_{p f e d_{j} t}}
    {\sum_{t} G^{(0)}_{f t}\, \left|\Psi_{p f e}(\theta_{j}, \phi_{t} - \alpha_{j})\right|^{2}},
\end{equation}
where $G^{(0)}_{ft}$ is 1 for valid samples and 0 for missing or masked samples (see \Cref{eq:flags}), and the sum runs over a time window centered on the source’s transit and spanning the extent of the primary beam at that declination.  In the absence of instrumental systematics, the delay filter would remove the point source contribution entirely, and $\widehat{\delta g}$ would be consistent with noise. In practice, residual high-delay structure in the bandpass gain couples with bright foreground flux and leaks through the delay cut, producing a measurable $\widehat{\delta g}$ for the brightest sources (see \cref{fig:hyfores_fig}, top panel).

We find that the frequency dependence of the multiplicative leakage coefficients $\widehat{\delta g}^{(s)}$ is largely common across different sources $j$, suggesting that a single correction per polarization, frequency channel, and baseline can account for most of the leakage. We interpret this shared frequency dependence as arising from an effective bandpass error introduced during daily real-time calibration. Instead of estimating these coefficients from only a limited set of bright sources, we use the Hybrid Foreground Residual Subtraction (HyFoReS) algorithm to estimate the leakage coefficients by summing over the full sky \citep{hyfores_2022, hyfores_2025a, hyfores_2025b}.

HyFoReS is a class of foreground subtraction algorithms designed to remove residual foregrounds in linearly filtered \tcm data. The key assumption underlying HyFoReS is that foreground residuals in the filtered data are correlated with the true foreground emission. By cross-correlating the filtered data with a suitable foreground template, one can estimate both the foreground residuals and any parametrizable systematics (such as bandpass-related leakage) that produce them.

A natural choice for the foreground template would be the prefiltered hybrid visibilities $\hybrid$, since they are dominated by foregrounds. However, the prefiltered data also contain \tcm signal and noise components that can correlate with the filtered data and bias the estimated leakage coefficients. To avoid this, we construct the foreground template by subtracting the filtered visibilities from the prefiltered visibilities, $\hybrid^{\mathrm{lpf}} = \hybrid - \hybrid^{\mathrm{hpf}}$, thereby removing the cosmological signal and noise contributions at high delay. In addition, noise cross-talk is subtracted from both the foreground template and the filtered data using the procedure described in \secref{sec:crosstalk_removal}.

HyFoReS then estimates the leakage coefficients by cross-correlating the filtered data with the template, effectively averaging over all sky positions:
\begin{equation}
    \label{eq:hyfores_coeff}
    \widehat{\delta g}_{p f e} = 
    \frac{\sum_{d} \sum_{t} G^{(0)}_{f t}\, \hybrid^{\mathrm{lpf}, *}_{p f e d t}\, \hybrid^{\mathrm{hpf}}_{p f e d t}}
         {\sum_{d} \sum_{t} G^{(0)}_{f t}\, \hybrid^{\mathrm{lpf}, *}_{p f e d t}\, \hybrid^{\mathrm{lpf}}_{p f e d t}} \, .
\end{equation}
These coefficients are estimated independently for each sidereal day. The averages are taken over nighttime data on that day, excluding telescope $y$ coordinates that experience significant \ns aliasing ($|y| \gtrsim 0.4$). Within the nighttime data, we use uniform weighting, but exclude all samples that are masked in either time or frequency.  The resulting $\widehat{\delta g}_{p f e}$ thus represents a daily, sky-averaged estimate of the residual bandpass error for each polarization, frequency channel, and \ew baseline.

The estimated leakage coefficients are then used to subtract bandpass-induced foreground residuals from the filtered visibilities:
\begin{equation}
    \label{eq:hvis_cal}
    \hybrid^{\mathrm{cal}}_{p f e d t} 
    = \hybrid^{\mathrm{hpf}}_{p f e d t} 
    - \sum_{f'} H_{t f f'} \, \widehat{\delta g}_{p f' e} \, \hybrid_{p f' e d t} \, ,
\end{equation}
where $H_{tff'}$ is the delay filter operator introduced in \cref{eq:delay_filter} and $\hybrid^{\mathrm{cal}}_{pfedt}$ denotes the bandpass-calibrated, foreground-filtered hybrid beamformed visibilities.

To assess the impact of HyFoReS, we re-estimate the leakage coefficients from the bandpass calibrated hybrid visibilities following the point-source fitting procedure described above.  The results are shown in the bottom panel of \cref{fig:hyfores_fig}, demonstrating that a single sky-averaged correction provides a modest reduction in bandpass-induced foreground leakage.

\subsection{Residual RFI Detection and Masking}
\label{sec:residualrfi}

After applying the pipeline steps described above, we observe residual contamination in the foreground-filtered data with frequency–time morphology consistent with RFI. This is most apparent when we form a $\chi^{2}$-like metric by summing the squared magnitude of the foreground-filtered, hybrid beamformed visibilities with inverse-variance noise weighting over telescope $y$ coordinates. To address this, we apply a final masking stage that follows the overall logic of the \emph{Spectral Filtering Method} in \appref{app:rfi_spec_filt}, but is modified to operate directly on the hybrid beamformed visibilities used for power spectrum estimation. For this stage, we also adopt the more advanced outlier identification algorithm originally developed for the \emph{Radiometer Test Method}, as described in \appref{app:rfi_radiometer}.

We operate directly on the main data product, $\hybrid^{\mathrm{cal}}_{pfedt}$, which employs a less aggressive delay cut (\SI{200}{\nano\second}) compared to that used in the previous masking stage (\SI{400}{\nano\second}), as discussed in \secref{sec:fg_filter_timeavg} and \appref{app:rfi_spec_filt}. The $\chi^{2}$ test statistic is evaluated by summing over intercylinder \ew baselines and telescope $y$ coordinates near zenith ($|y| \lesssim 0.40$), consistent with the subset of the data that will be used for power spectrum estimation. Because the noise is correlated as a function of telescope $y$, we calibrate the background level and scatter of the test statistic using robust, median-based estimators rather than assuming idealized Gaussian statistics.

Masking is performed using a multi-pass iterative procedure identical to that in \appref{app:rfi_radiometer}, but with three modifications: the threshold $\eta^{(k)}$ is lowered by a factor of 1.5 at each iteration from an initial value $\eta^{(0)} = 15.1875$ to a final value $\eta^{(4)} = 3$; the maximum convolution kernel length in the \texttt{SumThreshold} algorithm is set to $l_{\max} = 16$ samples; and the targeted flagging of digital TV channels is omitted. The resulting mask is denoted $Q^{\text{f-filt-2}}_{pft}$.

Unlike earlier stages, the mask is allowed to vary with polarization, minimizing unnecessary data loss. The resulting mask is incorporated into the global good-data flag:
\begin{equation}
    \label{eq:flags_v1}
    G^{(1)}_{pft} = G^{(0)}_{ft} \times (1 - Q^{\text{f-filt-2}}_{pft}),
\end{equation}
and is used in all downstream analyses.

\subsection{Time Averaging}
\label{sec:time_averaging}

The next step is to further average the bandpass-calibrated, foreground-filtered hybrid beamformed visibilities given by \cref{eq:hvis_cal}, which are sampled at the native \dtraw cadence, over longer time scales in order to reduce thermal noise. This is done in four stages.  First, we apply a time-domain data-quality mask that excludes known contaminated intervals, including all periods when the Sun is above the horizon.  Second, we subtract a daily estimate of the noise cross-talk. Third, we rebin each sidereal day onto a fixed grid in local ERA. Finally, we average over sidereal days. Because the bright, spectrally smooth foregrounds have already been filtered at high time resolution using a high-pass delay filter, these subsequent averaging steps no longer risk coupling temporal variations in the foregrounds into high-delay modes. This allows us to adopt a simpler averaging scheme than in \citet{chimestacking}, as described below.

\subsubsection{Daytime and Data-Quality Masking}
\label{sec:time_masking}

We construct two masks that identify time intervals to be excluded from the analysis. The daytime mask $Q_t^{\rm day}$ takes the value 1 for time samples when the Sun is above the horizon and 0 otherwise. Similarly, the data-quality mask $Q_t^{\rm bad}$ takes the value 1 during intervals of poor data quality and 0 otherwise. These include intervals associated with correlator or acquisition restarts, rainfall, anomalous autocorrelation behavior, or poor complex gain calibration.  Both masks are obtained following the procedures described in Section~3.2.7 of \citet{chimestacking}.  In contrast to that analysis, however, we do not mask times when the Moon is transiting, as it remains at telescope $y$ coordinates outside the range considered in this work. We combine these time-domain masks with the current good-data flag $G^{(1)}_{pft}$ (defined in \cref{eq:flags_v1}) to obtain:
\begin{equation}
    G^{(2)}_{pft} = G^{(1)}_{pft} \times (1 - Q^{\rm day}_t) \times (1 - Q^{\rm bad}_t) \,
\end{equation}
so that $G^{(2)}_{pft} = 1$ only for nighttime samples that are not flagged as RFI-contaminated by any of our four tests and that pass all other time-domain quality cuts, and $G^{(2)}_{pft} = 0$ otherwise.

\subsubsection{Noise Cross-talk Removal}
\label{sec:crosstalk_removal}

CHIME visibilities contain a quasi-stationary additive component arising from instrumental noise cross-talk. As discussed in Section~3.3.4 of \citet{chimestacking}, this cross-talk exhibits a complex spatial and spectral structure, but its temporal variability is relatively modest. In particular, the RMS intraday variations are typically $\lesssim 5\%$ of the mean level for zero-cylinder separations, increasing gradually at larger separations, while the interday variations are only slightly larger. This low level of temporal variation motivates a simplified treatment of the cross-talk in our analysis.

Physically, cross-talk arises when receiver noise from one feed is broadcast and subsequently picked up by neighbouring feeds, producing an additive offset in their cross-correlation. The frequency dependence of this offset is set by a small number of discrete delays corresponding to different propagation paths between the feeds, sometimes including one or more reflections off the cylinder surface.

In our analysis, we represent the cross-talk as a single additive term per sidereal day, with full dependence on polarization, frequency, \ew baseline, and telescope $y$ coordinate. Unlike \citet{chimestacking}, where cross-talk estimation and subtraction were performed directly on the visibilities prior to foreground filtering, we estimate and subtract our cross-talk estimate in hybrid beamformed visibility space \emph{after} the foreground filter has been applied. This ensures that the subtraction targets only the residual cross-talk at high delay and is not biased by the bulk of the bright sky signal, which has already been removed by the filter.

We estimate the additive cross-talk component during nighttime to avoid bias from the Sun. However, the portion of the sky that transits at night changes over the course of the year (see Figure~4 of \citet{chimestacking}). Since any constant component of the sky within the estimation interval will be absorbed into the cross-talk estimate and subtracted, it is important to apply this procedure consistently throughout the year.

To accomplish this, we divide the year into four quarters, each covering a successive three-month period. For each quarter, we assign a fixed reference range in $\phi$ that will be used to estimate and subtract the additive cross-talk component. These ranges are denoted $\mathcal{Q}_1$–$\mathcal{Q}_4$ and are each one hour wide. They are chosen to satisfy two key criteria: (1) they must transit during the nighttime throughout the associated quarter, and (2) they must avoid bright sources such as Cygnus~A and Cassiopeia~A. Specifically:
\begin{align}
\mathcal{Q}_{1} &: 150^\circ \leq \phi < 165^\circ \nonumber \\
\mathcal{Q}_{2} &: 240^\circ \leq \phi < 255^\circ \nonumber \\
\mathcal{Q}_{3} &: 315^\circ \leq \phi < 330^\circ \nonumber \\
\mathcal{Q}_{4} &:  15^\circ \leq \phi <  30^\circ \nonumber .
\end{align}
These ranges differ slightly from those used in \citet{chimestacking}, where $\mathcal{Q}_{1}$ and $\mathcal{Q}_{4}$ were defined as $165$--$180^\circ$ and $45$--$60^\circ$ respectively; these modifications were made to avoid antipodal transits of Cassiopeia~A and other bright sources.

For each sidereal day, we identify the time intervals during which the relevant reference range in $\phi$ is transiting overhead and compute, for every polarization, frequency, \ew baseline, and telescope $y$ coordinate, the \emph{weighted median} of the hybrid beamformed visibilities over this interval. The weights are given by the flag dataset $G^{(2)}_{pft}$. The weighted median is applied separately to the real and imaginary components of the visibilities.

The cross-talk–subtracted visibilities are thus given by
\begin{align}
    \hybrid^{\rm xtalk}_{pfedt} =
    \hybrid^{\rm cal}_{pfedt}
    & - \operatorname{Med}_{t \in \mathcal{Q}_{q}}\big(\operatorname{Re}\{\hybrid^{\rm cal}_{pfedt}\}\big) \nonumber \\
    & - j\, \operatorname{Med}_{t \in \mathcal{Q}_{q}}\big(\operatorname{Im}\{\hybrid^{\rm cal}_{pfedt}\}\big),
    \label{eq:xtalk_sub}
\end{align}
where $\operatorname{Med}_{t \in \mathcal{Q}_{q}}\left(\cdot\right)$ denotes the weighted median evaluated over times $t$ within the reference range $\mathcal{Q}_q$ appropriate for quarter $q$. This procedure ensures that all visibilities within a quarter are referenced to the same local ERA range.

If less than 30\% of the time samples in the reference range $\mathcal{Q}_q$ are marked as good for a given frequency and polarization, the cross-talk estimate is considered unreliable and the corresponding data are masked entirely.  This is implemented by updating the good-data flag as
\begin{equation}
    G^{(3)}_{pft}
    = G^{(2)}_{pft} \times
      \Theta\!\left(
        \sum_{t \in \mathcal{Q}_q} G^{(2)}_{pft}
        - 0.3 \sum_{t \in \mathcal{Q}_q} 1
      \right),
    \label{eq:flag_v3}
\end{equation}
where $\Theta(u)$ is the Heaviside step function, equal to 1 when $u > 0$ and 0 otherwise.

\subsubsection{Sidereal Rebinning}

We rebin the hybrid beamformed visibilities onto a fixed grid in $\phi$ (local ERA).  We refer to the interval between successive zero-crossings of $\phi$ as a sidereal day.  The rebinning is accomplished with a simple fractional-overlap averaging procedure.  This approach was motivated by issues we encountered when using the Wiener-filter-based regridder from \citet{chimestacking}, which has a tendency to introduce baseline-dependent artifacts at the edges of RFI-masked regions and is sensitive to the choice of sky covariance.  The rebinning method avoids both issues. However, because the RFI mask is frequency dependent, the number of native time samples contributing to a given $\phi$-bin varies with frequency. Since the sky signal evolves across the width of a $\phi$-bin due to Earth’s rotation, this frequency dependence converts intra-bin temporal variation into small-scale spectral structure, which leaks power to high delays upon Fourier transforming along frequency. We therefore apply a foreground filter prior to rebinning to suppress the slowly evolving component within each bin, as discussed in \secref{sec:fg_filter_timeavg}.

Let $t$ index the raw time samples of width $\Delta_{t} = \dtraw$, $r$ index the local ERA grid samples, and $s$ index the sidereal day. For each time sample $t$, we define the local ERA interval covered by the integration as $[\phi_t^{-},\, \phi_t^{+}]$, where $\phi_t^\pm$ are determined from the known mapping between time and local ERA. For each grid point $r$, we define the corresponding local ERA bin interval $[\phi_{r}^{-},\, \phi_{r}^{+}]$ with $\phi_{r}^{\pm} \equiv \phi_{r} \pm \frac{d_\phi}{2}$.  The grid spacing is $\Delta_{\phi} = \SI{5.27}{\arcminute}$, a factor of roughly 2 larger than the raw integration length, yielding $N_{\phi} = 4096$ grid points over \SI{360}{\degree}.

The overlap length between the time-sample interval $t$ and the local-ERA-bin interval $r$ is
\begin{equation}
\ell_{rt} \equiv 
\max\!\left[\,0,\; 
\min\{\phi_t^{+}, \phi_{r}^{+}\} - \max\{\phi_t^{-}, \phi_{r}^{-}\}
\right].
\end{equation}
We then define the fractional-overlap kernel
\begin{equation}
K_{rt} \equiv \frac{\ell_{rt}}{\phi_t^{+} - \phi_t^{-}}, \qquad 0 \le K_{rt} \le 1,
\end{equation}
which equals 1 if the time interval is fully contained within the local-ERA bin, 0 if there is no overlap, and otherwise the appropriate fractional value.

The effective number of valid raw time samples in each local-ERA bin is
\begin{equation}
   \nsample_{pfrs} = \sum_{t\in\mathcal{T}_{s}} K_{rt} G^{(3)}_{pft},
\end{equation}
where $G^{(3)}_{pft}$ is the flag indicating good ($1$) or bad ($0$) data, and $\mathcal{T}_{s}$ denotes the set of all raw time samples whose integration interval overlaps with the local ERA interval corresponding to sidereal day $s$. For bins with $\nsample_{pfrs}=0$, all rebinned quantities defined below are set to zero.

The rebinned visibilities are computed as
\begin{equation}
    \hybrid^{\rm rebin}_{pfedrs} = 
    \frac{1}{\nsample_{pfrs}} 
    \sum_{t\in\mathcal{T}_{s}} K_{rt} G^{(3)}_{pft} \hybrid^{\rm xtalk}_{pfedt}.
\end{equation}
The filter is propagated through the same linear operation,
\begin{equation}
    H^{\rm rebin}_{rsff'} = 
    \frac{1}{\nsample_{pfrs}} 
    \sum_{t\in\mathcal{T}_{s}} K_{rt} G^{(3)}_{pft} H_{tff'},
\end{equation}
as is the noise covariance,
\begin{equation}
    \Sigma^{\rm rebin}_{persff'} = 
    \frac{1}{\nsample_{pfrs} \ \nsample_{pf'rs}} 
    \sum_{t\in\mathcal{T}_{s}} K_{rt}^{2} G^{(3)}_{pft} G^{(3)}_{pf't} \Sigma_{petff'},
\end{equation}
which is assumed to be diagonal in $r$.

\subsubsection{Sidereal Day Averaging}
\label{sec:sidereal_day_averaging}

\begin{figure}
   \centering \includegraphics[width=\linewidth, keepaspectratio]{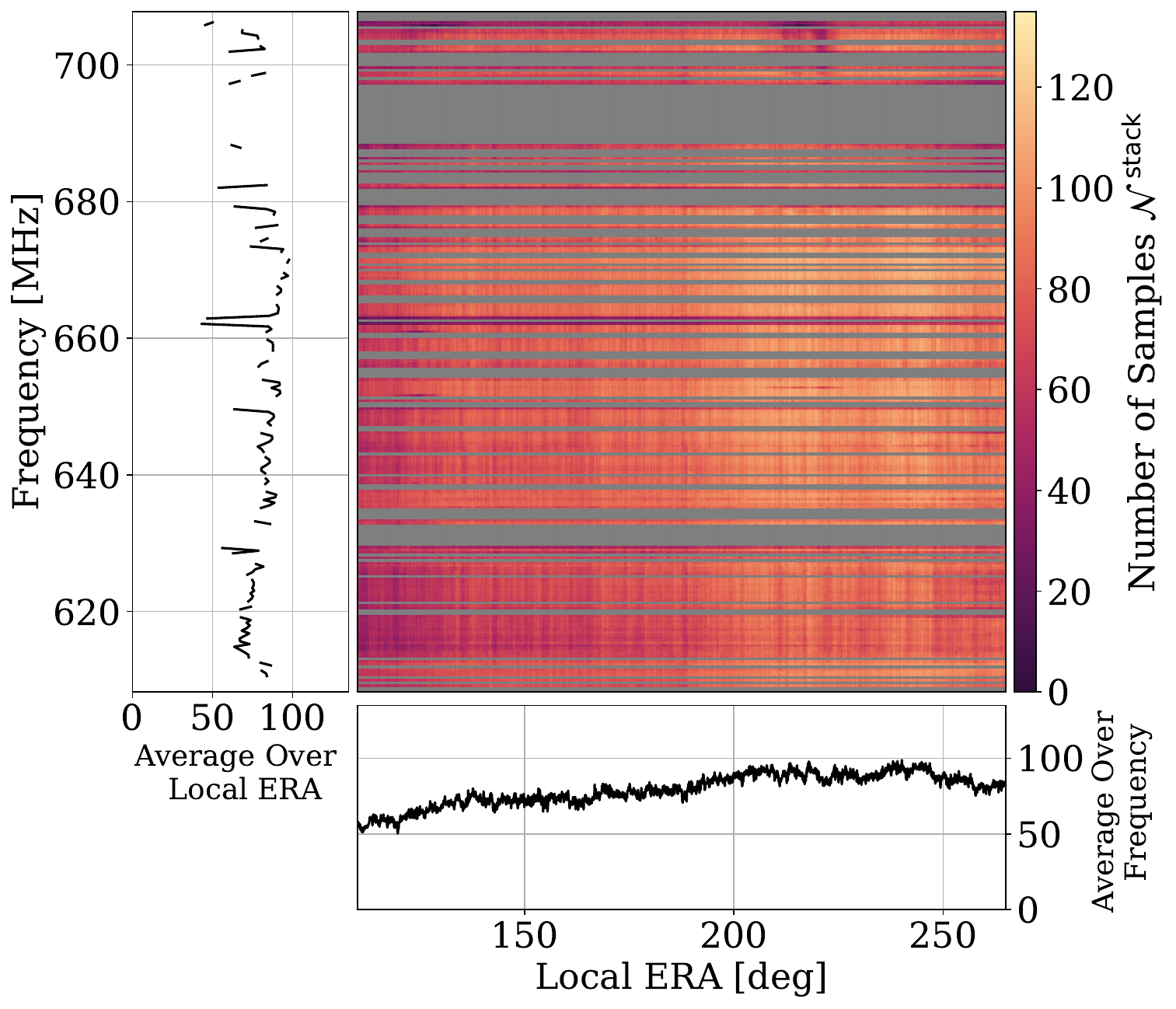}
\caption{Number of $\dtraw$ integrations contributing to each local-ERA–frequency bin in the stack over all \SI{94}{\day} of data. The center panel shows the two-dimensional distribution as a function of local ERA and frequency, restricted to the local-ERA range defining the field used in this work. Frequency channels that are entirely masked are shown in gray. The left panel shows the average over local ERA, while the bottom panel shows the average over frequency, excluding masked channels from the average.  Summing the bottom panel over local-ERA bins yields a total integration time of \SI{385}{\hour} on this field.}
    \label{fig:number_of_samples}
\end{figure}

Finally, we average the hybrid beamformed visibilities over 94 sidereal days acquired between \datestart and \dateend (see \secref{subsec:data_selection}).  Each visibility is weighted by $\nsample$, the number of raw integrations contributing to that local ERA bin, corresponding to an effectively \emph{uniform weighting} scheme.  As discussed in \secref{sec:fg_filter_timeavg}, the data have been foreground-filtered on a per-day basis.  The filter will strongly attenuate frequencies near large gaps of missing frequencies in order to suppress foregrounds to the required level.  This attenuation affects both the \tcm signal and the noise, but an inverse-variance weighting scheme would only reflect the noise suppression, effectively upweighting these heavily attenuated data points.  Uniform weighting avoids this imbalance.

Hybrid beamformed visibilities are stacked over sidereal days via
\begin{equation}
    \hybrid^{\rm stack}_{pfedr} 
    = \frac{1}{\nsample^{\rm stack}_{pfr}} 
    \sum_{s} \nsample_{pfrs}\,\hybrid^{\rm rebin}_{pfedrs}.
    \label{eq:vstack}
\end{equation}
where $s$ indexes sidereal days, and the sum runs over all days that pass the data quality tests described in Section~3.3.1 of \citet{chimestacking}.  Here $\nsample^{\rm stack}$ is the total number of raw integrations and is given by
\begin{equation}
    \label{eq:nsample}
    \nsample^{\rm stack}_{pfr} = \sum_{s} \nsample_{pfrs} \ .
\end{equation}
This quantity is shown in \cref{fig:number_of_samples}.  The same procedure is applied to the filter,
\begin{equation}
    \label{eq:stacked_filter}
    H^{\rm stack}_{rff'} 
    = \frac{1}{\nsample^{\rm stack}_{pfr}} 
    \sum_{s} \nsample_{pfrs}\,H^{\rm rebin}_{rsff'},
\end{equation}
and to the noise covariance,
\begin{equation}
    \label{eq:stacked_noise_cov}
    \Sigma^{\rm stack}_{perff'} 
    = \frac{1}{\nsample^{\rm stack}_{pfr}\,\nsample^{\rm stack}_{pf'r}} 
    \sum_{s} \nsample_{pfrs}\,\nsample_{pf'rs}\,\Sigma^{\rm rebin}_{persff'}.
\end{equation}
In the staged averaging described below, these expressions are applied repeatedly to the appropriate subset of days.

To construct the final sidereal-day average, we apply the stacking rules in \cref{eq:vstack,eq:nsample,eq:stacked_filter,eq:stacked_noise_cov} in several stages. First, the valid sidereal days are sorted by date and, within each quarter of the year, divided into two partitions that we label ``even'' and ``odd'' by assigning consecutive days alternately to each partition. This procedure yields two stacks with approximately equal numbers of days while keeping the temporal separation between days in each stack small. Within each partition and quarter, visibilities are then averaged over sidereal days using \cref{eq:vstack}. Next, the quarterly averages for the first, second, and third quarter are aligned to a common reference by subtracting the median value over $\mathcal{Q}_{2}$ (as described in \secref{sec:crosstalk_removal}). These three quarterly averages are then combined into a single stack, which we reference to $\mathcal{Q}_{4}$ before averaging with the fourth-quarter stack. This staged referencing ensures that all visibilities are placed on a common sky-relative reference while maximizing the amount of data available to determine that reference level. Frequencies for which more than $30\%$ of the time samples in the relevant reference interval are missing are masked entirely. The end result is a sidereal-day stack for each of the even and odd partitions.

\begin{figure}
   \centering \includegraphics[width=1\linewidth, keepaspectratio, ]{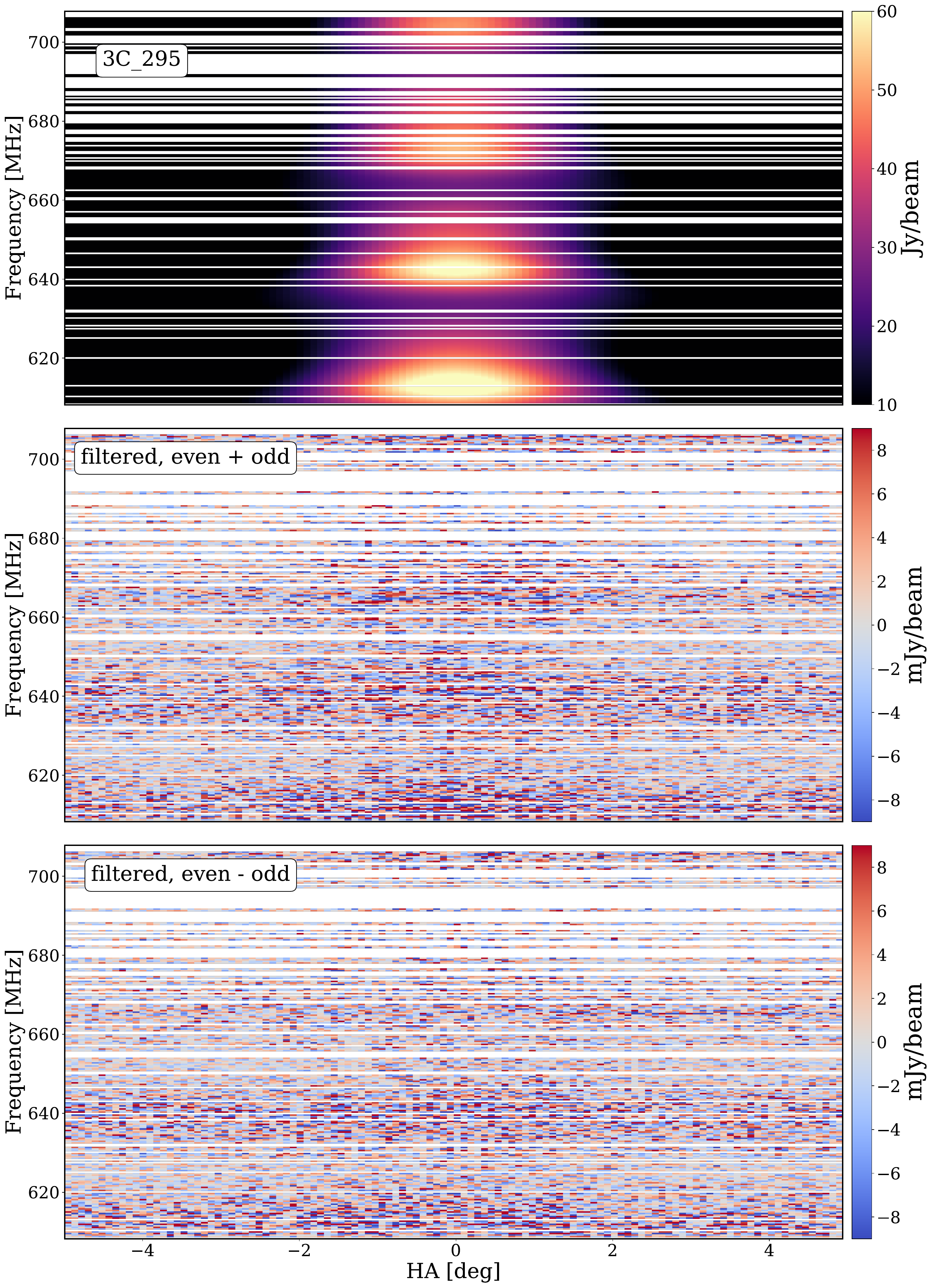}
\caption{
\emph{Top:} Hybrid beamformed visibility as a function of frequency and time for the XX polarization and \SI{22}{\meter} \ew baseline, evaluated at the telescope $y$ coordinate closest to that of 3C~295 (the brightest radio source in the field considered in this analysis) at transit. The time axis is expressed in degrees of 3C~295 hour angle, and the color scale shows the real component of the visibility after fringestopping. The top panel shows the average of unfiltered data over the full 94-day dataset, with features corresponding to the intrinsic spectrum of 3C~295 modulated by the primary beam response.
\emph{Middle:} Same quantity after applying the HyFoRes bandpass correction and a time-dependent high-pass filter along the frequency axis to the raw time series prior to averaging over sidereal days. This is the primary dataset used in our analysis, in which the foregrounds are suppressed by a factor of $\gtrsim 10^{4}$.
\emph{Bottom:} Difference between the even and odd sidereal-day stacks, which cancels foreground power that is stable across sidereal days. The close agreement with the middle panel demonstrates that foreground leakage is sub-dominant compared to thermal noise at this stage of the data processing. The frequency dependence of the noise level, including the prominent $\sim \SI{30}{\mega\hertz}$ ripple, reflects the primary beam response at the transit of Cygnus~A, which is imprinted onto the data through the complex gain calibration procedure.
}
    \label{fig:hybrid_vis_bright_source}
\end{figure}

\Cref{fig:hybrid_vis_bright_source} presents the stacked hybrid beamformed visibility $\hybrid^{\rm stack}_{pfedr}$ for a representative polarization and EW baseline, evaluated at the declination nearest to 3C~295, the brightest source in our field.  We show both the full sidereal-day average (middle panel) and the difference of the even and odd partitions (bottom panel).  Given CHIME’s night-to-night gain stability at the $\sim\!1\%$ level \citep{chimeoverview}, foreground power should remain coherent across days and therefore cancel in the difference, leaving predominantly thermal noise. The close agreement in morphology between the average and even–minus-odd maps, together with their comparable variances (within 5\%), indicates that any residual foreground leakage in this data product is sub-dominant to the noise after per-day foreground filtering and bandpass correction—even in the vicinity of the brightest source.

\subsection{Map Making}
\label{sec:mapmaking}

\begin{figure*}
   \centering \includegraphics[width=1\linewidth]{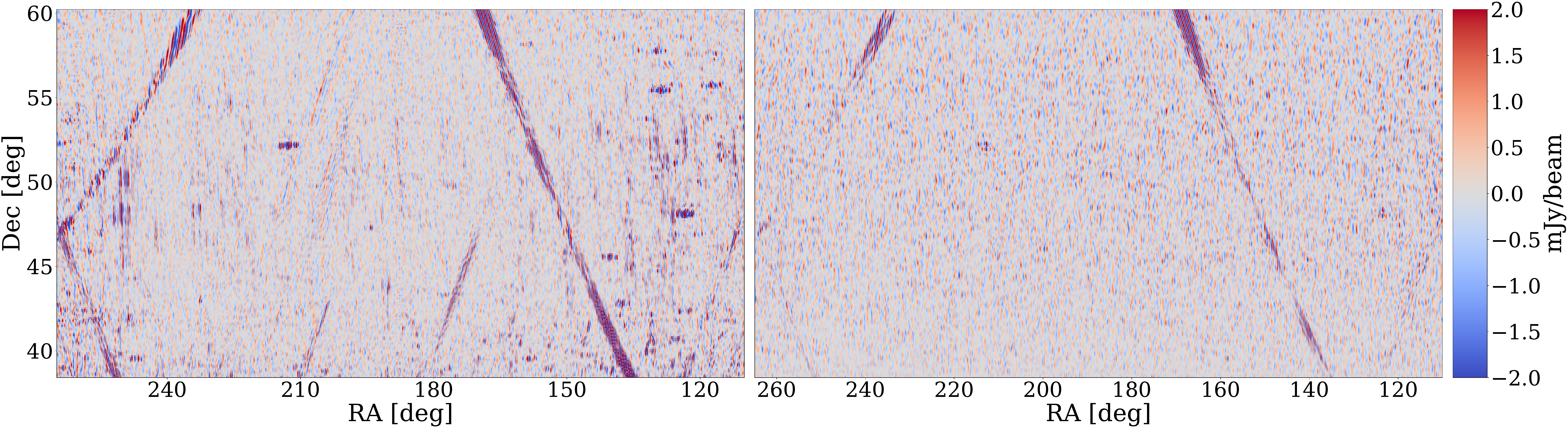}
\caption{Foreground-filtered intensity maps near the North Galactic Cap at \SI{678.5}{\mega\hertz} in YY polarization. \emph{Left:} Map produced using the original \citet{chimestacking} pipeline, which formed the basis of the high signal-to-noise cross-correlation detection with eBOSS galaxies and quasars. \emph{Right:} The same dataset reprocessed with the updated pipeline described in \secref{sec:pipeline}, showing a significant reduction in artifacts due to RFI and foreground leakage, which yields a cleaner input for auto–power spectrum analysis. Bright continuum sources in the far sidelobes still produce high-delay signal that passes through the foreground filter, appearing as characteristic ``U''-shaped tracks that are masked for this analysis (see \secref{sec:spatial_mask}).}
    \label{fig:ringmap}
\end{figure*}

The next stage of the pipeline constructs sky maps from roughly year-long stacks of hybrid beamformed visibilities for the even and odd partitions. Since the \ns direction has already been beamformed, map-making requires only the combination of measurements from the different \ew cylinder separations. We perform this combination in the $m$-mode domain \citep{shaw2014,shaw2015} — the Fourier conjugate of local ERA — where each \ew baseline is sensitive to a narrow, largely disjoint set of azimuthal modes $m$. Our procedure follows the general framework described in Section 4.3 of \citet{chimestacking}, with several modifications detailed below.

We first mask frequency channels without full sidereal-day coverage for both polarizations to avoid leakage in the subsequent Fourier transform over local ERA due to missing data.  This is achieved by constructing a final good-frequency flag
\begin{equation}
    \label{eq:freq_flag_stack}
    G^{\rm stack}_{f} =
    \prod_{p,r} \Theta\!\left(\nsample^{\rm stack}_{pfr}\right),
\end{equation}
which equals one if and only if every $(p,r)$ bin at frequency $f$ contains at least one unflagged sample in the stack, and zero otherwise.  Frequency channels with $G^{\rm stack}_{f} = 0$ are excluded from the subsequent analysis.

For each polarization $p$, frequency channel $f$, \ew baseline index $e$, declination grid point $d$, and Fourier mode $m$, we take a discrete Fourier transform over local ERA of both the stacked hybrid beamformed visibilities and the beam transfer function:
\begin{equation}
    \hybrid^{\mathrm{m\text{-}mode}}_{pfedm} = \sum_{r} \hybrid^{\mathrm{stack}}_{pfedr}\, e^{-j m \phi_{r}},
\end{equation}
\begin{equation}
    B^{\mathrm{m\text{-}mode}}_{pfedm} = \sum_{r} B_{pfe}(\theta_{d}, \phi_{r})\, e^{-j m \phi_{r}},
\end{equation}
where $\phi_r$ refers to the local ERA grid points and $B_{pfx}(\theta_{d}, \phi_{r})$ is our model for the beam transfer function (see \cref{eq:beam_transfer}), evaluated at declinations and hour angles given by the (declination, local-ERA) grid of the hybrid beamformed visibilities.

The hybrid beamformed visibilities from different \ew baselines are then combined in $m$-mode space to form maps:
\begin{equation}
    \label{eq:map_maker}
    \map^{\mathrm{m\text{-}mode}}_{pfdm} = \sum_{e} W^{\rm EW}_{e}\, \bigl(B^{\mathrm{m\text{-}mode}}_{pfedm}\bigr)^{*}\, \hybrid^{\mathrm{m\text{-}mode}}_{pfedm},
\end{equation}
where $W^{\rm EW}_{e} = [0.0,\, 0.5,\, 0.333,\, 0.166]$ for $e = [0,1,2,3]$, which masks intracylinder baselines and weights the intercylinder baselines according to their redundancy. The intracylinder (i.e., $e = 0$) baselines are excluded in this analysis because they exhibit both enhanced noise cross-talk and a much stronger response to diffuse Galactic emission than the intercylinder baselines. Masking them results in maps in which the foregrounds are dominated by compact extragalactic sources, which are substantially easier to model and mask in the subsequent power-spectrum analysis. A more comprehensive treatment of these short baselines is deferred to later iterations of the analysis.

Each \ew baseline samples a largely disjoint set of $m$-modes, with the central mode determined by the baseline length and the width of the sensitive band set by the diameter of the cylinder. 
\cref{eq:map_maker} combines the baselines in the $m$-mode domain by multiplying each baseline’s visibility by the complex conjugate of its beam transfer function $(B^{\mathrm{m-mode}}_{pfedm})^*$ and weighting by the redundancy factor $W^{\rm EW}_e$. The multiplication by $(B^{\mathrm{m-mode}}_{pfedm})^*$ applies a matched filter using the known instrumental response and automatically retains only the $m$-modes to which that baseline is sensitive (setting insensitive modes to zero).  The additional redundancy-based weights $W^{\rm EW}_e$ approximate optimal inverse-variance weighting across baseline types.  Together, these operations yield maps with the full $m$-mode coverage of the array and close-to-optimal thermal noise; substantially lower than that of the individual hybrid-beamformed visibilities.

The resulting $m$-mode maps are transformed back to local ERA via
\begin{equation}
    \map_{pfdr} = \frac{1}{a_{pf}} \sum_{m} \map^{\mathrm{m\text{-}mode}}_{pfdm}\, e^{j m \phi_{r}},
\end{equation}
where the polarization- and frequency-dependent normalization factor
\begin{equation}
    a_{pf} = \sum_{m} \sum_{e} W^{\rm EW}_{e} \left| B^{\mathrm{m\text{-}mode}}_{pfed_{\mathrm{ref}}m} \right|^{2}
\end{equation}
ensures that a point source located at the reference declination $d_{\mathrm{ref}}$ has a peak value equal to its flux density. We adopt the declination of Cygnus~A as the reference declination, consistent with the normalization convention used throughout the pipeline.

The synthesized beam along the $\phi$ direction is given by
\begin{align}
    \label{eq:synth_ew}
    B^{\rm EW}_{\text{synth},pfd}(\phi - \phi_{r}) = \frac{1}{a_{pf}} \sum_{m} \sum_{e} & W_{e}^{\rm EW} |B^{\text{m-mode}}_{pfedm}|^{2} \notag \\
     & \times e^{j m \left( \phi - \phi_{r}\right)}
\end{align}

The sky signal is attenuated by the primary beam during observation and is then multiplied by the known beam model in the matched-filtering step (\cref{eq:map_maker}). Consequently, the reconstructed sky brightness in our maps is modulated as a function of declination by the square of the primary beam response on the meridian (i.e., at zero hour angle), relative to its value at the reference declination. In \citet{chimestacking}, we corrected for this modulation by dividing the maps by the square of this meridian beam model, thereby normalizing the response across declination. This normalization reduced beam-induced spectral structure in the data and allowed us to use a less aggressive foreground filter. In this analysis, we do not implement this correction for two reasons. First, when we later Fourier transform along the declination axis during power spectrum estimation, the meridian beam response provides desirable down-weighting of declinations where the primary beam response is low. Optimal power spectrum estimation requires weighting by the beam response and the inverse noise variance, as is done here. Second, correcting for the beam response at this stage is unnecessary because foreground filtering has already been applied and we will forward model the effect of the beam on the \tcm signal. This second reason also allows us to omit the step of deconvolving the \ew beam response along local ERA, which was done in \citet{chimestacking}.

We employ an updated primary beam model (\texttt{rev\_03}) relative to the model used in \citet{chimestacking} (\texttt{rev\_02}) when calculating the beam transfer function.  Construction of this model follows the procedure described in Appendix~B of \citet{chimestacking}, with several improvements.  We applied enhanced RFI excision and a revised \ns beamforming window function when forming the dataset used to fit the sky model. We then fit for a beam model over the full CHIME band, enabling better correction of flux errors in the input sky model.

For the remainder of this work, we restrict our analysis to a region of sky near the North Galactic Cap (NGC), which transits close to zenith and lies at high Galactic latitude, far from the Galactic plane. Specifically, we examine a field with $N_\phi = 1763$ local ERA bins spanning $\phi \in [110^\circ, \ 265^\circ]$ and $N_y = 231$ telescope $y$ bins spanning $y \in [-0.19, \ 0.19]$, corresponding to declinations $\theta \in [38.4^\circ, \ 60.2^\circ]$. The survey area of this field is
\begin{equation}
\label{eq:survey_area}
\mathcal{A}_{\rm s} \simeq 0.67~\mathrm{sr} \simeq 2.2\times 10^3~\mathrm{deg}^2 \, .
\end{equation}

Figure~\ref{fig:ringmap} shows an example of the resulting map at \SI{678.5}{\mega\hertz} in YY polarization for this field.  Both panels are derived from the same underlying dataset, frequency channel, and polarization; the difference lies solely in the analysis pipeline. The left panel uses the processing from \citet{chimestacking}, which yielded a high signal-to-noise detection of \tcm emission in cross-correlation with the eBOSS ELG, LRG, and QSO samples, while the right panel shows the same data reprocessed with the improved pipeline described in this work. The measured noise in a single frequency channel is on average \SI{0.5}{\milli\jansky\per\beam}. The updated map exhibits markedly fewer artifacts from RFI and foreground leakage, providing a substantially cleaner dataset for subsequent auto–power spectrum analysis.

\subsection{Forming (Pseudo) Stokes $I$ and $Q$}
\label{sec:stokesIQ}

We combine the linearly polarized maps derived from the co-polar visibilities ($p\in\{\mathrm{XX},\mathrm{YY}\}$) to form pseudo-Stokes maps ($q\in\{I,Q\}$). For each frequency channel $f$, declination grid point $d$, and local ERA sample $r$, 
\begin{equation}
    \map^{\mathrm{Stokes}}_{qfdr}
    \;=\;
    \sum_{p} T_{qp}\,\map_{pfdr},
    \label{eq:pseudoIQ}
\end{equation}
with
\begin{equation}
    \mathbf{T}\;\equiv\;\tfrac{1}{2}
    \begin{bmatrix}
      1 & 1\\
      1 & -1
    \end{bmatrix},
\end{equation}

We compute only $I$ and $Q$ (not $U$ or $V$). These are \emph{pseudo}-Stokes maps because no primary beam deconvolution is applied; they are beam-weighted combinations of the co-polar maps. Note that Stokes maps are masked at frequency $f$ if either XX and YY are masked at that $f$ (see \cref{eq:freq_flag_stack}).

\subsection{Detection and Masking of Narrow-band Absorption Features}
\label{sec:absorbers}

\begin{figure}
   \centering \includegraphics[width=\linewidth, keepaspectratio, trim = 0 0 0 0]{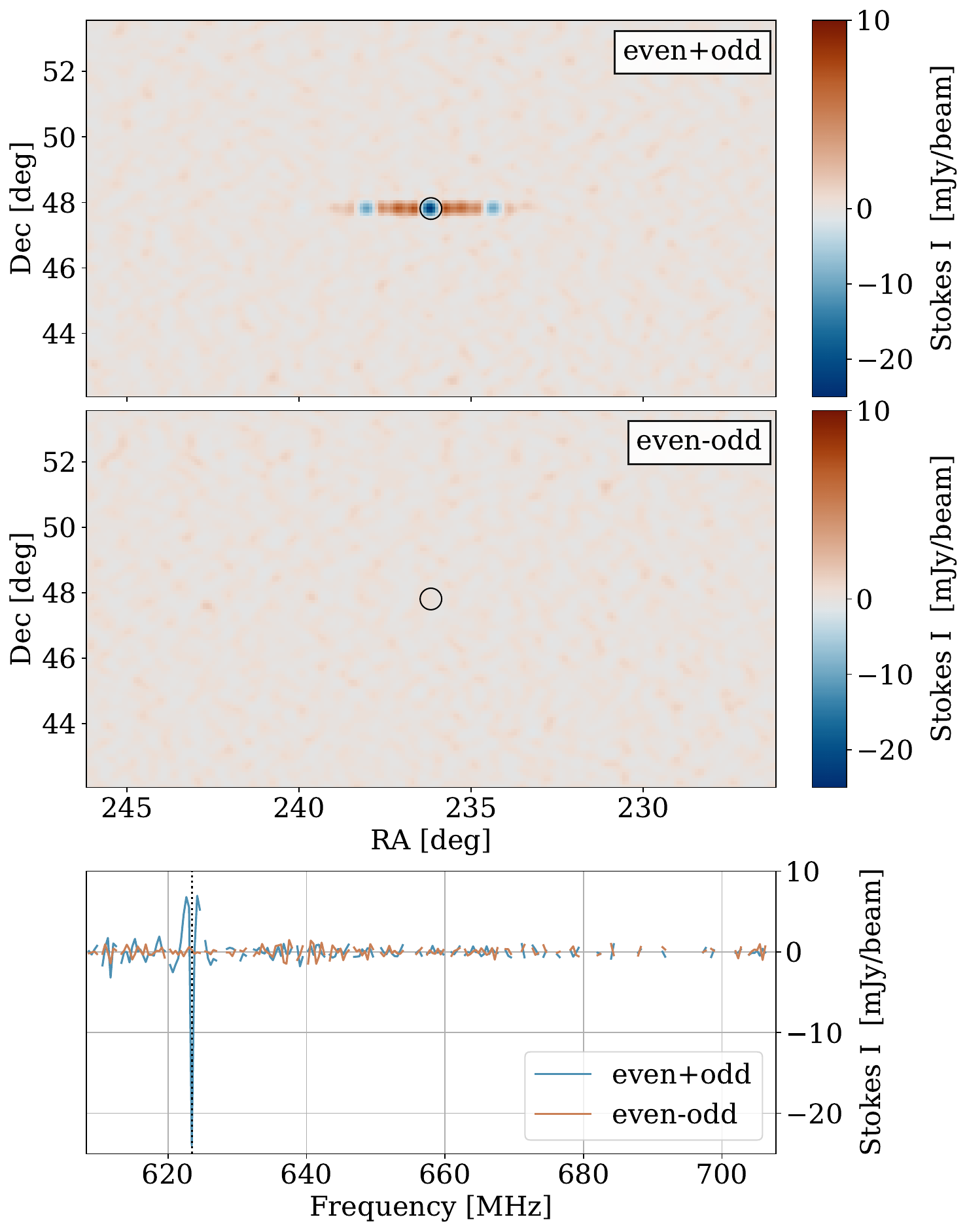}
\caption{CHIME detection of the known \tcm absorption system B1543+480 at $z=1.277$ \citep{exabsorber}.
\emph{Top:} Foreground-filtered Stokes~$I$ map at \SI{623.4}{\mega\hertz} in the vicinity of B1543+480. The source is clearly seen across the full extent of the CHIME primary beam in right ascension (RA). In the RA direction, the PSF response to this source shows a strong negative excursion at the central lobe; immediately outside it, symmetric positive shoulders caused by masked intra-cylinder baselines; and, farther out, symmetric negative responses at the grating lobes.  A diverging color scale with unequal slopes for positive and negative values is used to enhance contrast.
\emph{Middle:} Foreground-filtered Stokes‑$I$ even–odd jackknife map, obtained by differencing the sidereal-day stacks from the even and odd partitions. The absorption signal disappears in the difference, indicating its presence in both partitions and supporting a celestial origin.
\emph{Bottom:} Spectrum at the map pixel nearest the source location, showing the full sidereal-day stack and the even–odd jackknife. A narrow absorption line is evident in the stack but not in the jackknife. Positive shoulders bracketing the line arise from high-pass filtering along frequency to remove smooth foregrounds.}
    \label{fig:absorber_example}
\end{figure}

\begin{figure}
   \centering \includegraphics[width=\linewidth, keepaspectratio, trim = 0 0 0 0]{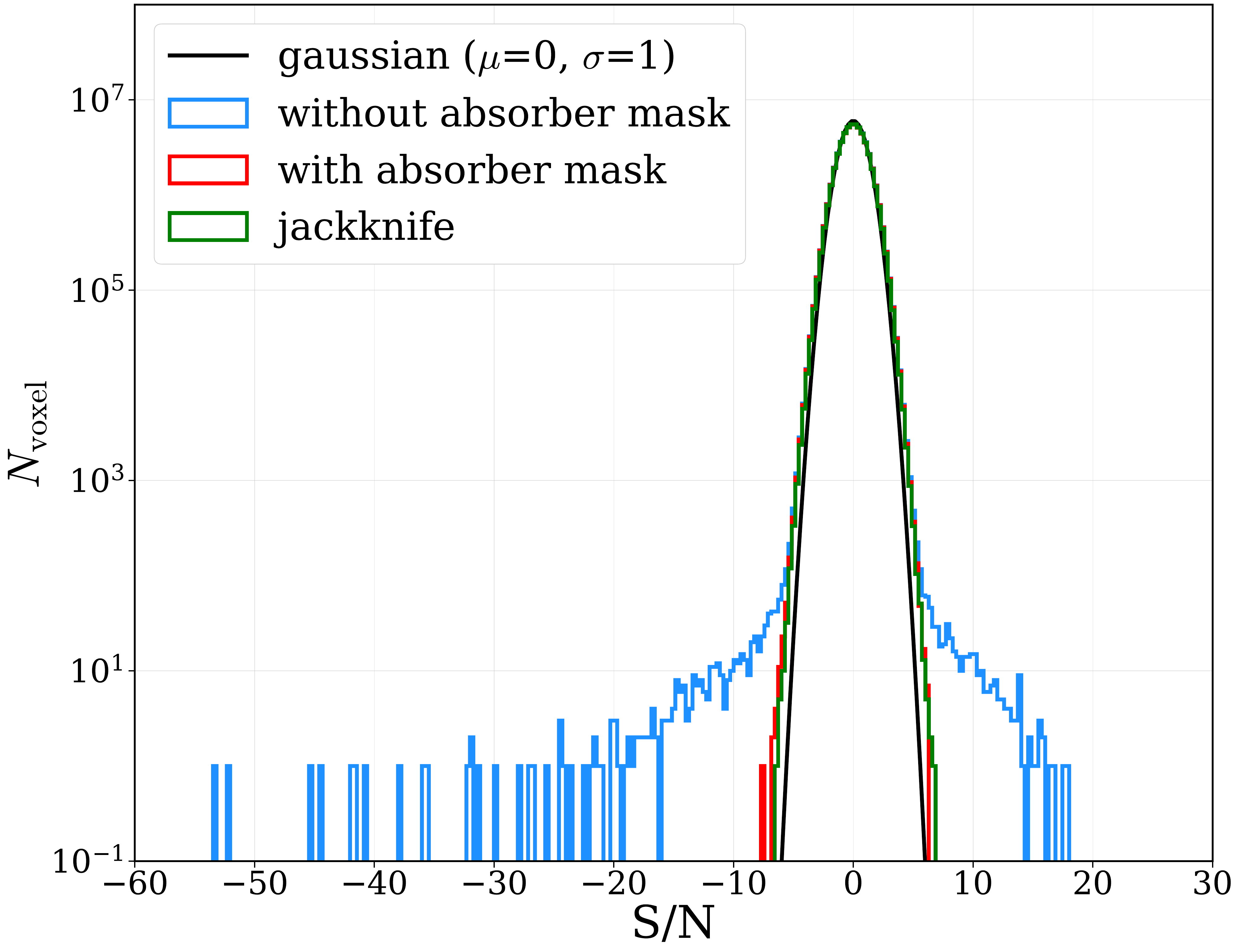}
    \caption{Histogram of voxel values in the foreground-filtered map after normalizing by the per-voxel standard deviation derived from the fast-cadence noise estimate. The blue histogram shows the distribution before masking \hi\ absorbers, revealing pronounced negative tails (down to $\mbox{S/N} \approx -60$) from absorption systems and corresponding positive tails (up to $\mbox{S/N} \approx 20$) produced by the high-pass filtering applied both spectrally and spatially (i.e., through the exclusion of intracylinder baselines), which generates adjacent positive excursions around strong negative features. The red histogram shows the distribution after masking absorbers, which is consistent with the jackknife map (green), obtained by differencing the even and odd sidereal-day stacks. A Gaussian with mean $\mu = 0$ and standard deviation $\sigma = 1$ is shown for comparison. Both the masked and jackknife distributions exhibit a $10$–$15\%$ excess width relative to the Gaussian, as expected because the noise estimate does not account for sky temperature variations. The agreement between the masked map and jackknife validates the absorber identification and masking procedure.}
    \label{fig:histogram_voxel}
\end{figure}

We have identified numerous narrow-band absorption-like features in our data that are stable across sidereal days and localized in both sky position and frequency. These features are candidate \tcm absorbers that require further confirmation, though one corresponds to a previously known \tcm absorber system.
A detailed methodology for the identification and characterization of these systems across the full CHIME band, along with their population statistics, will be presented in a forthcoming publication (CHIME Collaboration, in prep.). Here, we briefly outline the procedure used to identify and mask these features for the power spectrum analysis.

We generate a foreground-filtered map from the full 94 sidereal-day stack and, in parallel, a jackknife map by differencing sidereal-day stacks formed from the even and odd partitions.  The voxels in both the combined and jackknife maps are then normalized by the per-voxel standard deviation derived from the propagated fast-cadence estimate of noise (see \appref{app:rfi_radiometer}).  If the maps contained only noise, the distribution of voxel values in both of these 3D maps would be Gaussian with unit standard deviation. Instead, we find many outliers that appear in the combined map but not in the jackknife, indicating that these signals are stable in sidereal coordinates and likely associated with celestial sources.

Manual inspection of these outliers shows that they are localized in the sky, appear as point sources in the spatial domain, and exhibit narrow-band absorption-like features in their frequency spectra (see \cref{fig:absorber_example} for an example).   In the observed region of the NGC between 608–708 MHz, we identify 40 such candidates with a signal-to-noise ratio (S/N) $>7$.  Note that we see evidence of bright foreground sources leaking through far sidelobes of the primary beam and contaminating the NGC region (see \cref{fig:ringmap}). To mitigate this, we apply a spatial mask as described in \secref{sec:spatial_mask}, which removes approximately $33\%$ of the sky area. All 40 candidates are detected in the remaining unmasked regions. Future improvements in primary beam modeling and foreground filtering will enable searches across substantially larger sky areas without requiring such aggressive spatial masking. Cross-matching the 3D positions of our candidates with existing catalogs of known absorber systems, we find one match with the known \tcm absorption system B1543+480 at $z=1.277$, previously reported in \citet{exabsorber} (shown in \cref{fig:absorber_example}), which provides confidence in our detection methodology. The remaining 39 features are new candidate systems that will require further confirmation.

These narrow-band features are typically confined to a single 390 kHz-wide coarse channel of CHIME data. They also exhibit spectral sidelobes due to convolution with the foreground filter, similar to the stacked \hi\ \tcm emission signal as shown in \citet{chimestacking}. For each candidate absorber $s$, we generate a three-dimensional (3D) taper $T_{\rm abs}^{s}(\nu, y, \phi)$ that accounts for both spatial location and spectral position, masking 13 channels on either side of the peak absorption channel to cover spectral sidelobes. We apply this taper to the map, which zeroes $1.29\%$ of the voxels in our data cube. The details of the construction of the taper are described in Appendix~\ref{app:masking}.  We also update the global mask to reflect any zeroed data,
\begin{equation}
G^{\rm abs}_{fdr} = G^{\rm stack}_{f} \times \Theta\!\left(\prod_{s}T_{\rm abs}^{s}(\nu_{f}, y_{d}, \phi_{r}) \right)
\end{equation}
where the product runs over all candidate absorbers $s$, so that $G^{\rm abs}_{fdr}$ vanishes exactly where the combined absorber taper is zero.

The effect of masking these candidates on the distribution of normalized voxel values is shown in \cref{fig:histogram_voxel}. Note that we have already applied the spatial mask of the point source foregrounds to the data (see \secref{sec:spatial_mask}) and study the effect of additional masking of these narrow-band features here.  Before masking (blue histogram), the combined map shows a substantial excess at high S/N due to the features, with some systems exceeding $\rm S/N > 10$. After masking (red histogram), the distribution becomes consistent with the jackknife map (green histogram), both showing approximately Gaussian profiles centered at S/N = 0. We note that both the masked combined map and the jackknife distribution are $10$–$15\%$ broader than the theoretical unit Gaussian, which is expected because our noise model does not incorporate sky temperature variations (see \secref{sec:power_spectrum_estimation:noise_covariance_estimation}).  The close agreement between the masked map and jackknife distributions confirms that our masking procedure effectively removes these narrow-band features while preserving the noise properties of the data.

\section{Stacking on the \lowercase{e}BOSS QSO Catalog}
\label{sec:stacking}

While the improvements to the offline pipeline described in Section~\ref{sec:pipeline} successfully remove many systematics present in our earlier processing, it is essential to verify that the new pipeline does not inadvertently suppress or remove the cosmological \tcm signal. To test this, we repeat the cross-correlation (stacking) analysis of \citet{chimestacking} using the newly processed dataset that will later be used for the auto–power spectrum measurement. This allows us to quantify any signal loss introduced by the updated pipeline.

In \citet{chimestacking}, we reported the detection of the cosmological \tcm emission by stacking CHIME’s foreground-filtered maps on the angular and spectral positions of LRGs, ELGs, and QSOs from the eBOSS \citep{dawson2016} clustering catalog, achieving detection significances ranging from $7\sigma$ to $11\sigma$ depending on the tracer. In the current analysis, we first apply a spatial mask, as described in \secref{sec:spatial_mask}, for consistency with the auto-correlation analysis, and perform stacking on the QSO sample \citep{lyke2020,ross2020}, which provides the highest source density (33,119 objects) within our redshift range ($1.01 < z < 1.34$), compared to the LRG and ELG catalogs (see Figure~2 of \citealt{chimestacking}).

\begin{figure}
    \centering
    \includegraphics[width=0.98\linewidth,keepaspectratio]{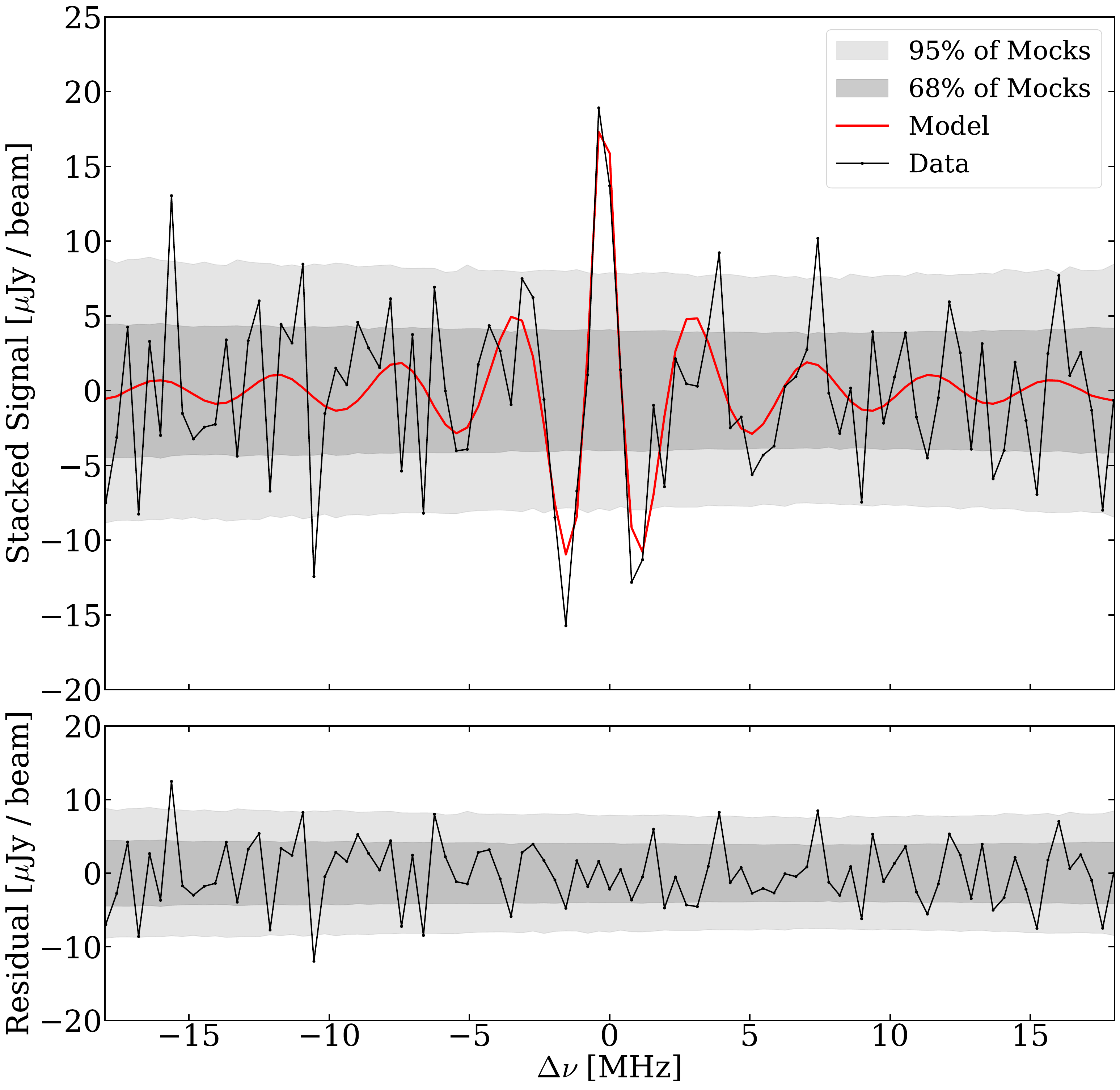}
    \caption{\emph{Top:} Stacked signal as a function of frequency offset for the eBOSS QSO catalog. The data are shown in black, and the best-fit model is shown in red. \emph{Bottom:} Residuals obtained by subtracting the best-fit model from the data. In both panels, the dark- and light-gray bands indicate the central 68\% and 95\% of values, respectively, obtained by applying the same stacking procedure to 10,000 random mock catalogs. }
    \label{fig:stacking_1D}
\end{figure}

\Cref{fig:stacking_1D} shows the stacked signal as a function of frequency offset. The black line shows the data, and the red line shows the best-fit signal model. The dark- and light-gray bands represent the 68\% and 95\% confidence intervals derived from stacking on 10,000 random mock catalogs. The bottom panel shows residuals after subtracting the best-fit model. The signal model, fitting procedure, and estimation of detection significance are described in \citet{chimestacking}. Following the same procedure, we again detect the \tcm signal in cross-correlation with the eBOSS QSO sample, with a detection significance of $9.1\sigma$.

Despite the substantial improvements in data quality, the resulting detection significance is slightly lower than that reported in \citet{chimestacking}. This is expected for several reasons. First, we analyze a narrower frequency range (\SIrange{608}{708}{\mega\hertz}) compared to the band (\SIrange{585}{800}{\mega\hertz}) used previously, resulting in fewer objects to stack on and a smaller cosmological volume.  Second, we apply a spatial mask that excludes approximately $33\%$ of the sky area (see \secref{sec:spatial_mask}), which was not applied in the previous analysis. Third, the foreground filtering differs between the two analyses: the previous analysis employed a declination- and polarization-dependent delay cutoff ranging from $\sim 120 - 180$ ns in the declination range analyzed here (see Figure 10 of \citealt{chimestacking}), while the current analysis uses a uniform 200 ns cutoff, resulting in greater signal attenuation.

To quantify the impact of these changes, we reanalyzed the dataset of \citet{chimestacking} using the same processing steps: applying a 200~ns delay cutoff for foreground filtering, selecting the frequency range \SIrange{608}{708}{\mega\hertz}, applying the absorber mask and spatial masks of bright point-source foregrounds, and then performing stacking on the QSO catalog. This yields a detection significance of $8\sigma$, compared to $9.1\sigma$ achieved with the newly processed data using identical analysis methods. This improvement demonstrates the enhanced sensitivity achieved by the current pipeline despite the more conservative foreground filtering. 

The successful recovery of the \tcm signal in cross-correlation with the updated dataset demonstrates that the improved processing pipeline preserves the cosmological signal while removing systematic contamination. This provides confirmation that the pipeline does not introduce significant signal loss, giving us confidence in proceeding with the \tcm auto–power spectrum measurement presented in the following sections.

\section{Power Spectrum Estimation}
\label{sec:power_spectrum_estimation}

In this section we describe how we estimate the power spectrum from the foreground–filtered maps.

\subsection{Delay Spectrum Estimation}
\label{subsec:delayspectrum}

We first form delay spectra by Fourier transforming each map pixel along the frequency axis. Because many channels are masked and a foreground filter has been applied along frequency, a naive discrete Fourier transform that zero-fills masked samples and ignores the filter will result in additional mixing between delay modes, inflating covariance between neighboring $k_{\parallel}$ bins. Our goal is to construct delay estimates with compact, well-behaved window functions in $k_{\parallel}$ so that subsequent binning remains efficient. We therefore estimate delay spectra using a linear matched-filter, inverse-covariance–weighted transform. This transform explicitly accounts for the mask and foreground filter and uses the propagated post-filter frequency–frequency covariance to weight frequency channels, thereby mitigating mode mixing and reducing the variance of the delay estimates relative to a naive transform.

Let $\map^{\rm Stokes}_{qfdr}$ denote the foreground–filtered map, with indices
$q$ (polarization), $f$ (frequency channel), $d$ (declination), and $r$ (local ERA).
In what follows we consider each map pixel $(q,d,r)$ independently and, to simplify notation, suppress these indices throughout this subsection.  We define the frequency–domain data vector for this pixel as
\begin{equation}
\label{eq:map_vector}
[\mathbf{m}]_{f} = \map^{\rm Stokes}_{qfdr} \qquad \text{for } f \in \left\{ 1 \ , \ldots \ , N_{\nu} \right\},
\end{equation}
where $N_{\nu}$ is the total number of frequency channels.  We model this data vector as
\begin{equation}
\label{eq:map_vector_forward_model}
\mathbf{m} = \mathbf{R}\,\tilde{\mathbf{m}} + \mathbf{n} \ ,
\end{equation}
where $\tilde{\mathbf{m}}$ denotes the underlying \tcm signal in delay space,
$\mathbf{n}$ is the (post–filter) instrumental noise in frequency space, and
$\mathbf{R}$ is the forward operator that maps delay to frequency and applies the same
frequency–axis operations as our pipeline,
\begin{equation}
\label{eq:fourier_projection_operator}
\mathbf{R} = \mathbf{W}\,\mathbf{\mask}\,\mathbf{H}\,\mathbf{F}.
\end{equation}
It consists of the following components:
\begin{itemize}
  \item $\mathbf{F}$: delay-to-frequency Fourier operator, given by
        \begin{equation}
            [\mathbf{F}]_{fa}= e^{j 2\pi \nu_f \tau_a} / \sqrt{N_{\nu}}
        \end{equation}

  \item $\mathbf{H}$: $\phi$-dependent, stacked foreground filter (see \cref{eq:stacked_filter}).

  \item $\mathbf{\mask}$:  $(y, \phi)$-dependent masking operator that zeroes channels that are missing or masked by an earlier stage of the pipeline,
        \begin{equation}
          [\mathbf{\mask}]_{ff'}=G^{\rm abs}_{frd}\,\delta_{ff'} \ .
        \end{equation}

  \item $\mathbf{W}$: operator that selects the sub-band of interest,
        \begin{equation}
            [\mathbf{W}]_{ff'} =
            \begin{cases}
              \delta_{ff'}, & \text{if } \nu_{\rm lower} \le \nu_f \le \nu_{\rm upper} \ , \\
              0, & \text{otherwise.} \\
            \end{cases}
        \end{equation}

\end{itemize}

Our estimate of the delay–domain \tcm signal for the pixel is obtained by applying a fixed linear operator to the frequency–domain data vector,
\begin{equation}
\mathbf{\hat{\tilde{m}}} = \mathbf{D} \,\mathbf{m} \ .
\end{equation}
The operator $\mathbf{D}$ implements inverse-covariance weighting followed by the adjoint frequency-to-delay projection,
\begin{equation}
    \label{eq:delay_operator}
\mathbf{D} \equiv \mathbf{R}^\dagger\, \bigl(\mathbf{C}_{\rm reg} + \mathbf{N}\bigr)^{-1} \ .
\end{equation}
Here $\mathbf{N} \equiv \langle \mathbf{n}\mathbf{n}^\dagger\rangle$ is the frequency–frequency noise covariance,  which can be expressed as
\begin{equation}
\mathbf{N} = \mathbf{W}\,\mathbf{\mask}\,\mathbf{\Sigma}\,\mathbf{\mask}\,\mathbf{W},
\end{equation}
with $\mathbf{\Sigma}$ obtained by propagating the sidereal-day averaged covariance from \cref{eq:stacked_noise_cov} through the map-making procedure.

In addition to the propagated noise covariance, we include a small regularisation term $\mathbf{C}_{\rm reg}$ in the total frequency–frequency covariance entering \cref{eq:delay_operator}. This matrix is chosen to approximate the expected \tcm covariance, calibrated from end-to-end simulations, and is small compared to the propagated noise covariance in the well-measured subspace.  We have verified that the measured power spectrum is insensitive to the overall normalisation of $\mathbf{C}_{\rm reg}$ over a wide range around this calibration (see \appref{app:varying_prior} for details).

Because fully masked channels produce zero rows and columns, $\mathbf{C}_{\rm reg}+\mathbf{N}$ in \cref{eq:delay_operator} is singular. We therefore perform the inversion on the unmasked frequency subspace (which is symmetric positive definite), compute the inverse via a Cholesky decomposition, and embed the result into the full matrix with zeros on masked rows and columns. This is equivalent to the Moore--Penrose pseudoinverse under masking \citep{barata:2012}.

This procedure is strictly linear in the data. In earlier work, we instead used a Gibbs-sampling algorithm to infer both the delay power spectrum and the corresponding delay–transform Wiener filter from the data \citep[][Appendix~A]{chimestacking}. While viable, this type of non-linear, data-dependent filter estimation can bias power-spectrum estimates if not implemented with care \citep{Ali2018,Cheng2018}. Here we avoid that class of systematics by constructing the delay transform operator from the propagated noise covariance and a simple, simulation-inspired model for the \tcm covariance.

We repeat this for every pixel $(q,d,r)$ and assemble
\begin{equation}
    \tilde{\map}^{\rm Stokes}_{qadr} \equiv \bigl[ \hat{\tilde{\mathbf{m}}} \bigr]_a  \ ,
\end{equation}
where $a$ indexes delay. The pixel-dependent delay–transform operators $\mathbf{D}_{qdr}$ are also saved and applied to the \tcm and noise simulations in order to interpret the measured power spectrum.

\subsection{Spatial Mask}
\label{sec:spatial_mask}

Residual contamination in the delay-filtered maps is spatially associated with the brightest continuum sources in the sky. Because our beamforming provides good localization in the telescope $y$ direction and the positions of these sources are well known, we can excise the affected regions using three types of smoothly tapered spatial masks.

\begin{enumerate}

\item \textbf{Transit Mask:} To suppress contamination from bright sources near transit, we apply an elliptical tapered mask, $T_{\rm transit}^{(s)}(y, \phi)$, to each continuum source $s$ in the \emph{specfind v3} catalog \citep{Stein2021} with flux density greater than \SI{10}{\jansky} at \SI{600}{\mega\hertz}. The extent of each mask is set by the primary beam width in $\phi$ and the synthesized beam width in $y$, both evaluated at the lowest frequency (\SI{608.2}{\mega\hertz}).

\item \textbf{Track Mask:} We observe residual foreground contamination from the brightest sources in the sky (e.g., Cygnus~A, Cassiopeia~A, Taurus~A, Virgo~A) even when they lie in the far sidelobes of the primary beam, appearing as ``U''-shaped tracks in the foreground-filtered maps (see \cref{fig:ringmap}). This arises from sparse sampling of east–west spatial modes by the array, which prevents precise localization of sources in the \ew direction and allows bright off-axis sources to contribute high-delay power that survives foreground filtering. To suppress this contamination, we construct extended track masks, $T_{\rm track}^{(s)}(y,\phi)$, for each catalog source $s$ with flux density greater than \SI{60}{\jansky} at \SI{600}{\mega\hertz}, following the full path of the source across the sky.

\item \textbf{Galactic Plane Mask:} For a specific range of local ERA in our field, we observe residual contamination in the foreground-filtered maps that is spatially associated with the track of the inner Galactic plane as it moves through the far sidelobes. This contamination is apparent at delay modes just below the threshold used in our final analysis, appearing at $\tau \lesssim 280~\mathrm{ns}$, which is within the transition region of the delay filter (see \secref{subsec:ps_estimation}). Nevertheless, to guard against possible leakage into the higher-delay modes used for power spectrum estimation, we conservatively apply a spatial mask, $T_{\rm gal}(y, \phi)$, to exclude these regions. We later verified that removing this mask has a negligible effect on the measured power spectrum.

\end{enumerate}

The combined spatial mask for bright continuum foreground sources is constructed by multiplying the individual mask functions:
\begin{equation}
\label{eq:final_mask}
\begin{aligned}
T_{\rm base}(y, \phi)
 &= T_{\rm gal}(y,\phi) \\
    &\quad \times \prod_{s \in \mathcal{S}_{10}} T_{\rm transit}^{(s)}(y,\phi) \\
 &\quad \times
    \prod_{s^{\prime} \in \mathcal{S}_{60}} T_{\rm track}^{(s^{\prime})}(y,\phi) \, .
\end{aligned}
\end{equation}
where $\mathcal{S}_{10}$ and $\mathcal{S}_{60}$ denote the sets of catalog sources above the \SI{10}{\jansky} and \SI{60}{\jansky} flux-density thresholds, respectively. This mask excludes approximately $33\%$ of the sky area.  The details of the mask construction, including the mathematical formulation and parameter choices, are provided in \appref{app:masking}.

\subsection{Power Spectrum Estimation}
\label{subsec:ps_estimation}
The final stage of the analysis is the estimation of the cosmological power spectrum from the delay-space map.  The procedure involves transforming the data into a 3D cosmological wavenumber space, forming a cross-power spectrum to remove noise bias, and then averaging the result into cylindrical and spherical bins to produce the 2D and 1D power spectra.

We first define a final spatial window function by multiplying the smoothly tapered mask $T_{\rm base}(y, \phi)$ constructed in \secref{sec:spatial_mask} with a 2D apodization $T_{\rm edge}(y,\phi)$ to reduce edge effects in the subsequent Fourier transform. The apodization is separable, $T_{\rm edge}(y,\phi) = \operatorname{Tukey}(y)\,\operatorname{Tukey}(\phi)$, where $\operatorname{Tukey}$ is a 1D Tukey window function with cosine fraction 0.5 (i.e., a flat central region over 50\% of the field with a cosine roll-off to zero at the edges):
\begin{equation}
    T(y,\phi) = T_{\rm base}(y,\phi) \times T_{\rm edge}(y,\phi) \, .
\end{equation}
We apply the same window function identically to all delay modes of the map:
\begin{equation}
    \tilde{\map}^{\rm windowed}_{qadr} = T(y_d,\phi_r) \, \tilde{\map}^{\rm Stokes}_{qadr} \, .
\end{equation}

We then transform the map from its native coordinates $(\phi, y)$ to spatial Fourier modes $(u, v)$ by applying a discrete 2D Fourier transform to the map for each delay mode $\tau_{a}$:
\begin{align*}
    \tilde{V}_{q}(u, v, \tau_{a})
    = \frac{1}{N_{y} N_{\phi}}
    \sum_{d} \sum_{r} &
       \tilde{\map}^{\rm windowed}_{qadr}
       e^{j 2\pi \big(u \,\Delta x_{r} + v \,\Delta y_{d}\big)} \,,
       \numberthis
\label{eq:spatial_transform}
\end{align*}
where
\[
\Delta x_{r} \equiv \cos\bar{\theta}\,(\phi_{r} - \bar{\phi}),
\qquad
\Delta y_{d} \equiv y_{d} - \bar{y}
\]
are local flat-sky Cartesian coordinates, defined as offsets from the centre of the field $(\bar{\phi}, \bar{y})$. The factor $\cos\bar{\theta}$ uses the average declination of the field to convert offsets in local ERA into approximate angular distances in the \ew direction for the entire field.  We denote the result by $\tilde{V}$ to emphasize that the 2D Fourier transform of the delay–space map returns us to a visibility-like representation in delay space.

The discrete Fourier transform in \cref{eq:spatial_transform} invokes the flat-sky approximation, which is valid for our analysis because we exclude the intracylinder baselines (as mentioned in \secref{sec:mapmaking}) \citep{Datta_2007,Eastwood_2019}. This exclusion removes sensitivity to large angular scales probed by the shortest pure \ns baselines, restricting our measurements to  $k_\perp$  modes beyond 0.08 $\ihMpc$, or $\ell > 200$, (see Figure~\ref{fig:ps_2D}) where the curvature of the sky is negligible over the relevant field of view.

The instrumental coordinates $(u, v, \tau)$ are then converted to comoving cosmological wavenumbers $\mathbf{k} = (k_x, k_y, k_{\parallel})$ using the standard relations for the central redshift of our observation as \citep{Morales2004ApJ...615....7M, Liu&Shaw2020}: 
\begin{align} k_x = \frac{2\pi u}{D_c(z)}, \quad k_y = \frac{2\pi v}{D_c(z)}, \quad k_\parallel = \frac{2\pi \nu_{21} H_0 E(z)}{c (1+z)^2} \tau, 
\end{align} 
where  $\nu_{21} = \SI{1420.40575}{\mega\hertz}$ is the rest-frame frequency of the \tcm line, $D_c(z)$ is the comoving transverse distance, $E(z) = \sqrt{\Omega_m(1+z)^3 + \Omega_\Lambda + \Omega_{k}(1+z)^{2}}$, $z$ corresponds to the central redshift of the band,  and $H_{0}$  is the Hubble constant \citep{Hogg1999}.

As described in \secref{sec:sidereal_day_averaging}, to remove noise bias, we split the 94 sidereal days into two interleaved partitions, labeled ``even'' and ``odd''.  Each partition is processed independently through an identical analysis pipeline. The power spectrum is then computed by cross-correlating the resulting data cubes. Since the noise is uncorrelated between different sidereal days, this produces an estimate of the power spectrum without any noise bias.  The 3D power spectrum is then given by \citep{Mertens2020}
\begin{align}
\label{eqn:PS3D}
\hat{P}_{q}(k_x,k_y,k_{\parallel})
= \mathcal{V}_{\rm s} \
\tilde{V}^{\rm even}_{q}(k_x, k_y, k_{\parallel}) \ 
\tilde{V}^{\mathrm{odd}, *}_{q}(k_x, k_y, k_{\parallel}) \ , \notag \\[+0.5ex ]
\end{align}
where $\mathcal{V}_{\rm s}$ is the observed comoving volume of the data cube in units of ($h^{-3} ~ \rm Mpc^{3}$), given by
\begin{equation}
\label{eq:survey_volume}
\mathcal{V}_{\rm s} = \frac{\mathcal{A}_{s} D_c^2(z) N_{\nu} \Delta_{\nu} \Delta D}{f_{\rm window}} \ .
\end{equation}
Here $\mathcal{A}_{\rm s}$ is the survey area defined in \cref{eq:survey_area}, $N_{\nu}\Delta_\nu$ is the total bandwidth of the observation, and $\Delta D = \frac{c(1+z)^2}{H(z)\nu_{21}}$ converts that bandwidth into comoving line-of-sight depth. The factor $f_{\rm window} = \frac{\left(\sum_{dr} T_{dr}\right)^{2}}{\sum_{dr} T_{dr}^{2}}$ accounts for the reduction in survey area due to the window function.

Assuming isotropic and translationally invariant signal statistics, we average the 3D power spectrum into cylindrical and spherical bins. The cylindrically averaged 2D power spectrum $\hat{P}_{q}(k_\perp, k_\parallel)$ is obtained by averaging $\hat{P}_{q}(k_x,k_y,k_{\parallel})$ in annular bins of constant perpendicular wavenumber $k_\perp = \sqrt{k_x^2 + k_y^2}$.

\begin{figure}
    \centering
    \includegraphics[width=0.98\linewidth,keepaspectratio]{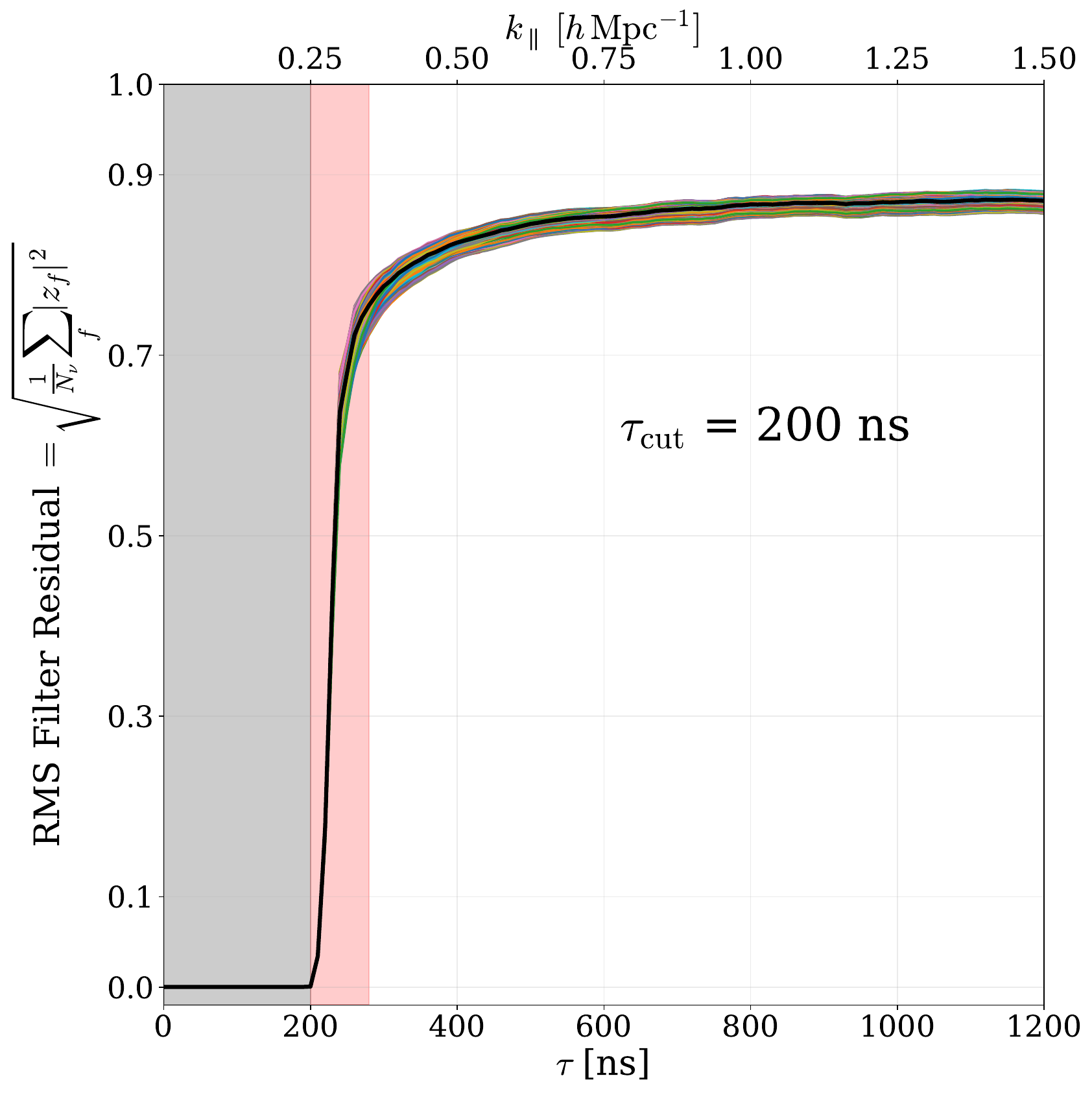}
    \caption{RMS of the residual after applying the stacked foreground filter (\cref{eq:stacked_filter}) to a sinusoid of unit amplitude and delay $\tau$. At the top, we show the $k_{\parallel}$ modes corresponding to the delays estimated at the central redshift of the band ($z = 1.16$). The filter depends on RFI mask, which varies over local ERA and ranges from $15-21\%$. Different colors correspond to different local ERA bins, and the black line is the median over local ERA. There is an overall local-ERA--dependent attenuation, reflecting the effect of varying the RFI mask. However, it does not bias the power spectrum estimate as the application of the filter is a linear operation and applied to data and \tcm simulation identically.
    We see that below the filter cutoff (\SI{200}{\nano\second}), the delay modes are strongly attenuated (gray shaded region) and there is a transition zone (pink shaded region)  where the filter response gradually increases from near-zero and reaches a plateau at high delay modes. Beyond \SI{280}{\nano\second} the attenuation is less than $10\%$ relative to the high-delay plateau. In this analysis, we exclude all delay modes below \SI{280}{\nano\second}.} 
    \label{fig:filter_rms}
\end{figure}

The foreground filter strongly attenuates power at delays below the cutoff, $\tau_{\rm cut} = \SI{200}{\nano\second}$, where spectrally smooth foregrounds are expected to dominate. At higher delays ($\tau > \tau_{\rm cut}$), the attenuation decreases only gradually with delay, resulting in residual suppression of the cosmological signal across a broad transition region (see \S3.5 of \citealt{ewall-wice2021} for further discussion on signal attenuation). To quantify the effective attenuation outside the nominal cutoff, we compute the root-mean-square (RMS) amplitude of the filter residual as a function of delay, following the methodology of \citet{ewall-wice2021}. In \cref{fig:filter_rms}, we plot the RMS of the filter residual, $\vec{z}^{\tau} = \vec{H}\vec{x}^{\tau}$, where $\vec{x}^{\tau}$ is a complex sinusoid with delay $\tau$ and amplitude equal to unity. Here $\vec{H}$ is the sidereal day averaged filter described in \cref{eq:stacked_filter}. The filter depends on the RFI mask and can vary with local ERA, as different local ERA bins have different flagged frequency channels. The frequency flagging fraction varies between $15-21\%$ over local ERA bins. We compute the filter RMS for different local ERA bins and show them in different colors in Figure~\ref{fig:filter_rms}, with the median over all local ERA bins shown in black. The filter exhibits an overall attenuation, and the scatter across local ERA bins reflects variations induced by the changing RFI mask.  Crucially, this local-ERA--dependent overall attenuation does not bias our power spectrum estimates because the filter is applied as a linear operator identically to the data and \tcm simulations. Any suppression of the signal is therefore properly forward modelled, ensuring an unbiased comparison. We find that beyond $\tau \sim \SI{280}{\nano\second}$, the median filter attenuation is less than $10\%$ relative to the high-delay plateau as shown in \cref{fig:filter_rms}. To avoid these attenuated modes beyond the filter cut-off ($\tau_{\rm cut}$), we generate a conservative 2D mask $W(k_\perp, k_\parallel)$, where we exclude all delay modes $|\tau| < \SI{280}{\nano\second}$, which corresponds to $k_{\parallel,\rm min} = 0.35 ~h ~\rm Mpc^{-1}$.

Finally, we compute the spherically averaged power spectrum by averaging the masked 2D power spectrum in logarithmically spaced spherical shells of constant wavenumber magnitude, $k = |\mathbf{k}| = \sqrt{k_\perp^2 + k_\parallel^2}$, given by:
\begin{equation}
\label{eq:PS1d}
\hat{P}_{q}(k_i) = \frac{\sum_{j} W(k_{\perp,j}, k_{\parallel,j}) \ \hat{P}_{2D}(k_{\perp,j}, k_{\parallel,j})}{\sum_{j} W(k_{\perp,j}, k_{\parallel,j}) },
\end{equation}
where the sum is over all $(k_\perp, k_\parallel)$ bins that fall within the $i$'th $k$-shell. Since the data are already optimally weighted in the map-making and delay-transform estimation stages, we use a uniform weight while averaging the power spectrum into spherical bins. 

To convert the power spectrum estimates into physical units of $\mathrm{K}^{2} ~h^{-3} ~\mathrm{Mpc}^{3}$, we employ a simulation-based approach. First, we simulate a sky map corresponding to a flat power spectrum with unit variance ($P({k}) = 1 ~\mathrm{K}^{2} ~h^{-3} ~\mathrm{Mpc}^{3}$). From this map, we generate mock visibility data following the procedure outlined in \citet{chimestacking} (see also \citealp{shaw2015}).  This simulated visibility data is processed through the same analysis pipeline used for the actual data to estimate its power spectrum. The resulting power spectrum gives the conversion factor, in both 2D ($\hat{P}_{c}(k_{\perp}, k_{\parallel})$) and 1D ($\hat{P}_{c}(k)$). We apply this single factor to the estimated power spectra from the observed data, the cosmological \tcm signal simulation, and the noise realization. This procedure places all quantities in consistent physical units, enabling a direct, one-to-one comparison.

\subsection{Noise Covariance Estimation}
\label{sec:power_spectrum_estimation:noise_covariance_estimation}

Our goal in this section is to obtain an accurate estimate of the noise covariance, i.e., the covariance between the power measured in different $k$–bins that arises solely from instrumental and sky noise. This quantity is needed to assess the detection significance of the measured power spectrum, to perform internal consistency and null tests, and to propagate uncertainties to any fitted model parameters. At the visibility level we have a reliable estimate of the noise variance from fast–cadence, even–odd differencing (see \secref{app:rfi_radiometer}), and this variance is propagated through the subsequent linear operations in the pipeline. However, those same operations (NS beamforming, foreground filtering, sidereal rebinning and stacking, mapmaking, delay transforming, spatial masking, spatial Fourier transforming, and $k$–space binning) also couple samples along multiple axes, such that the noise acquires covariances in frequency, local ERA, telescope $y$ coordinate, etc. Tracking those covariances would require storing an array with an additional axis for every data dimension, which is computationally prohibitive for our data volumes. Instead, we estimate the covariance by forward (Monte Carlo) propagation: we generate random noise realizations at the visibility level, process them using the same pipeline that was applied to the data, and take the sample covariance of the resulting power spectra. This is complicated by the fact that the daily beamforming and foreground filtering introduce covariance at the individual-day level, and it is not feasible to generate and process noise realizations for every day. Instead, we require a stack–level noise description that already incorporates the correlations produced by those daily operations.

To provide this, the pipeline propagates and outputs the frequency–frequency noise covariance of the sidereal–stacked, foreground–filtered, hybrid–beamformed visibilities, $\Sigma_{p e r f f'}^{\rm stack}$, where $p$ is the polarization index, $e$ is the east–west baseline index, $r$ is the local ERA index, and $f,f'$ are frequency indices (see \cref{eq:stacked_noise_cov}). This quantity already contains the correlations introduced by the daily foreground filtering, but it no longer carries an explicit \ns baseline index because the beamforming step has collapsed that axis. We applied the foreground filter after beamforming so that a single frequency–frequency operator could be used for all telescope $y$ coordinates. This both reduced the computational cost of constructing the filter and made it feasible to propagate the resulting frequency–frequency noise covariance, since it is identical for every telescope $y$ coordinate. Consequently, $\Sigma^{\rm stack}$ tells us how the noise is correlated across frequency after beamforming, but not how that noise should be apportioned among the individual \ns baselines before beamforming. This apportioning is important because the noise assigned to each \ns baseline will determine the noise variance as a function of $k_y$.

To recover a visibility–level description, we distribute the propagated frequency-frequency noise covariance over the individual \ns baselines. We assume that the noise covariance on a \ns grid point $n$ scales with the redundancy $\redundancy_{p e n}$ of that baseline,
\[
    \sigma^2_{p f e n r} \propto \frac{1}{\redundancy_{p e n}},
\]
which captures the dominant trend seen in the data, with residual structure at the $\lesssim 10\%$ level (see \cref{fig:beamforming_weights}). Under this assumption, we define a scaling factor
\begin{equation}
    \label{eq:scale_NS}
    J_{p e f f'} \equiv \sum_{n} \frac{W^{\rm NS}_{f n} W^{\rm NS}_{f' n}}{\redundancy_{p e n}} \, ,
\end{equation}
where $W^{\rm NS}_{f n}$ is the beamforming window given by \cref{eq:normalized_beamforming_weights,eq:beamforming_weights}.  This quantity is used to construct a visibility–level covariance
\begin{equation}
    \label{eq:cov_scaled}
    \Sigma_{p e r f f'}^{\rm vis} = \frac{\Sigma_{p e r f f'}^{\rm stack}}{J_{p e f f'}} \, .
\end{equation}
This is precisely the scaling factor required so that, if each baseline $n$ is assigned noise with covariance $\Sigma_{p e r f f'}^{\rm vis} / \redundancy_{p e n}$ and we apply the same beamforming window $W^{\rm NS}_{f n}$ used on the data, the beamformed noise will have the measured frequency–frequency covariance $\Sigma_{p e r f f'}^{\rm stack}$.

The matrix $\boldsymbol{\Sigma}_{p e r}^{\rm vis}$ is Hermitian and positive-definite, so we take its Cholesky factorization
\begin{equation}
    \boldsymbol{\Sigma}_{p e r}^{\rm vis} = \mathbf{L}_{p e r} \, \mathbf{L}_{p e r}^{T},
\end{equation}
where $\mathbf{L}_{p e r}$ is lower triangular with positive diagonal entries. We generate independent, circularly symmetric complex Gaussian noise,
\begin{equation}
    Z_{p f e n r} \sim \mathcal{G}(0,1),
\end{equation}
and impose the desired frequency correlations and redundancy scaling via
\begin{equation}
    \label{eq:vis_noise}
    V^{\rm noise}_{p f e n r} = \frac{1}{\sqrt{\redundancy_{p e n}}} \sum_{f'} L_{p e r f f'} \, Z_{p f' e n r} \, .
\end{equation}
Applying the north–south beamforming operator (\cref{eq:beamform_ns}) to $V^{\rm noise}_{p f e n r}$ then produces hybrid–beamformed noise visibilities whose frequency–frequency covariance matches the propagated  $\Sigma^{\rm stack}$, while retaining a baseline dependence consistent with the assumed redundancy model.  For each noise realization, we construct a map, apply the absorber mask, perform the delay transform, and apply the spatial mask, using exactly the same pipeline that was applied to the data (see \secref{sec:pipeline}). This is done separately using the covariance matrix $\Sigma^{\rm stack}$ from the even and odd partition of sidereal days.

\paragraph{Spatially Varying Sky/Self Noise}
The construction above assumes that visibility noise is uncorrelated between baselines.   While receiver noise, which originates independently in each antenna's electronics, is indeed uncorrelated between baselines, the contribution from sky temperature is correlated, leading to covariance between different baselines. This effect, often termed ``self-noise'', results in a non-uniform noise distribution in the final synthesized map, where the noise power is modulated by the sky brightness distribution \citep{Kulkarni1989}. Consequently, when we beamform (Section \ref{sec:beamforming}) to a particular location in the sky, certain regions will be noisier than others depending on the local sky intensity.

To account for this spatial variation, we construct a jackknife map by differencing the data map constructed from the even and odd partition of sidereal days. This cancels the sky signal, which is coherent across sidereal days, leaving a realization of the noise that captures the spatially-dependent contribution from $T_{\rm sky}$.  We scale this difference by the square root of the inverse variance propagated from the fast-cadence estimate of the noise.  We then compute the root-mean-square (RMS) of this normalized difference map along the frequency axis to produce a 2D spatial map, $\sigma_{\rm spatial}(\theta,\phi)$, characterizing the noise variation across the field relative to our expectation. We apply this spatial scaling factor to the delay transform of each noise realization map before computing its power spectrum. This enforces the correct large–scale variation of the noise across the field.  The assumption here is that the sky–noise modulation is spectrally smooth, so a single factor per sky pixel is adequate; in practice there is a residual frequency dependence in the sky noise at the $\lesssim 10\%$ level, which we ignore.

\begin{figure}
   \centering
   \includegraphics[width=1.0\linewidth]{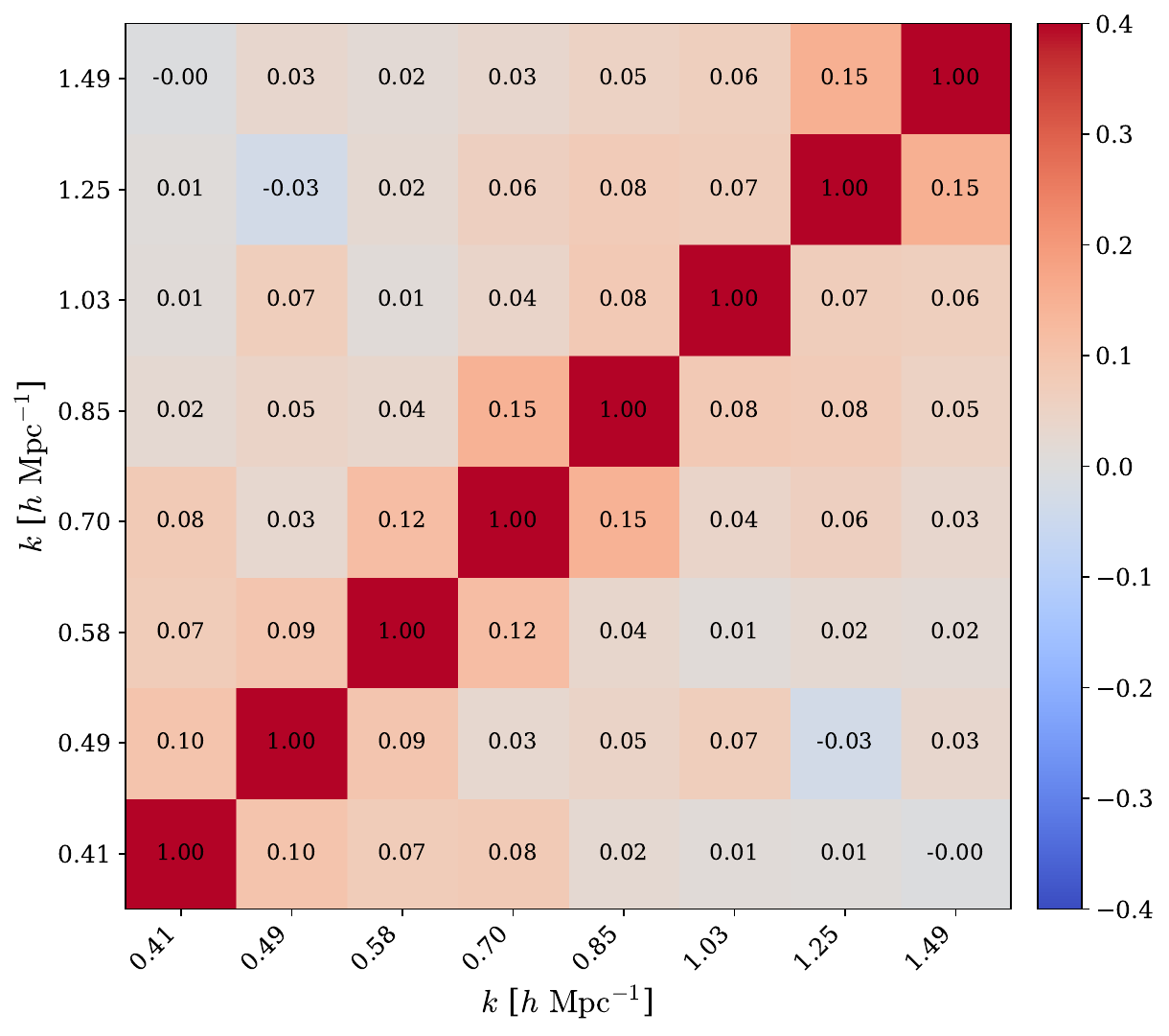}
   \caption{Correlation matrix of the power–spectrum $k$–bins, estimated from 1000 noise-only simulations. The matrix is nearly diagonal, with off–diagonal correlations largest for neighboring bins but $\lesssim 15\%$. Such correlations are expected due to the windows and masks applied by the pipeline in the map and frequency domain. These correlations are accounted for when fitting the signal model to the data (\secref{sec:results:theoretical_interpretation:model_fitting}).}
   \label{fig:noise_cov}
\end{figure}

\paragraph{Monte Carlo Estimate of the Noise Covariance}
Next, the spatial transform of each noise realization is taken, the even and odd partitions are cross–correlated to remove the overall bias, and the resulting power spectrum is binned in an identical manner as the data.  One pass through this pipeline produces one noise power spectrum (2D and 1D).  We repeat this $N_{\rm samp} = 1000$ times with independent draws of $Z$ to build up an empirical estimate of the power–spectrum noise covariance:
\begin{equation}
    \hat{\mathbf{C}}_{\rm N} = \frac{1}{N_{\rm samp} - 1} \sum_{m=1}^{N_{\rm samp}}
    \bigl( \vec{p}_{m} - \bar{\vec{p}} \bigr)
    \bigl( \vec{p}_{m} - \bar{\vec{p}} \bigr)^{T},
\end{equation}
where $\vec{p}_{m}$ is the vector of bandpowers for the $m$th realization and $\bar{\vec{p}}$ is their mean. \Cref{fig:noise_cov} shows the corresponding correlation matrix. It is nearly diagonal, with off–diagonal correlations $\lesssim 15\%$.  These correlations arise primarily from multiplicative operations applied by the pipeline in the spectral/spatial domain (windowing, RFI masking, absorber and point–source masking), which mix Fourier modes.  We use $\hat{\mathbf{C}}_{\rm N}$ when fitting the signal model to the measured power spectrum (see \secref{sec:results:theoretical_interpretation:model_fitting}). The uncertainties on each bandpower are estimated from the standard deviation of 1000 noise power spectra.

\paragraph{Limitations}
This procedure is not exact, for several reasons. (i) The real dependence of the noise variance on NS baseline does not follow $1/\redundancy_{p e n}$ perfectly, so individual $k_y$ bins can differ from the true variance at the $\lesssim 10\%$ level even though the average trend is captured. (ii) Correlations introduced by the time–to–$\phi$ rebinning, where each time sample can contribute to two adjacent $\phi$ bins, are not explicitly modeled. (iii) Sky/self noise is enforced via a single spatial template that is assumed constant over frequency, whereas the true sky noise has a mild frequency dependence that is not captured in our noise model. Nevertheless, the resulting noise realizations agree with the jackknife–measured noise level to better than $\lesssim 10\%$ in any domain, and, by construction, capture the dominant correlations introduced by the pipeline, while remaining unbiased on average.

\section{Results}
\label{sec:results}
Building on the data reduction pipeline described above, we now present the resulting measurement of the cosmological \tcm power spectrum and its theoretical interpretation.

\begin{figure*}
   \centering \includegraphics[width=1\linewidth, keepaspectratio, trim = 0 0 0 0]{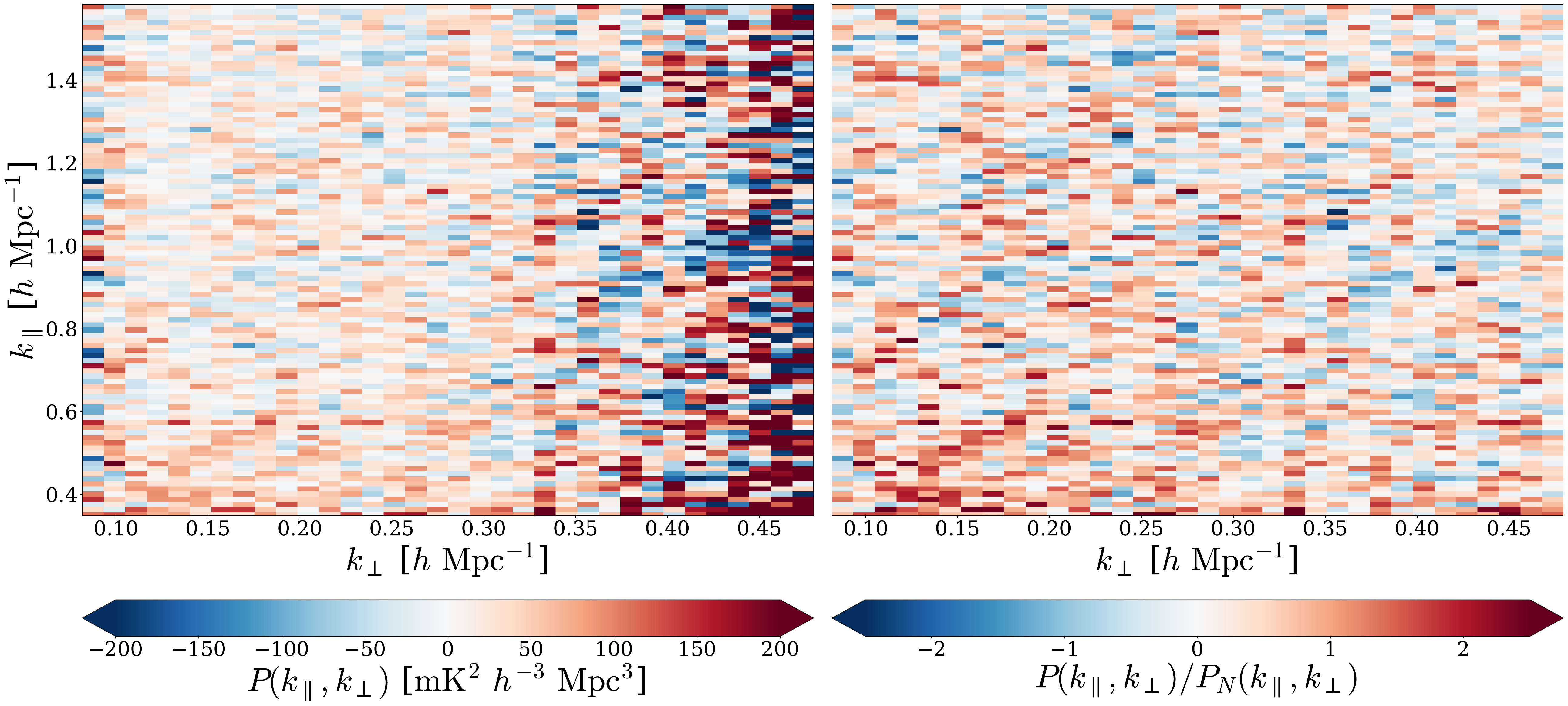}
    \caption{\emph{Left panel:} The real part of the cylindrically averaged power spectrum for Stokes-I data.  \emph{Right panel:} the power spectrum normalized by the power spectrum of expected thermal noise. The measured 2D power spectrum is consistent with the expected thermal noise and shows no evidence of significant contamination from residual foregrounds or other systematics.}
    \label{fig:ps_2D}
\end{figure*}

\subsection{Power Spectrum Measurement}
\label{sec:power_spectrum}

The cylindrically averaged 2D power spectrum, $P(k_\perp, k_\parallel)$, is shown in Figure~\ref{fig:ps_2D}. The power spectrum is estimated from Stokes-$I$ map over the frequency range \SIrange{608.2}{707.8}{\mega\hertz} ($z=1.34$ to $1.01$), corresponding to a central redshift of $z \sim 1.16$.  The left panel shows the real component of the power spectrum, and the right panel shows the signal-to-noise ratio (SNR),  defined as $P(k_\perp, k_\parallel)/P_{N}(k_\perp, k_\parallel)$. 
Although we do not yet have the sensitivity for detection of the \tcm power spectrum in this $(k_\perp, k_\parallel)$ domain with high SNR, which would enable us to constrain line-of-sight effects, the cylindrically-binned 2D power spectrum remains an effective diagnostic tool for identifying any residual systematic contamination. Note that we employ very fine binning along the line-of-sight direction $k_{\parallel}$, without further averaging beyond the native spectral resolution ($\Delta k_{\parallel} = 0.012 \ihMpc$). A coarser binning in $k_{\parallel}$ would reduce the noise variance and could reveal tentative evidence of the 21 cm signal in this 2D $(k_{\perp}, k_{\parallel})$ space. However, we defer such an analysis, as well as the coherent integration of multi-year observations to further improve sensitivity, to forthcoming work.

The measured power spectrum is complex-valued as we cross-correlate two data cubes with independent noise and hence can fluctuate to negative values. The power spectrum of a coherent sky signal is positive by construction and confined to the real component, leaving the imaginary component to be noise-dominated. \cref{fig:ps_2D} shows the real component of the 2D power spectrum fluctuating between positive and negative values, with SNR varying between  $\pm 3 \sigma$. The 2D power spectrum shows excellent agreement with expected thermal noise and shows no evidence of any significant systematic contamination.

\begin{figure*}
   \centering \includegraphics[width=1\linewidth, keepaspectratio, trim = 0 0 0 0]{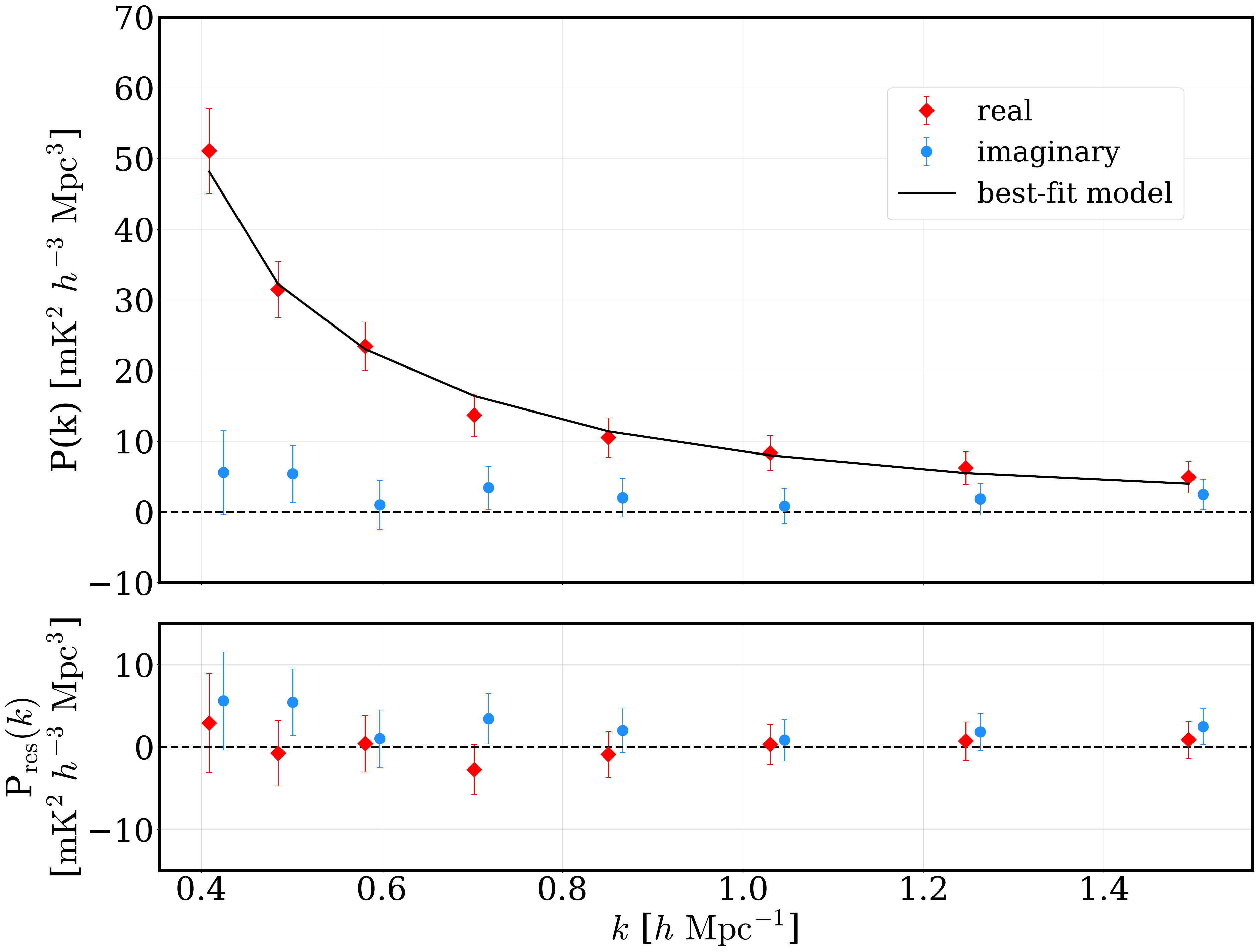}
    \caption{\emph{Top:} The real component of the spherically averaged 1D Stokes-$I$ power spectrum (in red) estimated over a frequency range \SIrange{608.2}{707.8}{\mega\hertz} ($z=1.34$ to $1.01$), corresponding to a central redshift of $z \sim 1.16$. The best-fit model to the data points is shown in black. The uncertainties on each measured bandpower are derived from the standard deviation of 1000 noise power spectra (see \secref{sec:power_spectrum_estimation:noise_covariance_estimation}).
    The best-fit model, described in \secref{sec:results:theoretical_interpretation}, {yields a $p$-value of $0.92$}, indicating an excellent fit.  \emph{Bottom:} The residual power spectrum, after subtracting the best-fit model from the data (in red). In both panels, we show the imaginary components of the power spectrum (in blue), slightly offset in $k$ for clarity. The power spectrum is complex-valued, and a coherent signal from the sky will be confined to the real component, while the imaginary component should be noise-dominated (see \secref{sec: validation:imaginary_component}).  Both residuals and the imaginary component are consistent with the estimated noise level.
    }
    \label{fig:ps_1D}
\end{figure*}

We further average the 2D power spectrum in spherical bins of constant wavenumber $k$, assuming isotropy of the cosmological signal. The spherically binned 1D power spectrum is shown in Figure~\ref{fig:ps_1D}. The top panel shows the real part of the power spectrum in red, and the black line is the best-fit model of the cosmological \hi\ signal, as discussed in \secref{sec:results:theoretical_interpretation}. The bottom panel shows the result of subtracting the best-ﬁt model from the data.   We find that the residual power spectrum is consistent with noise. Since $k_\perp$ is limited to $<0.48 \ihMpc$, while the $k_\parallel$ extends to $1.58 \ihMpc$ (see Figure~\ref{fig:ps_2D}), the spherically-binned 1D power spectrum at $k > 0.4 \ihMpc$ is primarily sensitive to the line-of-sight modes.

\begin{figure*}
   \centering \includegraphics[width=1\linewidth, keepaspectratio, trim = 0 0 0 0]{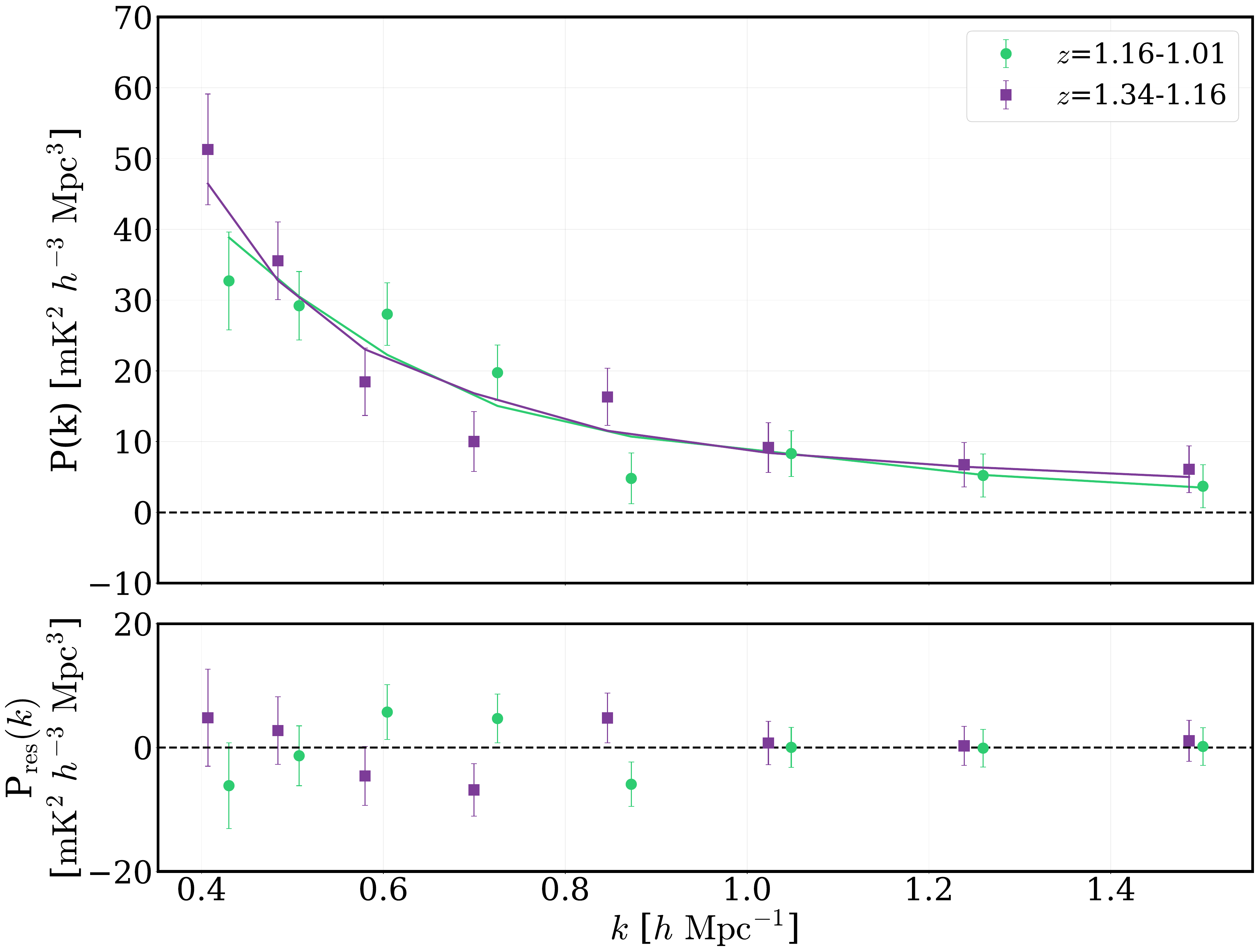}
    \caption{The power spectrum of the upper sub-band spanning 658.2–707.8 MHz ($z = 1.16$-$1.01$) and lower sub-band spanning 608.2–658.2 MHz ($z = 1.34$-$1.16$). The top panel shows the real component of the power spectrum and the best-fit model for each sub-band. The bottom panel shows the residual obtained after subtracting the best-fit model from the data. In both sub-bands, the model agreement is excellent, and the residuals are consistent with the expected noise levels. The imaginary component of the power spectrum for both sub-bands is also consistent with noise and is not shown for clarity.
}
    \label{fig:ps_freq_bins}
\end{figure*}

We also divide the full bandwidth into two independent sub-bands of approximately 50 MHz each.  The upper sub-band spans $658.2$-$707.8$ MHz ($z=1.16$-$1.01$), centered at $z \sim 1.08$, whereas the lower sub-band spans $608.2$-$658.2$ MHz ($z=1.34$-$1.16$), centered at $z \sim 1.24$.  Although the spatial mask remains identical to that used for the full band, the RFI masks are different. Our RFI flagging pipeline removes approximately $50\%$ of frequency channels in the upper sub-band but only $17\%$ in the lower sub-band. Figure \ref{fig:ps_freq_bins} shows the measured power spectrum for both sub-bands.  The top panel shows the real component of the power spectrum along with the best-fit model, while the bottom panel shows the residuals.  Note that we generate an independent signal model for each of these sub-bands and fit them to the data. We find an excellent fit to both the sub-bands, as well as for the full frequency band.  The description of the signal model and fitting procedures, including detection significance, for both the full band and the sub-bands, is described below.

\subsection{Theoretical Interpretation}
\label{sec:results:theoretical_interpretation}

Our companion paper \citep{chime-auto-interpretation-paper} describes our theoretical model for interpreting the measured power spectrum, along with how this model is evaluated, the constraints we have obtained on its parameters, and the information this provides about the clustering of \hi. In this paper, we use the modelling approach from \citet{chime-auto-interpretation-paper} to determine the best-fit power spectrum model that is plotted alongside our measurements, and to assess the significance of our detection. Therefore, in this subsection we briefly describe the relevant aspects of the modelling, referring the reader to \citet{chime-auto-interpretation-paper} for further details.

\subsubsection{Power Spectrum Model}
\label{sec:results:theoretical_interpretation:power_spectrum_model}

We interpret the measured power spectrum using the following model:
\begin{align*}
P_{21}(k, \mu; z) 
	&= \Tbar(z)^2
	\left[ \bHI(z) + f(z) \mu^2 \right]^2 \\
&\quad \times
	P_{\rm m}(k, z)
	\DFoG{\sHI}(k\mu; z)^2\ ,
\numberthis
\label{eq:P21-theory}
\end{align*}
where
\begin{align*}
P_{\rm m}(k, z) 
	&= \left[ \frac{D^+(z)}{D^+(z_{\rm fid})} \right]^2
	\bigg\{ \PL(k, z_{\rm fid})  \\
&\quad\;\;+ \alphaNL \left[ 
		\PNL(k, z_{\rm fid}) - \PL(k, z_{\rm fid}) 
	\right] \bigg\}\ .
\numberthis
\label{eq:Pm-model}
\end{align*}
The \tcm power spectrum is proportional to the square of the mean brightness temperature $\Tbar(z)$, which is itself proportional to the mean \hi\ density $\OmegaHI(z)$. The linear \hi\ bias $\bHI(z)$ relates the large-scale clustering of \hi\ to the clustering of matter, and is added to the Kaiser term $f\mu^2$ \citep{kaiser1987} to incorporate the leading effect of redshift-space distortions on large scales. We take $\OmegaHI(\zeff)$ and $\bHI(\zeff)$, with $\zeff=1.16$, as free parameters in our analysis, assuming that the redshift dependence of $\OmegaHI(z)$ and $\bHI(z)$ follow the \citet{crighton2015} fitting function and the IllustrisTNG $\bHI(z)$ fitting function from \citet{chimestacking}, respectively.

We model the scale-dependence of the power spectrum using two factors. The first is a matter power spectrum, \cref{eq:Pm-model}, that uses a single parameter $\alphaNL$ to interpolate between the linear version and the nonlinear model from \citet{mead2021}, using the linear growth factor $D^+(z)$ to rescale a fiducial model evaluated at $z_{\rm fid}=1$. 
We use the \texttt{CAMB} package \citep{lewis1999} to compute these matter power spectra.
The second is the square of a Lorentzian damping factor $\DFoG{\sHI}(k\mu; z)$, with damping scale $\sigma_{\rm FoG}(z)$, that approximates ``Finger-of-God'' redshift-space distortions \citep{jackson1972} from small-scale \hi\ velocities. Along with $\alphaNL$, we use $\alphaFoG \equiv \sigma_{\rm FoG}(\zeff)/ \sigma_{\rm FoG}^{\rm (fid)}(\zeff)$ as a free parameter, where $\sigma_{\rm FoG}^{\rm (fid)}(z)$ is the fitting function from \citet{chimestacking}, based on simulation measurements from \citet{sarkar2019}. In summary, the power spectrum model in Equations~\eqref{eq:P21-theory}-\eqref{eq:Pm-model} has four free parameters: $\OmegaHI$, $\bHI$, $\alphaNL$, and $\alphaFoG$.

This model is undoubtedly simplistic, and we emphasize that $\alphaNL$ and $\alphaFoG$ should be viewed as phenomenological shape parameters, intended to capture a range of plausible shapes for the nonlinear \hi\ power spectrum without having easily-interpretable physical meanings.
There are several reasons why this model is appropriate for our analysis:
\begin{enumerate}
\item At the signal-to-noise ratio and spatial scales of our measurement, any parameters describing the shape of the power spectrum (in our case, $\alphaNL$ and $\alphaFoG$) are highly degenerate with parameters describing the amplitude ($\OmegaHI$ and $\bHI$). Therefore, using a more detailed model with a larger number of parameters likely would not yield further useful information.
\item In \citet{chime-auto-interpretation-paper}, we describe several simulation-based tests that demonstrate that, at the signal-to-noise ratio of our measurement, our model allows us to obtain unbiased constraints on the power spectrum amplitude parameter $\AHIps$ (see below) when the shape parameters $\alphaNL$ and $\alphaFoG$ are marginalized over.
\item Any power spectrum model must be propagated through a transfer function that is specific to the analysis we have performed. As described in \citet{chime-auto-interpretation-paper}, this propagation can be carried out efficiently for the specific model from Equations~\eqref{eq:P21-theory}-\eqref{eq:Pm-model}, using a simulation-based template approach based on the CHIME-eBOSS stacking analysis in \citet{chimestacking}.
\end{enumerate}

\subsubsection{Model Fitting}
\label{sec:results:theoretical_interpretation:model_fitting}

We use the \texttt{emcee} package \citep{foreman-mackey2013} to perform Markov Chain Monte Carlo sampling from the posterior distribution for the four parameters in the model from \secref{sec:results:theoretical_interpretation:power_spectrum_model}. We use a Gaussian likelihood,
\beq
\calL(\vec{\theta}) 
	= \frac{1}{|2\pi \CNoise|^{1/2}} \exp\!\lp -\frac{\chi^2(\vec{\theta})}{2} \rp\ ,
\eeq
with
\beq
\label{eq:chi2theta}
\chi^2(\vec{\theta}) 
	\equiv \left[ \vec{d} - \vec{s}(\vec{\theta}) \right]^{T} 
	\CNoise^{-1} 
	\left[ \vec{d} - \vec{s}(\vec{\theta}) \right]\ ,
\eeq
where $\vec{d}$ is the data vector, $\vec{s}(\vec{\theta})$ is the model prediction evaluated at parameter values $\vec{\theta}$, and $\CNoise$ is
the noise covariance described in \secref{sec:power_spectrum_estimation:noise_covariance_estimation}. 
When evaluating \cref{eq:chi2theta}, we multiply $\CNoise^{-1}$ by a ``Hartlap factor" $(n_{\rm mocks} - n_{\rm data} - 2)/(n_{\rm mocks} - 1) \approx 0.99$ to remove bias due to the finite number of noise realizations used to estimate $\CNoise$ \citep{hartlap2007}.
We run chains with 12 walkers for 5,000,000 samples each, discarding the first 15\% of samples for each walker as burn-in and verifying that the chains have converged using the multivariate~$\hat{R}$ statistic from \citet{BrooksGelman1998}, with~$\hat{R}-1 < 0.01$. We use broad, uniform priors for the parameters: $\alpha_{\Omega} \sim \mathcal{U}(0, 20)$, $\alpha_{b} \sim \mathcal{U}(0,10)$, $\alphaNL\sim \mathcal{U}(0,5)$, and $\alphaFoG \sim \mathcal{U}(0,10)$, where for $\OmegaHI$ and $\bHI$, we have defined priors in terms of 
$\alpha_{\Omega} \equiv \OmegaHI / \Omega_{\mathrm{HI}}^{\mathrm{fid}}$ and 
$\alpha_{b} \equiv \bHI / b_{\mathrm{HI}}^{\mathrm{fid}}$.

We take the best-fit parameters to be those from the maximum a posteriori sample from the MCMC chains. 
{For the analysis of the full band, this yields 
$( \alpha_{\Omega}, \alpha_{b}, \alphaNL, \alphaFoG) = (2.8, 0.24, 0, 0.29)$.
For the upper sub-band, we obtain 
$(0.17, 8.8, 5.0, 1.2)$
while for the lower sub-band, we obtain 
$(0.42, 4.1, 0.084, 0.048)$.
}
We use these parameters to generate the best-fit model curves that are plotted alongside the measurements in this paper. 
Note that some of these parameter values are quite extreme ({compared to expected values of order unity}), and/or close to the boundary of the prior. However, these best-fit values do not correspond to maximum values of the marginalized posteriors for each parameter, and there are many other points in parameter space where the values are closer to expectations and the posterior is comparably high. Therefore, the appearance of extreme best-fit values is not a cause for concern.
For a complete picture of the parameter values inferred from the data, one should refer to the discussion of the marginalized posteriors in \citet{chime-auto-interpretation-paper}.

{The best-fit parameters in the previous paragraph yield $\chi^2$ values of $1.6$, $7.1$, and $6.0$ for the full band, upper sub-band, and lower sub-band, respectively.
We convert these into $p$-values using a Monte-Carlo-based procedure, whereby we re-fit the signal model to 1000 mock datasets comprising the best-fit signal prediction and a random noise realization, and compare the $\chi^2$ values from the data to the ensemble of mock-based values (see Appendix~\ref{app:detection_significance} for details). This procedure yields $p$-values of $0.92$, $0.25$, and $0.31$ for the three bands, indicating acceptably good fits in all cases.
}

\subsubsection{Power Spectrum Amplitude}
\label{sec:results:theoretical_interpretation:power_spectrum_amplitude}

The power spectrum model in \cref{eq:P21-theory} has a strong degeneracy between $\OmegaHI$ and $\bHI$, such that the marginalized posterior on either parameter is very broad. The form of the degeneracy can be read off from \cref{eq:P21-theory}: the amplitude of the power spectrum is determined by the product of $\OmegaHI^2$ (since $\Tbar \propto \OmegaHI$) and $(\bHI + f\mu^2)^2$. This implies that the angle-averaged power spectrum $P_{21}(k;z)$ is approximately proportional to the amplitude parameter\footnote{To be precise, the angle-averaged power spectrum is proportional to $\bHI^2 + 2\bHI \langle f\mu^2 \rangle + \langle f^2\mu^4 \rangle$, but we show in \citet{chime-auto-interpretation-paper} that the approximation $\langle f^2\mu^4 \rangle \approx \langle f\mu^2 \rangle^2$ is sufficiently accurate over the range of $\mu$ of our measurement.}
\beq
\label{eq:ps_ampltiude}
\AHIps \equiv 10^6\, \OmegaHI^2 \left( \bHI + \langle f\mu^2 \rangle \right)^2\ ,
\eeq
where the angle brackets surrounding $f\mu^2$ denote the appropriate angular average. 
Our notation reflects the fact that this parameter is the square of the $\AHI$ parameter used to describe the amplitude of the CHIME-eBOSS stacking signal in \citet{chimestacking}.
{We use $\langle f\mu^2 \rangle = 0.761$ to compute~$\AHIps$, since it corresponds to the best-constrained definition of this parameter and also agrees with a direct average of $f\mu^2$ over the set of $\mu$ bins used in our measurement.}

In \citet{chime-auto-interpretation-paper}, we perform a full Bayesian analysis to infer $\AHIps$ from the measured power spectrum, and find that 
$\AHIpsConstraintFullBand$
for the full frequency band, 
$\AHIpsConstraintUpperBand$
for the upper sub-band, and 
$\AHIpsConstraintLowerBand$ 
for the lower sub-band. 
The statistical uncertainties are determined from the 68\% highest posterior density intervals of the marginalized posteriors for~$\AHIps$, while the systematic uncertainty incorporates several contributions related to the estimated accuracy of our calibration and modelling framework (see \citealt{chime-auto-interpretation-paper} for details).

Adopting a literature-informed prior on $\bHI$ can enable our constraints on $\AHIps$ to be propagated into inferences about $\OmegaHI$, and vice versa; we discuss this further in \citet{chime-auto-interpretation-paper}. We also compare the constraining power of the auto spectrum and stacking the same data on eBOSS quasars in \secref{sec:validation:stacking}.

\subsection{Detection Significance}
\label{sec:results:detection_significance}

We use a Monte-Carlo-based likelihood ratio test to assess the detection significance of our power spectrum measurements in the full band and each sub-band. For each of the 1000 noise power spectra described in \secref{sec:power_spectrum_estimation:noise_covariance_estimation}, we compute
\beq
\label{eq:Deltachi2}
\Delta\chi^2 \equiv \chi^2_{\rm null} - \chi^2_\text{best-fit}\ ,
\eeq
where $\chi^2_{\rm null}$ is given by \cref{eq:chi2theta} with $\vec{s}=0$, corresponding to a noise-only model, and $\chi^2_\text{best-fit}$ is
\beq
\label{eq:chi2bestfit}
\chi^2_\text{best-fit} 
	\equiv \min_{\vec{\theta}} \chi^2(\vec{\theta}) \ .
\eeq
We then compare the $\Delta\chi^2$ value for the data, computed in the same manner, with an analytical probability distribution function fitted to the $1000$ values from the noise power spectra.

\cref{tab:sn} shows the results of three versions of this procedure. The most conservative version, which we use for our main results,
is based on fitting an $F$ distribution to the $1000$ $\Delta\chi^2$ values (after appropriate scaling). This is motivated by the fact that the noise covariance used in \cref{eq:Deltachi2} and \cref{eq:chi2bestfit} is estimated from simulations and therefore adds its own source of scatter to the $\chi^2$ values (see Appendix~\ref{app:detection_significance} for further discussion). The second S/N column in \cref{tab:sn} is based on fitting a $\chi^2$ distribution instead of an $F$ distribution, which is conceptually simpler but ignores scatter in the noise covariance estimate; therefore, the inferred distribution of $\Delta\chi^2$ values has a lighter tail than for the $F$ distribution, and this slightly inflates the detection significance inferred from the $\Delta\chi^2$ value for the data.

Finally, in the third S/N column of \cref{tab:sn}, we quote an even simpler S/N estimate, based on scaling the best-fit power spectrum prediction (with all model parameters fixed) with an amplitude prefactor $A$, and computing $\sigma_A/A$. As expected, these values are higher than those from the other methods, since they do not account for the freedom to determine the four model parameters from the data.

\begin{deluxetable}{c c CCC}[tb]
    \tablecaption{Power spectrum detection significance in each band.\label{tab:sn}}
    \tablecolumns{5}
    \tablehead{
    	\multicolumn{2}{c}{Band} &
        \multicolumn{3}{c}{S/N}
        \\
        \cmidrule(lr){1-2}
        \cmidrule(lr){3-5}
        \colhead{$\nu$ (MHz)} &
        \colhead{$z$} &
        \colhead{$F$-based} &
        \colhead{$\chi^2$-based} &
        \colhead{\shortstack[c]{Amplitude\\[0pt] -based}}
    }
    \startdata
    $608.2$-$707.8$ &
    	$1.34$-$1.01$ &
	{\bf 12.4} &
	13.0 &
	13.6 \\
   $658.2$-$707.8$ &
    	$1.16$-$1.01$ &
	{\bf 8.6} &
	8.8 &
	9.6  \\
    $608.2$-$658.2$ &
    	$1.34$-$1.16$ &
	{\bf 9.1} &
	9.4 &
	10.0   \\
    \enddata

    \tablecomments{We quantify the detection significance by computing $\Delta\chi^2 \equiv \chi^2_{\rm null} - \chi^2_\text{best-fit}$ for each of 1000 simulated noise power spectra, determining the best-fit distribution of those $\Delta\chi^2$ values, and using this distribution to compute a S/N value corresponding to the $\Delta\chi^2$ value for the data. An $F$ distribution provides a good fit to the $\Delta\chi^2$ values (after appropriate scaling; see Appendix~\ref{app:detection_significance} for details), and we use the associated S/N values for each band (shown in bold) as our main results. For comparison, we also quote S/N values based on the best-fit $\chi^2$ distribution, and based on scaling the best-fit model prediction with a free amplitude.}
\end{deluxetable}

\section{Validation}
\label{sec:validation}

In this section, we describe the comprehensive validation framework employed to establish the robustness and reliability of our power spectrum measurement. We designed several null tests, in which the data are processed such that any cosmological signal cancels out, and the resulting power spectrum should be consistent with thermal noise alone.   In complementary tests, we partition the data into independent subsets and assess whether the measured power spectra are statistically consistent across subsets or indicate contamination from residual systematics.

\begin{deluxetable*}{lcc}
\tablecaption{
Statistical $p$-values for the validation tests described in \secref{sec:validation},
which assess whether jackknife combinations and differences between data splits or
mask choices are consistent with our noise model.
\label{tab:validation}}
\tablewidth{0pt}
\tablehead{
  \multicolumn{3}{c}{\textbf{\textit{Validation tests}}} \\[3pt]
  \colhead{\textbf{Data Selection}} & \colhead{\textbf{Selection Criteria}} & \colhead{$\mathbf{p}$\textbf{-value}}
}
\startdata
Imaginary component & Im[$P(k)$] & 0.58 \\
\midrule
(Even--Odd) nights  & Re[$P(k)$] & 0.90 \\
                    & Im[$P(k)$] & 0.55 \\
\midrule
Stokes-$Q$          & Re[$P(k)$] & 0.21 \\
                    & Im[$P(k)$] & 0.33 \\
\midrule
RA bins             & (RA2 -- RA1) & 0.54 \\
\midrule
Dec bins             & (Dec2 -- Dec1) & 0.48 \\
\midrule
Baseline Selections  & [(1+2+3)-cylinder] &  \\
                     & -- [1-cylinder]    & 0.75\\ 
\tableline
\multicolumn{3}{c}{\textbf{\textit{Flux cuts}}} \\[3pt]
\textbf{Mask Type} & \textbf{Threshold [Jy]}  &  $\mathbf{p}$\textbf{-value}\\
\tableline
             & 20   & 0.20 \\
             & 25  & 0.16 \\
Track Mask   & 30   & 0.77 \\
             & 40   & 0.74 \\
\midrule            
             & 2  & 0.85 \\
             & 3  & 0.53 \\
Transit Mask & 4  & 0.42 \\
             & 6  & 0.68 \\
\midrule             
\enddata
\end{deluxetable*}

For each null test,  we construct the null bandpower vector  $\vec{d}_{\rm null}$, which should be statistically consistent with zero. We test this consistency  by computing  the $\chi^2$ under the null hypothesis that data are drawn from a zero-mean distribution with the appropriate noise covariance:
\begin{equation}
\label{eq:null_test}    
    \chi^2 = \vec{d}_{\rm null}^{T} \CNoise^{-1}  \vec{d}_{\rm null}.
\end{equation}
The noise covariance matrix $\CNoise^{-1}$ for each null test is estimated by applying the identical analysis pipeline on 1000 noise simulations as described before. We then compute the $p$-value from the $\chi^{2}$ statistics with 8 degrees of freedom, as we have 8 bandpower $k$-modes probed in this analysis.\footnote{In each case, we have verified with the Kolmogorov-Smirnov and Anderson-Darling tests that the 1000 $\chi^2$ values are well-described by a $\chi^2$ distribution. This justifies our usage of $\chi^2$ statistics, as opposed to the $F$ distribution we use for detection significance calculations.} We mention the $p$-values for all validation tests in Table~\ref{tab:validation}.

\subsection{Imaginary Component of the Power Spectrum}
\label{sec: validation:imaginary_component}

The power spectrum is complex-valued because we cross-correlate two independent maps, each constructed from a stack of either even or odd sidereal days.  If the sky is stationary and the instrument response is identical on even and odd sidereal days, then the cross-power from the sky is real-only and the imaginary part is consistent with noise. Departures from this expectation can arise from, for example, relative spectral shifts or small astrometric misalignments between the maps (due to pointing offsets introduced by phase calibration errors), which rotate real power into the imaginary component \citep{Kolopanis2019, HERA_I_2022}. The imaginary component of the Stokes-$I$ power spectrum is shown in the bottom panel of Figure~\ref{fig:ps_1D} and yields $\chi^{2}/\mathrm{dof}=0.82$ with a corresponding $p$-value of 0.58. This indicates excellent consistency with thermal noise and no evidence for residual phase-calibration–induced offsets. In the null tests below, we likewise evaluate the imaginary component against the noise-only expectation.

\subsection{Consistency Between Even and Odd Days}

As described previously (\secref{subsec:ps_estimation}),
we split the total 94 nights of data into two even and odd subsets and cross-correlate them to estimate the power spectrum.  To assess temporal stability and check for consistency between these subsets, we further subdivide each into two halves and construct differenced hybrid beamformed visibility datasets,   $\tilde{V}_{\rm diff,1}^{p}$ and $\tilde{V}_{\rm diff,2}^{p}$ (following \citealt{chimestacking}). We then process these datasets through the identical analysis pipeline (described in Section \ref{sec:pipeline} and Section \ref{sec:power_spectrum_estimation}) and generate the power spectrum by cross-correlating them.

\begin{figure}
   \centering \includegraphics[width=\linewidth,height=6in, keepaspectratio, trim = 0 0 0 0]{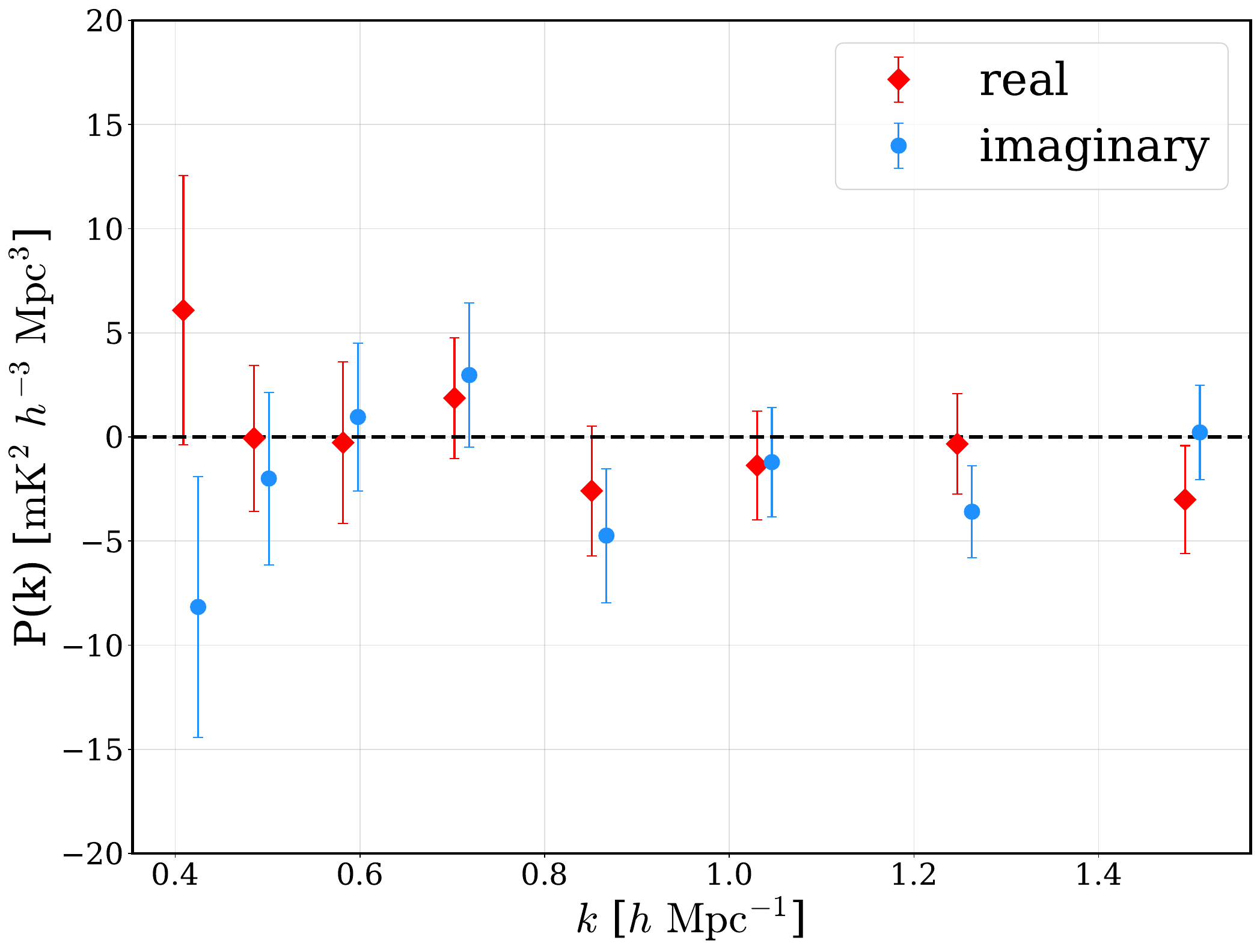}
    \caption{The real and imaginary components of the power spectrum estimated from the difference of interleaved nights. For visual clarity, we apply a small offset to the data points along $k$. These are consistent with the null hypothesis that the data are drawn from the noise covariance.}
    \label{fig:ps_even_odd}
\end{figure}

We expect the cosmological \tcm signal to remain constant across sidereal days and therefore cancel in the difference between subsets. The noise is independent between different days and hence will remain in each of the differenced visibility datasets. However, this will not result in any positive bias in a cross power spectrum between the two differenced data sets $\tilde{V}_{\rm diff,1}^{p}$ and $\tilde{V}_{\rm diff,2}^{p}$, because each will have an independent noise realization.  Transient RFI is expected to be independent between subsets and will persist in the difference. Systematic foreground residuals exhibit different behavior depending on their origin: residuals arising from chromatic instrumental effects or seasonal variations in the instrument response will cancel, as these remain stable on day-to-day timescales. However, foreground contamination driven by day-to-day instrumental variability will not cancel and may appear in the difference.  

Therefore, in the absence of transient RFI or temporal instrumental instability, the resulting power spectrum should be consistent with zero.  Figure \ref{fig:ps_even_odd} shows both real and imaginary components of the power spectrum from these differenced datasets.  Under the null hypothesis that the observed signal remains the same on different subsets,   we find $\chi^{2}/\rm dof = 0.43$ with a $p$-value of 0.90 for the real component and $\chi^{2}/\rm dof = 0.86$ with $p$-value of 0.55 for the imaginary component.   Both are fully consistent with noise, indicating the absence of transient RFI or day-to-day instrumental variations.

\subsection{Power Spectrum of Stokes Q Data}
A critical potential contaminant of the measurement of the \tcm power spectrum is polarized foreground emission. The cosmological \hi\ \tcm signal is intrinsically unpolarized. In contrast, polarized foreground emission can undergo Faraday rotation as it propagates through magnetized plasma in the interstellar medium, introducing a chromatic structure that varies as $\lambda^{2}$ and potentially mimics the frequency dependence of the cosmological signal \citep{Moore2013,shaw2015,Asad2015_I,Moore2017, Martinot2018ApJ}. This frequency structure can leak into Stokes-$I$ measurements through various instrumental non-idealities: non-orthogonal or rotated antenna feeds, cross-talk between feeds, and asymmetries between the \mbox{XX} and \mbox{YY} beam patterns.  These effects provide mechanisms for spectrally non-smooth, Faraday-rotated polarized emission to contaminate the unpolarized \tcm signal.

\begin{figure}
   \centering \includegraphics[width=\linewidth, keepaspectratio, trim = 0 0 0 0]{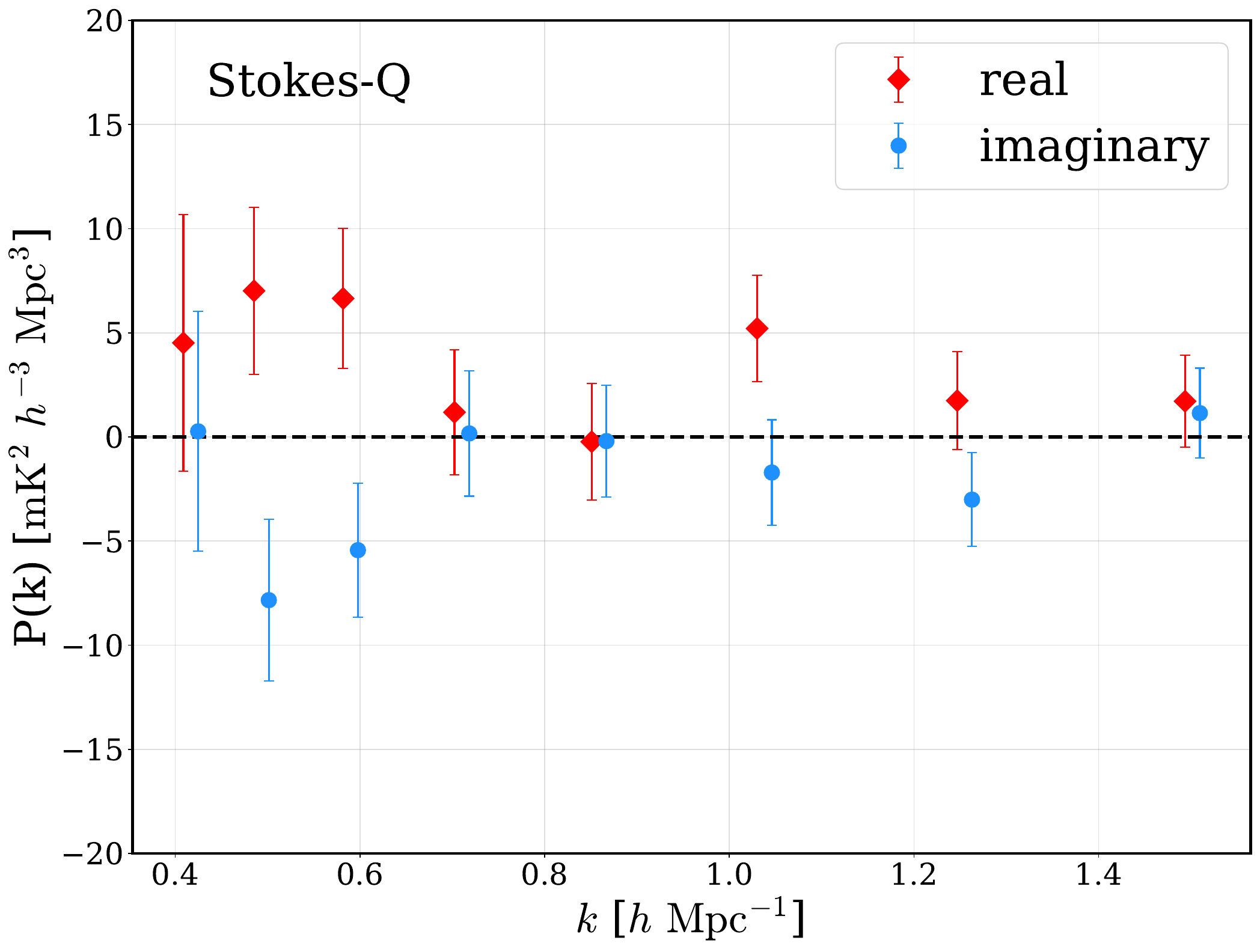}
    \caption{Here we show the power spectrum estimated from the Stokes-$Q$ maps. In the absence of significant deviation between \mbox{XX} and \mbox{YY} beam patterns, we expect that the power in Stokes-$Q$ should be consistent with noise fluctuations.
    We find that the $p$-value is greater than 0.2, under the null hypothesis that the Stokes-$Q$ power spectrum is drawn from the noise covariance and consistent with zero. This indicates that we do not detect any significant power in the Stokes-$Q$ component and rule out polarized foreground contamination in the Stokes-$I$ measurement through leakage from Stokes-$Q$ to Stokes-$I$. While the real and imaginary components show apparent structure, it is not statistically significant and remains consistent with noise fluctuation. The measured power spectrum in Stokes-$Q$ provides an upper bound on $Q \leftrightarrow I$ leakage and indicates that contamination in Stokes-$I$ measurement due to any polarized foreground leakage is subdominant.
    }
    \label{fig:ps_res_IQ}
\end{figure}

\citet{shaw2015} provides the mathematical formalism to study the leakage between all four Stokes parameters ($I, Q, U, V$)  for a cylindrical drift-scan telescope and establish the calibration accuracy and beam modeling requirements necessary to detect the cosmological signal. They showed that if the two orthogonal feeds satisfy two key constraints — equal normalized power patterns everywhere on the sky ($|\vec{A}_{\rm XX}|^{2}$ = $|\vec{A}_{\rm YY}|^{2}$) and orthogonal polarization orientations across the entire field of view ($\vec{A}_{\rm XX} \cdot \vec{A}_{\rm YY} = 0$)—then the measured visibility data is sensitive only to the total intensity Stokes-$I$ and insensitive to the other polarization modes.  In other words, if there are no significant deviations from these two constraints, the Stokes-$I$ component should measure the unpolarized cosmological \tcm signal without polarized contamination. Of course, a real instrument like CHIME does not perfectly satisfy these two conditions. We quantify the potential contamination from polarized leakage due to instrument non-idealities by measuring the power spectrum of Stokes-$Q$ maps.

We process Stokes-$Q$ maps for even and odd days datasets through the identical pipeline used for Stokes-$I$ and measure the power spectrum. The real and imaginary components of the power spectrum are shown in Figure \ref{fig:ps_res_IQ}. Under the null hypothesis that the Stokes-$Q$ power spectrum is drawn from the noise covariance and consistent with zero, we find $p$-values of 0.21 and 0.33 for the real and imaginary components, respectively.  The consistency of the Stokes-$Q$ power spectrum with noise, combined with the highly significant detection in Stokes-$I$, provides strong validation that the measured signal originates from intrinsically unpolarized \hi\ emission, and any polarized foreground leakage is subdominant. If polarized foregrounds were responsible for the Stokes-$I$ detection through instrumental leakage, they would necessarily produce detectable power in Stokes-$Q$ at a comparable or higher level, which we do not observe. The measured Stokes-$Q$ power spectrum serves as an upper bound on potential $Q \leftrightarrow I$ leakage.     

This test demonstrates that any polarized foreground leakage is negligible and subdominant to the detected Stokes-$I$ measurement of the \tcm signal.

\subsection{Power Spectrum of Two Independent RA Bins}
\label{sec:validation:ra_bins}
To test the presence of any RA-dependent systematics, we divide the data into two independent right ascension bins and measure the power spectrum separately for each subset.  The full NGC region spans 110–$265^{\circ}$ in RA (Figure \ref{fig:ringmap}), which we partition into RA1 (110–$190^{\circ}$) and RA2 (190–$265^{\circ}$). Each subset is processed through the identical analysis pipeline described in \secref{sec:power_spectrum_estimation}.

The frequency mask applied to the full RA range data is the same for both RA bins. However, the spatial mask, which depends on the location of the foreground sources in the sky,  is different between the two RA bins. Consequently, the flagging fraction is also different. The spatial mask excludes approximately $25\%$ of the sky area in RA1, compared to $42\%$ in RA2.

\begin{figure}
   \centering \includegraphics[width=1\linewidth, keepaspectratio, trim = 0 0 0 0]{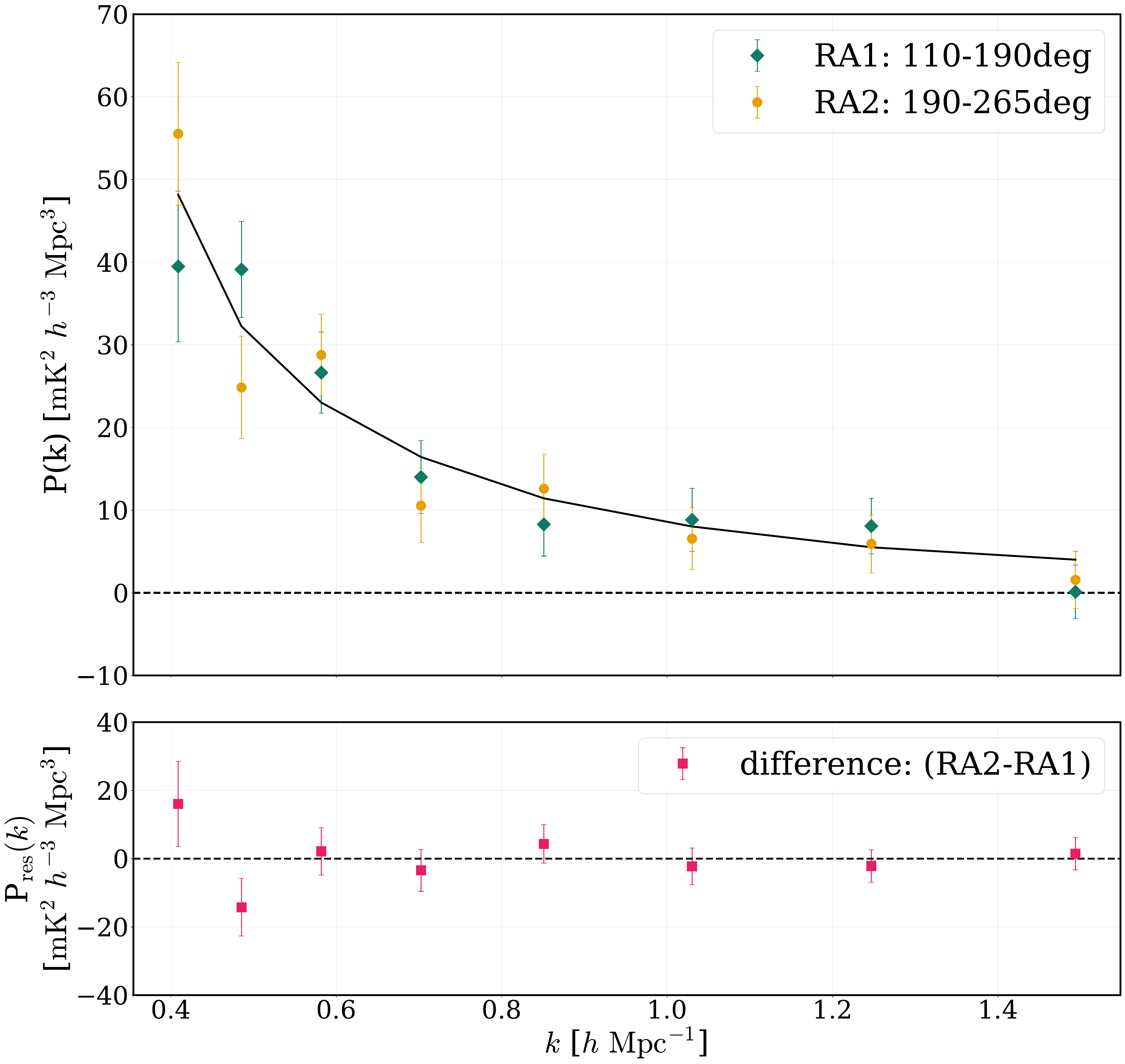}
    \caption{Top panel shows the real component of the power spectrum of two RA bins, RA1: 110-$190^\circ$ and RA2: 190-$265^\circ$. We also show the best-fit \hi\ model to the data of the whole RA range (the same best-fit model curve as shown in Figure~\ref{fig:ps_1D}).  We subtract the power spectrum of RA2 from RA1, both for data and noise realizations, and construct the noise covariance from the difference noise power spectra. The residual power spectrum, along with the $1\sigma$ uncertainty, is shown in the bottom panel. We find that the residual power spectrum is consistent with arising from noise fluctuations.}
    \label{fig:ps_ra_bins}
\end{figure}

Residual RFI and instrumental systematics that are localized in RA or LST will cause excess variance in the power spectrum for one RA bin but may not be present in the other. Additionally, contamination from Galactic foreground depends on RA and may cause systematic foreground leakage into high-delay modes that could manifest differently across the two RA bins. Simulations of Galactic foreground emission show more than $40\%$ higher variance in RA1 compared to RA2 due to the proximity to the Galactic plane. Even after excluding the intracylinder baselines, which are dominated by Galactic emission (see \secref{sec:mapmaking}), the differences persist between RA bins. 
However, the cosmological \tcm signal is isotropic and RA-independent and therefore should be identical across both RA bins.  Hence, for a true detection of the cosmological \tcm signal, we expect the measured power spectrum should be similar between different RA bins. 

Figure \ref{fig:ps_ra_bins} shows the measured power spectra for both RA bins. The top panels show the real component of the power spectrum.  In the bottom panel, we show the residual power spectrum estimated by taking the difference between RA2 and RA1. To assess whether this residual is consistent with noise, we construct a noise covariance matrix by taking the difference between 1000 noise power spectra for the two RA bins. The uncertainty on the residual power spectrum (shown as error bars in the bottom panel) is estimated as the $1\sigma$ scatter of these 1000 differenced noise realizations.
Using this covariance matrix, we compute the $\chi^2$ statistic for the residual power spectrum. Under the null hypothesis that the observed signal is identical in both RA bins and the residual is consistent with noise fluctuations alone, we obtain a $p$-value of 0.54. This indicates that the residual is fully consistent with noise, with no evidence for systematic differences between the two RA bins.

This test demonstrates that there is no evidence of RA-dependent systematics present in the data. Importantly, since Galactic emission is substantially brighter in RA1 compared to RA2, the observed consistency between RA bins provides strong evidence that our detected signal is not dominated by residual Galactic foregrounds.

\subsection{Power Spectrum of Two Independent Declination Bins}
\label{sec:validation:dec_bins}
To test for declination-dependent systematics, we divide the data into two 
independent declination bins, Dec1 ($38.4^\circ$–$49.2^\circ$) and Dec2 ($49.3^\circ$–$60.2^\circ$), and measure the power spectrum for each subset independently following the procedure described in \secref{sec:power_spectrum_estimation}.

The contamination from the point source foregrounds leaking through the mainlobe and far sidelobes of the telescope beam will be different between the two declination bins. We also expect the Galactic plane emission will have a declination-dependent structure (similar to \secref{sec:validation:ra_bins}). Additionally, \citet{chimestacking} demonstrated that 
the primary beam response contains large ($\sim 50\%$) ripples in frequency and declination due to multipath interference (see their Figure 9). If foreground contamination or any other declination-dependent systematics are present, we would expect excess variance in the measured power spectrum as a function of declination due to all of these effects. In contrast, the cosmological \tcm signal is isotropic and should yield identical power spectra in both bins.

\begin{figure}
   \centering \includegraphics[width=1\linewidth, keepaspectratio, trim = 0 0 0 0]{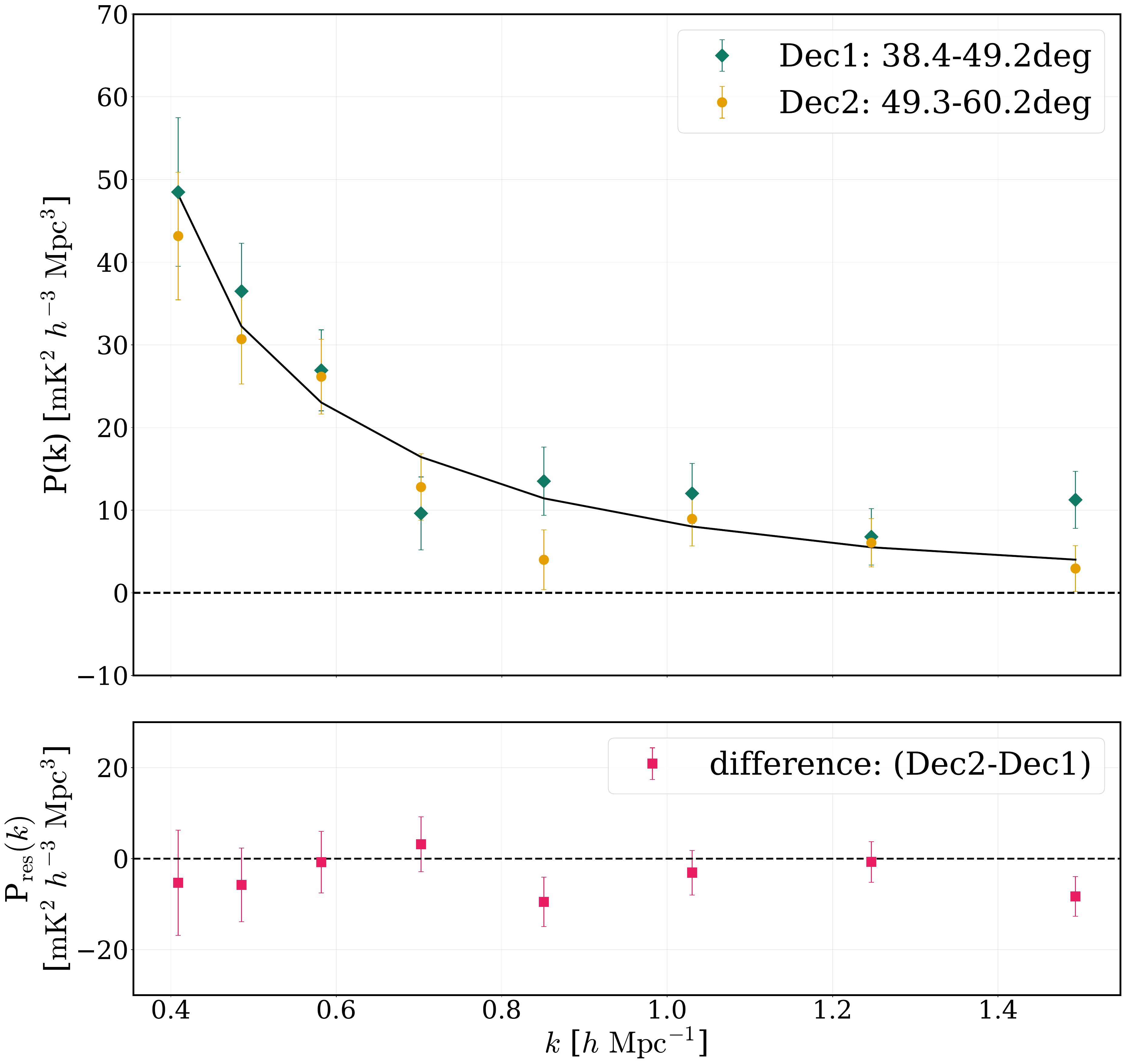}
    \caption{Top panel shows the real component of the power spectrum of two declination bins, Dec1: $38.4^\circ$–$49.2^\circ$ and Dec2: $49.3^\circ$–$60.2^\circ$, along with the best-fit \hi\ model to the data of the whole declination range (the same best-fit model curve as shown in Figure~\ref{fig:ps_1D}) in black.  We subtract the power spectrum of Dec2 from Dec1, both for data and 1000 noise realizations, and construct the noise covariance from the difference noise power spectra. The different power spectrum, along with the $1\sigma$ uncertainty, is shown in the bottom panel. We find that the residual power spectrum is consistent with the noise fluctuations.}
    \label{fig:ps_dec_bins}
\end{figure}

Figure~\ref{fig:ps_dec_bins} shows the real component of the measured power spectra for both declination bins in the top panel, while the bottom panel shows the difference between them.  We compute the noise covariance matrix by taking the difference between 1000 noise power spectra of these two declination bins. Under the null-hypothesis that the observed signal is identical in both bins and the difference between them is consistent with noise, we obtain a $p$-value of 0.46. This shows that the residual is in agreement with the noise. 

This test demonstrates that we measure the same cosmological \tcm signal in two declination bins, and there is no evidence of declination-dependent systematics.

\subsection{Power Spectrum With Different Baseline Selections}
\label{sec:validation:baseline_cuts}

In this section, we test for the presence of baseline-dependent foreground contamination in the measured power spectrum. Any residual foreground leakage is expected to exhibit baseline-dependent structure.  The projected geometric delay is larger for longer baselines, and because of this, spectrally smooth foregrounds will appear at larger delay or higher $k_{\parallel}$ modes for those baselines,  resulting in a characteristic ``foreground wedge" \citep{parsons2012-wedge}. Although we beamform along the NS direction, which removes this dependency for NS baselines, geometric delays remain associated with EW baselines (see \secref{sec:beamforming}).  This results in foreground contamination at higher delay modes for longer EW baselines, and this can potentially evade the foreground filter. Additionally, bright point sources drifting through the far sidelobes of the telescope beam can leak power at higher delays due to large projected delays associated with longer EW baseline separations (see \secref{sec:validation_track_mask}). 
Hence, if any potential foreground leakage contaminates our measurement, we would expect systematic differences between power spectra measured using different baseline subsets.

To test for this effect, we generate two maps using different baseline configurations: one using only 1-cylinder ($b^{\rm EW} \in {22}~\rm m$) baselines and another using the combined (1+2+3)-cylinder ($b^{\rm EW} \in {22, 44, 66}~\rm m$) baselines employed in our fiducial analysis. We then process both maps through an identical analysis pipeline and estimate the 2D and 1D power spectra.
The cylindrically binned 2D power spectrum for the 1-cylinder baseline map has more limited $k_{\perp}$ coverage compared to the (1+2+3)-cylinder map due to the shorter maximum baseline length. To enable a consistent comparison, we bin both power spectra to 1D spherical bins using identical bin edges corresponding to the 1D $k$-mode ranges probed by the 1-cylinder map.

\begin{figure}
   \centering \includegraphics[width=1\linewidth, keepaspectratio, trim = 0 0 0 0]{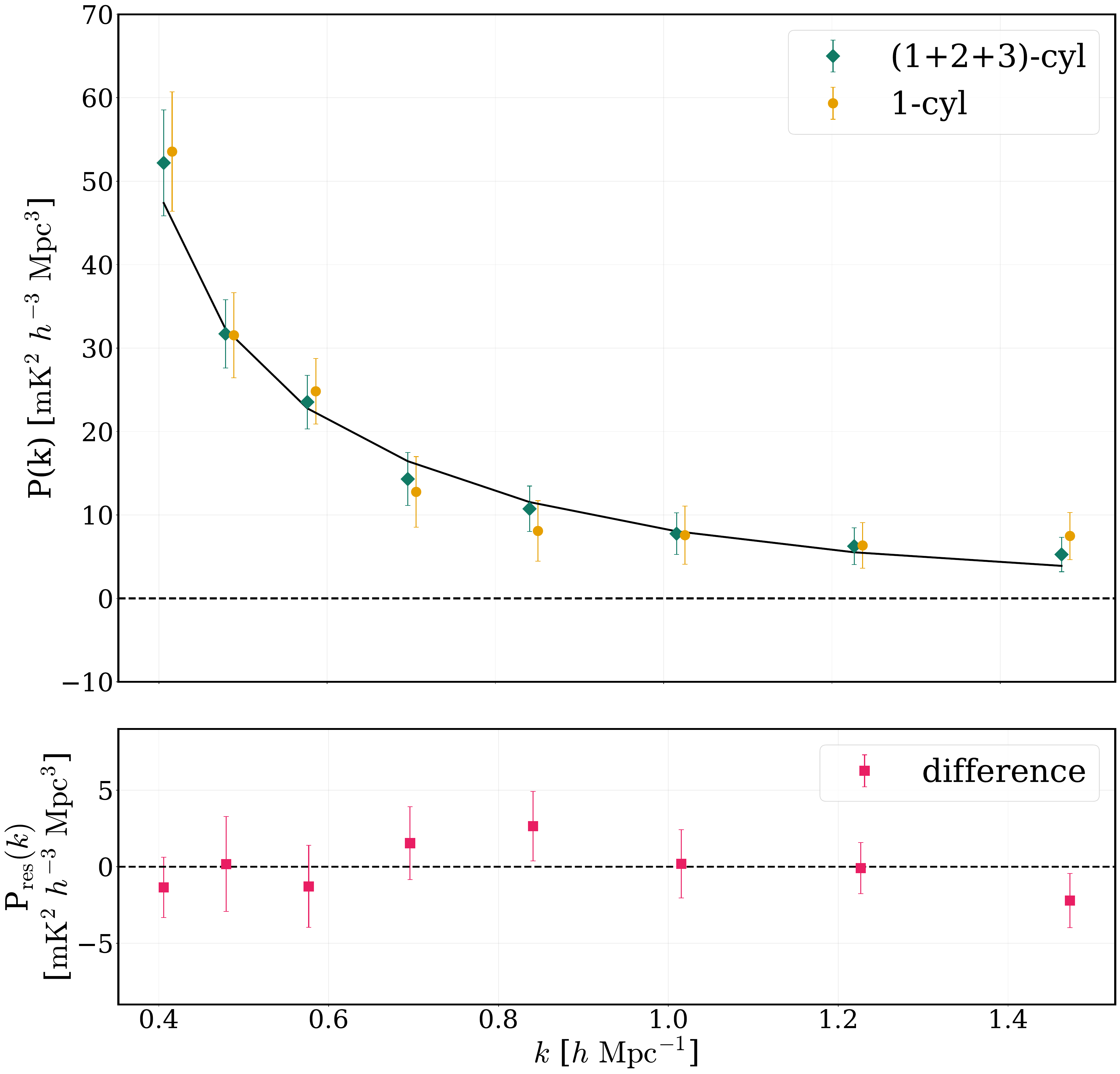}
    \caption{The top panel shows the real component of the spherically averaged power spectrum measured using 1-cylinder baselines only and the combined (1+2+3)-cylinder baselines employed in our fiducial analysis. The black curve shows the best-fit signal model rebinned to the same $k$-bins  corresponding to the 1-cylinder map. Data points are slightly offset in $k$ for visual clarity. Both baseline configurations yield consistent measurements across all scales.  In the bottom panel, we show the difference between these two power spectra. The uncertainty on each bandpower is estimated from the standard deviation of 1000 differenced noise power spectra.    
    The residuals yield a $p$-value of 0.75 under the null hypothesis that the difference is consistent with noise fluctuations. This indicates there is no baseline-dependent foreground contamination present in our measurement.}
    \label{fig:ps_baseline_cuts}
\end{figure}

The real component of the power spectrum for two baseline selections is shown in the top panel of Figure~\ref{fig:ps_baseline_cuts}, where in the bottom panel we show the difference between them. We also show the best-fit model after rebinning using the same 1D bin edges as used for the data. We offset the data points slightly along $k$ for visual clarity.  We construct a noise covariance matrix by taking the difference between 1000 noise power spectra of these two baseline selections, and the uncertainty on the residual power spectrum (bottom panel) is estimated by taking the standard deviation of these differenced noise power spectra. Under the null hypothesis that the observed signal is identical in the 1-cylinder and (1+2+3)-cylinder baseline configurations and the difference between them is consistent with noise fluctuations alone, we obtain a $p$-value of 0.75. This indicates that there is no systematic difference between power spectra measured using different subsets of baselines, and the difference between them is fully consistent with noise.

If residual foreground leakage contaminates our measurement, we would expect the (1+2+3)-cylinder baseline configuration, which includes the longest  \SI{66}{\meter} \ew baselines, to show enhanced power compared to the 1-cylinder baseline configuration that includes only the shortest \SI{22}{\meter} \ew baselines. We do not see any such evidence in the measured power spectrum for these two baseline selections. 
This test demonstrates that we measure the cosmological \tcm signal consistently using different baseline selections, with no indication of residual baseline-dependent systematics.

\subsection{Power Spectrum of Two Independent Sub-bands}
\label{sec:validation:sub_bands}

\begin{figure}
   \centering \includegraphics[width=1\linewidth, keepaspectratio, trim = 0 0 0 0]{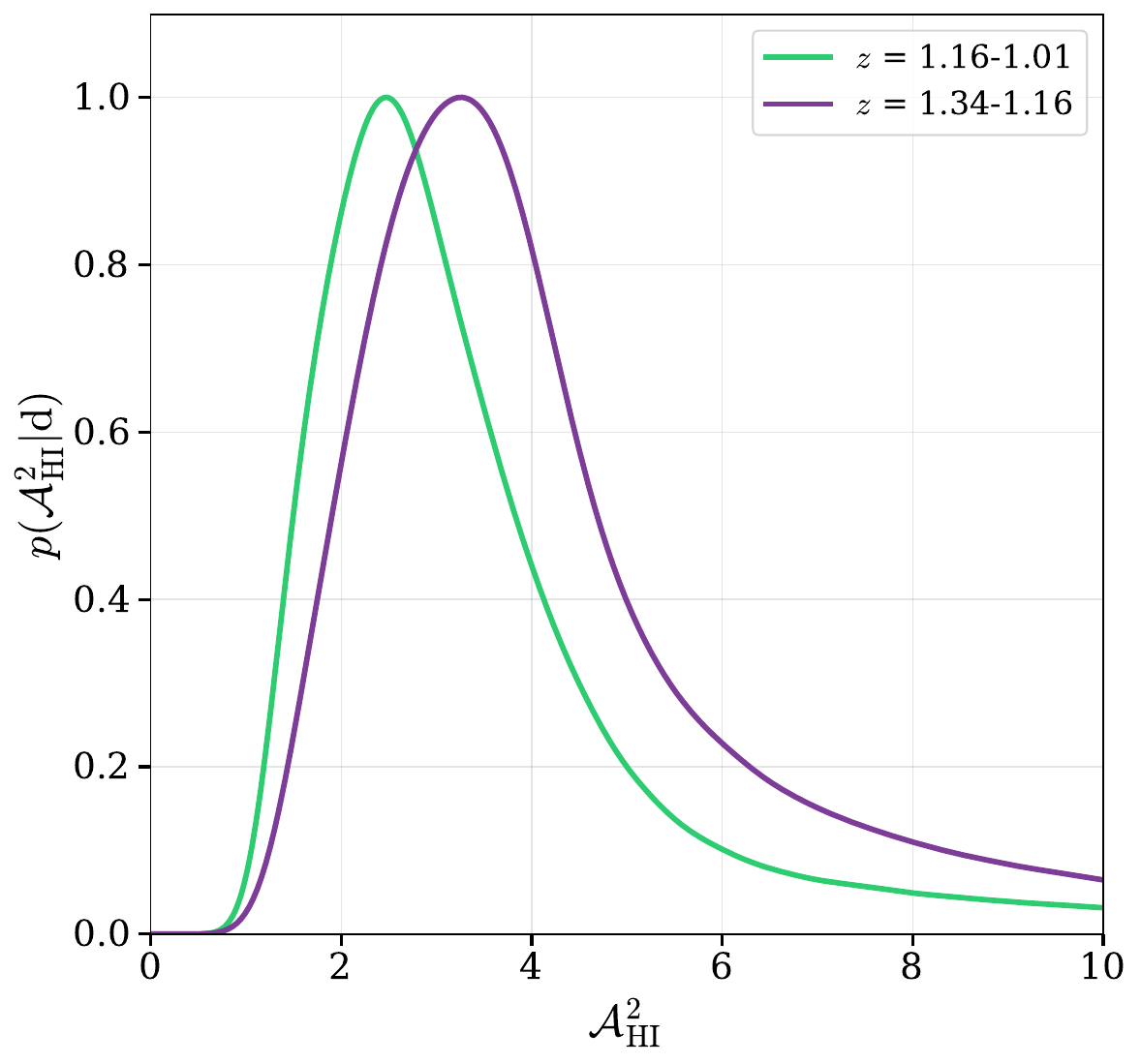}
    \caption{The marginalized 1D posterior of the power spectrum amplitude parameter ($\AHIps$) derived from the best-fit model for two sub-bands. The upper sub-band spanning 658.2–707.8 MHz ($z = 1.16$-$1.01$), centered at $z = 1.08$ and lower sub-band spanning 608.2–658.2 MHz ($z = 1.34$-$1.16$), centered at $z = 1.24$. We find that the amplitude parameters are consistent with each other within $1\sigma$.
    Although we expect mild evolution between two redshifts, the current SNR is insufficient to probe the redshift evolution conclusively.}
    \label{fig:A_HI_posterior_sub_band}
\end{figure}

We report, in Section~\ref{sec:power_spectrum}, the detection of the cosmological \tcm signal in two independent sub-bands: the upper sub-band centered at $z \sim 1.08$ and the lower sub-band centered at $z \sim 1.24$. The spherically averaged power spectrum for these sub-bands is shown in Figure~\ref{fig:ps_freq_bins}. We find that the power spectra are broadly consistent with each other.  Fitting a signal model independently to each sub-band yields best-fit power spectrum amplitudes ($\AHIps$) that are consistent within $1\sigma$, as discussed in Section~\ref{sec:results:theoretical_interpretation:power_spectrum_amplitude}. The one-dimensional marginalized posterior distributions for the power spectrum amplitude parameter for both sub-bands are shown in Figure \ref{fig:A_HI_posterior_sub_band}.

These independent measurements also provide a critical test for frequency-dependent systematic contamination. If residual foreground contamination were present, we would expect significantly large bias in the lower sub-band ($z = 1.24$), since foregrounds—dominated by synchrotron emission—follow a power-law scaling with frequency. However, the intrinsic \tcm signal is not expected to evolve substantially over this narrow redshift range. Based on the fitting functions for $\OmegaHI(z)$ and $\bHI(z)$ that we adopt in \secref{sec:results:theoretical_interpretation:power_spectrum_model}, we compute the theoretically predicted evolution of the power spectrum amplitude between the two redshifts. The model predicts that the amplitude at  $z = 1.24$ is only $\sim13\%$ higher than at $z = 1.08$.

{The measured best-fit power spectrum amplitudes are consistent with each other within $1\sigma$.} The current signal-to-noise ratio is insufficient to probe redshift evolution conclusively.  Future work with substantially more data can probe the evolution of the \tcm signal over redshifts. 
However, the consistency between the two sub-bands provides evidence against frequency-dependent foreground contamination. Significant foreground leakage would manifest as excess power at lower frequencies, which we do not observe, and demonstrates that there is no frequency-dependent foreground leakage. 

\subsection{Changing the Track Mask for Different Flux Cuts}
\label{sec:validation_track_mask}

The dominant contamination in our foreground-filtered maps arises from bright radio point sources drifting through the far sidelobes of the primary beam (see \cref{fig:ringmap}). At low elevation, the projection of the source direction onto the east–west baseline yields a large geometric delay, pushing sidelobe pickup into high-delay modes. Although most far-sidelobe signals are removed by the fringe-rate filter applied during the mapmaking process (see \cref{eq:map_maker}), there are low-elevation intervals for which the apparent fringe rate lies within its passband; during those intervals, the large east–west delay places power beyond our delay cutoff, allowing leakage to survive both filters and appear prominently at high delay. Because the far-sidelobe primary-beam response is low, this effect is visible only for the brightest sources. To mitigate it, we apply spatial masks that exclude the predicted sky tracks of these sources (see \secref{sec:spatial_mask}).

A critical concern is whether fainter sources—below our baseline flux threshold—contribute residual foreground power that could be misinterpreted as cosmological \tcm signal. To test this hypothesis, we investigate the impact of fainter sources that lie below our nominal flux threshold.  We generate increasingly aggressive track masks by progressively lowering the flux density threshold and measure the resulting power spectra. Each mask is defined by a different flux density cutoff, $f_{\rm cut}$, applied to an external source catalog.  This systematic variation allows us to directly assess the sensitivity of our power spectrum measurements to the exclusion of sources with progressively lower flux densities.

\begin{figure}
   \centering \includegraphics[width=1\linewidth, keepaspectratio, trim = 0 0 0 0]{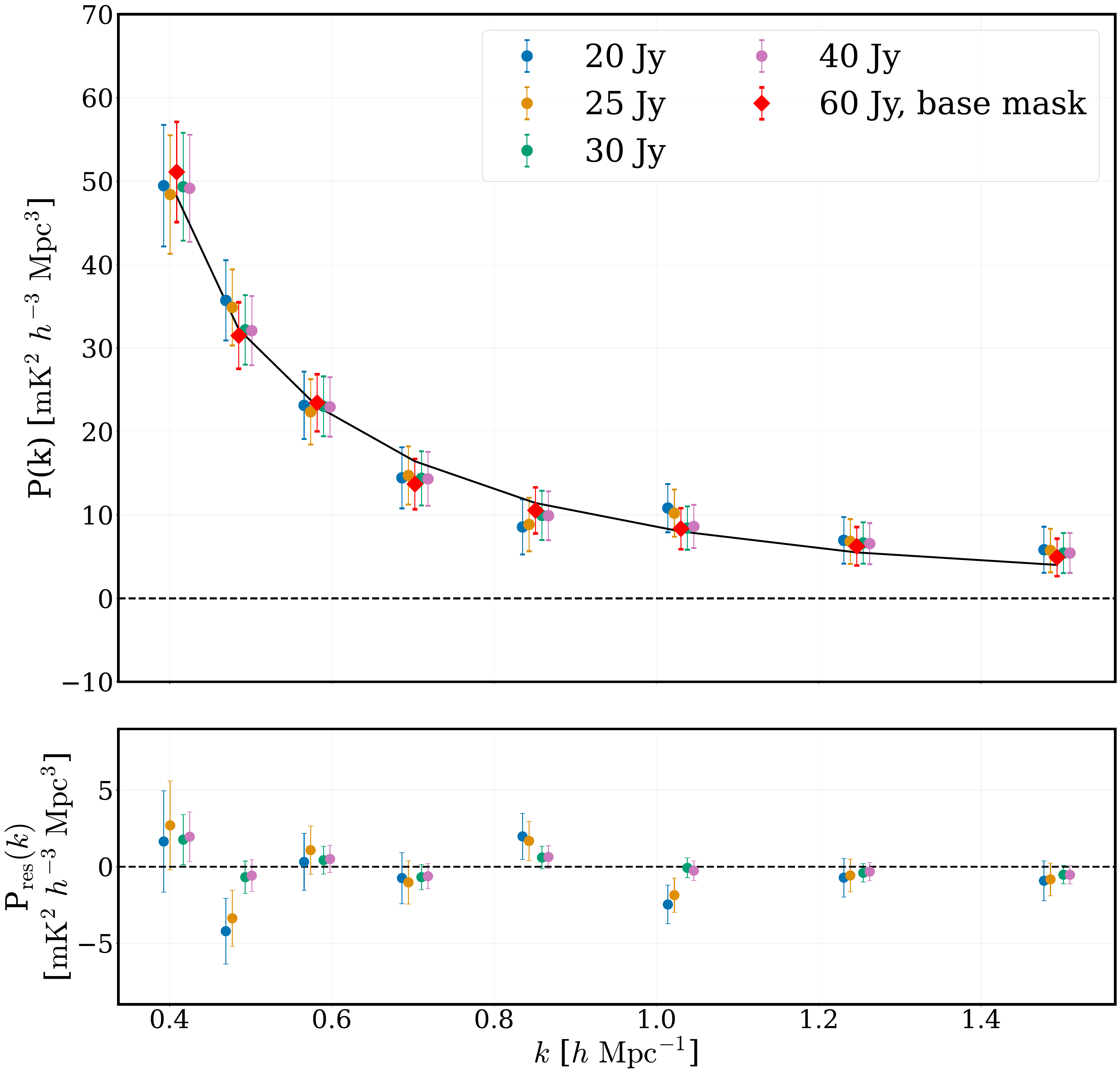}
    \caption{Power spectra made with different spatial sidelobe masks generated with different flux cuts. We show the real part of the power spectrum in the top panel, and in the bottom panel, we show the residual after subtracting the power spectrum for different cuts from the base mask. In all these different sidelobe masks, we kept the main lobe mask fixed and chose a flux cut for the main lobe mask to be $10\,{\rm Jy}$. For visual clarity, we apply a small offset to the data points along $k$. The residuals across different flux cuts are statistically consistent with zero, indicating no systematic dependence on the sidelobe masking threshold.}
    \label{fig:ps_track_mask}
\end{figure}

We construct the spatial mask for a given flux cutoff, by combining our base mask (Section \ref{sec:spatial_mask}) with the individual track masks of all sources exceeding that cutoff:
\begin{equation}
T_{\rm f_{cut}}(y, \phi) = T_{\rm base}(y, \phi)  \times \left[ \prod_{s \in S_{f_{\rm cut}}} T_{\rm track}^{s} \right],
\end{equation}
where $S_{f_{\rm cut}}$ is the set of catalog sources with flux densities exceeding $f_{\rm cut}$. For this test,  we generate masks using flux thresholds of $f_{\rm cut} = \{20, 25, 30, 40\}\,{\rm Jy}$, compared against our baseline threshold of $60\,{\rm Jy}$ (base mask) used for the main analysis. These progressively aggressive masks flag $58\%, 52\%, 43\%,$ and  $40\%$ of spatial pixels, respectively, compared to $33\%$ for the base mask. Note that the transit masks for sources above $10\,{\rm Jy}$ remain fixed across all tests.

For each mask configuration, we estimate the power spectrum following the procedure in  \secref{sec:power_spectrum_estimation}. If residual foreground contamination from fainter sources dominates our measurement, we expect the measured power to decrease systematically as we increase the spatial masking fraction (i.e., as we exclude more potential foreground sources). Conversely, if we are detecting the cosmological \tcm signal—which is statistically isotropic—the measured power should remain stable across different mask configurations, with only the statistical uncertainty increasing due to reduced sky coverage. The top panel of Figure~\ref{fig:ps_track_mask} presents the real component of  the power spectra for all mask configurations along with the base mask and the corresponding best-fit signal model (same as Figure~\ref{fig:ps_1D}). We generate 1000 noise realizations for each mask configuration separately and estimate the power spectrum uncertainty from these simulations.  We observe no systematic trend in the measured power as a function of masking fraction; all power spectra remain consistent within their $1\sigma$ uncertainties. 

Because most of the unmasked samples are identical for any two masks, the corresponding power spectra are computed from nearly the same underlying data; therefore the error bars in the top panel of \cref{fig:ps_track_mask} are strongly correlated across masks. To quantify consistency more rigorously, we form residual power spectra by subtracting the base-mask spectrum from each test-mask spectrum. We apply the same paired subtraction to 1000 noise realizations, using the same realization for both masks before differencing. We then estimate the covariance of these residuals from the ensemble of differenced noise realizations and use it in \cref{eq:null_test} to test consistency with zero. The bottom panel of \cref{fig:ps_track_mask} shows the residual spectra constructed in this way, with correspondingly smaller error bars. All residuals yield $p>0.1$, indicating that the small changes observed in the power spectra as we apply increasingly aggressive masks are noise-driven rather than due to residual foreground contamination.

This null test provides strong evidence that fainter sources  do not contribute measurable foreground leakage through sidelobes in our analysis. The stability of the measured power spectrum across varying degrees of spatial masking demonstrates that we are detecting the cosmological \tcm signal rather than residual sidelobe contamination. 

\subsection{Changing the Transit Mask for Different Flux Cuts}

In this section, we test for multiplicative systematics that could imprint high-delay structure on foreground emission in the NGC field. Several instrumental effects could produce such systematics: (i) narrow-band features in per-feed gain solutions caused by residual RFI during calibration; (ii) imperfect handling of the RFI mask during foreground filtering; (iii) chromatic primary-beam response due to mutual coupling between feeds; (iv) time-variable cable reflections (e.g., from the $50\,{\rm m}$ coax with round-trip delay of $390\,{\rm ns}$) that produce bandpass ripples at high delay. Our test is agnostic to the specific cause and instead targets their generic consequence: modulation of otherwise spectrally smooth foregrounds, resulting in high-delay leakage that scales with foreground brightness and is therefore largest for bright point sources in the main lobe.

\begin{figure}
   \centering \includegraphics[width=1\linewidth, keepaspectratio, trim = 0 0 0 0]{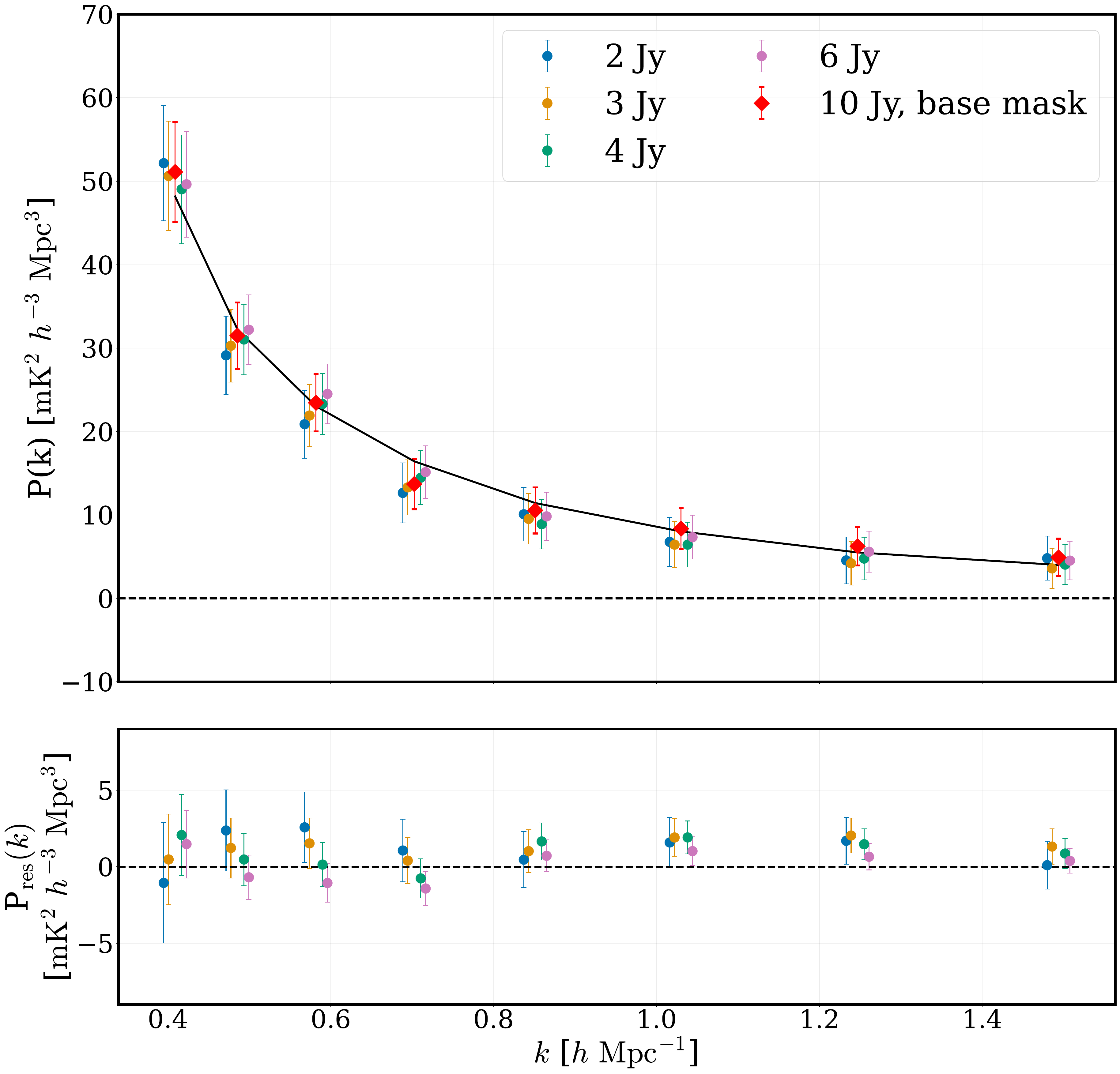}
    \caption{Power spectra made with different transit masks generated with different flux density thresholds. We show the real part of the power spectrum in the top panel, and in the bottom panel, we show the residual after subtracting the power spectrum for different cuts from the base mask. For visual clarity, we apply a small offset to the data points along $k$.  We find that for different cuts, the residual is consistent with noise. }
    \label{fig:ps_tran_mask}
\end{figure}

To accomplish this, we design a jackknife test similar to the one described in \secref{sec:validation_track_mask}, where we progressively mask fainter sources in the main lobe of the synthesized beam by lowering the flux density threshold when generating the transit mask (as described in \secref{sec:spatial_mask}). The combined mask is given by
\begin{equation}
T_{\rm f_{cut}}(y, \phi) = T_{\rm base}(y, \phi)  \times \left[ \prod_{s \in S_{f_{\rm cut}}} T_{\rm transit}^{s} \right],
\end{equation}
where $S_{f_{\rm cut}}$ is the set of catalog sources with flux densities exceeding $f_{\rm cut}$. We test four flux thresholds, $f_{\rm cut} = \{2, 3, 4, 6\}$ Jy, and then compare against our base mask, where a $10 \,{\rm Jy}$ threshold is used for the transit mask. These transit masks exclude  $53\%, 45\%, 42\%,$ and  $39\%$ of sky area, respectively. We then compute the power spectrum of the data with each of these masks applied.

The resulting power spectra are shown in Figure~\ref{fig:ps_tran_mask}. The top panel shows the real component of the power spectrum for all transit mask configurations, with the best-fit signal model to the base mask overlaid. We observe no systematic decrease in the measured power as a function of the transit mask flux threshold; all measurements remain consistent within their $1\sigma$ uncertainties. The bottom panel shows residuals relative to the base mask, computed following the same procedure as described in \secref{sec:validation_track_mask}. All residuals yield $p$-values $>0.3$, confirming consistency with zero.

If multiplicative high-delay systematics dominated the measured power, we would expect a predictable decrease as we lower the transit-mask threshold. For a multiplicative effect, the contaminant from each source scales with its flux $S$, so the power contributed by the source scales roughly as $S^{2}$. With typical near-Euclidean differential counts ($dN/dS\propto S^{-\beta}$ with $\beta\simeq2.5$, \citep{zotti_de_2010}), the residual leaked power after masking sources brighter than $S_{\rm c}$ scales approximately as $S_{\rm c}^{\,3-\beta}$. Thus, reducing the threshold from \SI{10}{\jansky} to \SI{2}{\jansky} would be expected to lower any such leakage by a factor $(2/10)^{\,3-\beta}\approx(0.2)^{0.5}\simeq0.45$ (a $\sim$55\% reduction).  We observe no such trend in our data, which argues against multiplicative high-delay leakage from main-lobe sources dominating the measured power. By contrast, if we are measuring cosmological \tcm signal, the power spectrum should be stable to these mask changes, aside from increased statistical uncertainty from reduced sky area, as observed. 

These results indicate that multiplicative high-delay systematics, if present, are sub-dominant at our sensitivity and do not materially affect the measured power spectrum.

\subsection{Consistency of \hi\ Clustering Amplitude Between Auto-correlation and Cross-correlation}
\label{sec:validation:stacking}

As mentioned in Section \ref{sec:stacking}, we perform the stacking analysis on the same data used for the auto-power spectrum measurement. We stack the masked foreground-filtered map on 33,119 QSOs from the eBOSS clustering catalog within the redshift range 1.01–1.34 (608–708 MHz), corresponding to the sky region analyzed in our auto-power spectrum measurement.  We detect the \tcm signal in this cross-correlation with QSOs, with detection significance of $9.1\sigma$.

Following the procedure of \citet{chimestacking}, we use this cross-correlation measurement to constrain the effective clustering amplitude of neutral hydrogen, defined as $\AHI \equiv 10^3 \, \OmegaHI \left(\bHI + \fmu \right)$. {We find  $\AHI = 1.93^{+2.50}_{-0.93}$ when non-linear parameters are free to vary, and $\AHI =1.63^{+0.19}_{-0.18}$, when those parameters are fixed to their fiducial values.} 
We then compare this effective clustering amplitude parameter with the value we derived from the auto-power spectrum measurement (see \cref{eq:ps_ampltiude}).

\begin{figure}
   \centering \includegraphics[width=1\linewidth, keepaspectratio, trim = 0 0 0 0]{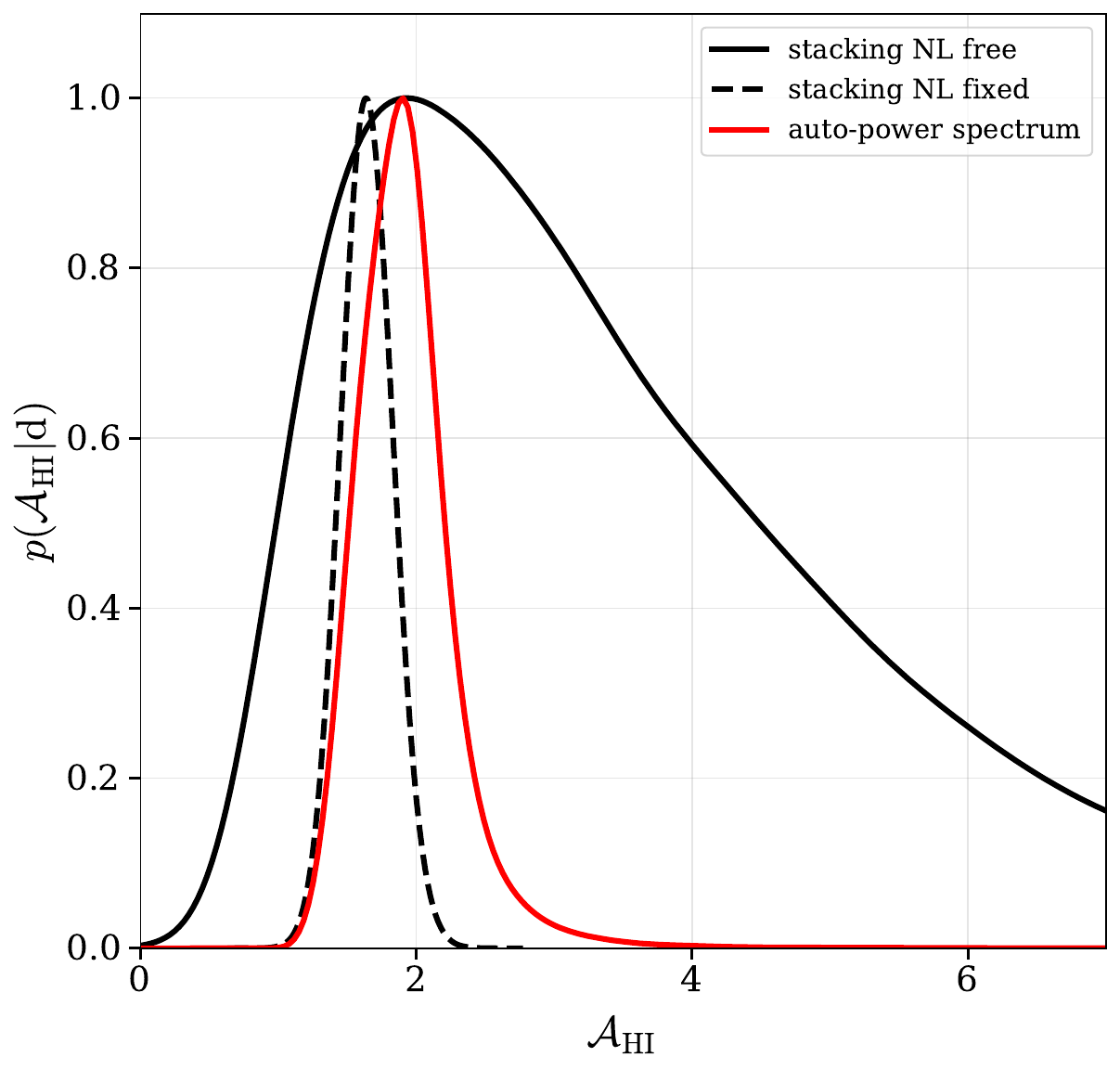}
    \caption{The posterior of the effective clustering amplitude parameter ($\AHI$) derived from stacking on the QSO catalog from eBOSS. We show the constraint for both models,  where all parameters are allowed to vary, and when the nonlinear parameters (NL) are fixed to their fiducial values (same as \citealt{chimestacking}). We also show the constraint on the power spectrum amplitude parameter derived from the auto-correlation measurement by taking the square root of the $\AHIps$ (see \cref{eq:ps_ampltiude}). We find that the derived amplitude parameters from cross-correlation and auto-correlation analyses are consistent with each other within $1\sigma$.} 
    \label{fig:A_HI_posterior}
\end{figure}

Figure~\ref{fig:A_HI_posterior} compares the marginalized posterior distributions of  $\AHI$ derived from both the stacking analysis and the auto-power spectrum analysis. We take the square root of the power spectrum amplitude parameter ($\AHIps$) for one-to-one comparison.  The two measurements are statistically consistent within $1\sigma$. This consistency provides strong evidence against contamination from residual systematics. Systematic effects that are uncorrelated between CHIME and eBOSS, such as residual foreground leakage, RFI, or other instrumental artifacts,  would bias the auto-correlation measurement but not the cross-correlation. Such contamination would result in a significantly larger amplitude in the auto-power spectrum relative to the stacking measurement. The observed agreement between the two independent analyses therefore demonstrates that there is no residual systematic contamination present in our data.

Furthermore, we find that the marginalized posterior on $\AHI$ is much narrower for the auto power spectrum than for the stacking analysis with all model parameters free to vary (NL-free). For diagnostic purposes, we also consider a variant of the stacking model in which the nonlinear parameters are fixed to fiducial values (NL-fixed)\footnote{As in \citet{chimestacking}, the mean of the NL-fixed posterior is determined by the fiducial values of the nonlinear parameters and should not be interpreted as an estimate of the true $\AHI$, since the actual values of those parameters are unknown.}.  In this case, the width of the stacking posterior becomes comparable to that for the auto power spectrum, highlighting the extent to which degeneracies between $\AHI$ and the nonlinear parameters contribute to the broadening of the NL-free stacking constraint.

Several factors likely contribute to the tighter constraints obtained from the auto–power–spectrum analysis relative to the NL-free stacking analysis. First, the auto–power–spectrum detection has a higher overall signal-to-noise ratio than the stacking measurement (13.1$\sigma$ versus 9.1$\sigma$). Second, the auto power spectrum scales quadratically with the amplitude parameter, $P_{\rm HI} \propto \AHIps$, whereas the stacked signal (or cross power spectrum) scales linearly, $P_{\rm HI,g} \propto \AHI$, so at fixed noise covariance the auto measurement is intrinsically more sensitive to $\AHI$. Third, the stacking analysis must marginalize over a larger set of nuisance parameters, including the linear bias of the QSO catalog $b_{g}$ (which is well constrained), the shot-noise contribution parameterized by $M_{10}$ (which is not), and two separate Fingers-of-God parameters (compared to a single Fingers-of-God parameter for the auto power spectrum). This larger nuisance-parameter space leads to more severe prior-volume effects on $\AHI$. Finally, although both analyses average over modes to form a one-dimensional statistic, in the current spectral-stacking implementation most of the signal-to-noise is concentrated at small line-of-sight separations, so only a small number of broad, highly overlapping linear combinations of $k_\parallel$ are well constrained. By contrast, the power-spectrum estimator retains several $k$-bins (dominated by different ranges of $k_\parallel$) with more distinct responses to the model parameters, providing additional effective scale information that helps to break parameter degeneracies.

\section{Conclusions}
\label{sec:conclusions}

In this paper, we have presented the first detection of the cosmological \tcm intensity mapping signal in auto-correlation at $z\sim 1$ using data from the Canadian Hydrogen Intensity Mapping Experiment (CHIME). Using 94 nights of observation from January through November 2019, we have measured the 21 cm auto power spectrum over a frequency range from 608.2 MHz to 707.8 MHz, corresponding to a mean redshift of $z \sim 1.16$ ($z$ = 1.34 to 1.01), {with a statistical significance of $12.4\sigma$.} We have also independently measured the \tcm signal in two sub-bands, {obtaining $9.1\sigma$ and $8.6\sigma$ detections} at mean redshifts of $z \sim 1.24$ and $z \sim 1.08$, respectively. This result marks a significant milestone, moving beyond the reliance on cross-correlation with external survey data and establishing the \tcm auto-power spectrum as a direct and powerful cosmological probe at these redshifts.

This detection was made possible by substantial advancements in our data processing pipeline, specifically designed to overcome the immense challenge of separating the faint cosmological signal from astrophysical foregrounds that are orders of magnitude brighter. Key methodological advances include: (1) novel RFI  detection and masking algorithms combining radiometer tests, fringe-rate filtering, and spectral filtering methods (Section~\ref{sec:rfi}); (2) implementation of achromatic beamforming techniques to maintain consistent synthesized beams across frequency (Section~\ref{sec:beamforming}); (3) application of foreground filtering before time averaging to minimize spectral leakage (Section~\ref{sec:fg_filter_timeavg}); and, (4) deployment of the HyFoReS algorithm to correct residual bandpass errors (Section~\ref{sec:hyfores}). 
With our current analysis choices, the foreground filtering removes sensitivity to linear cosmological scales related to BAOs, but enables high signal-to-noise-ratio measurements at non-linear scales $0.4\ihMpc \lesssim k \lesssim 1.5\ihMpc$ (see Figure~\ref{fig:ps_1D}).

The robustness of our detection has been established through a comprehensive suite of validation tests. Our validation framework includes: (1) the imaginary component of the power spectrum is consistent with noise, indicating the absence of phase calibration errors; (2) temporal stability through even-odd day comparisons; (3) the power spectrum derived from Stokes-$Q$ data is also consistent with noise, ruling out any significant polarized foreground leakage; (4) the consistency of the signal across independent right ascension bins rules out RA-dependent systematics; (5) the consistency of the signal across independent declination bins rules out any declination-dependent systematics; (6) the consistency of the signal across different baseline configurations (1-cylinder versus combined (1+2+3)-cylinder baselines) rules out baseline-dependent foreground contamination;  (7) the consistency of the signal across independent sub-bands rules out frequency-dependent systematic contamination; (8) the signal remains consistent with varying spatial track masks and rules out foreground leakage through sidelobes of the beam; (9) the consistency of the signal with varying transit masks rules out the narrow-band systematic contamination; and (10) the consistency of the auto-correlation measurement with the cross-correlation detection for the same dataset stacked on eBOSS quasars, providing strong evidence against contamination from systematics that are uncorrelated between CHIME and external tracers.

We also briefly discuss the theoretical interpretation of the measurements (Section~\ref{sec:results:theoretical_interpretation}), and the detailed modeling of the \tcm signal is outlined in our companion paper \citep{chime-auto-interpretation-paper}. Given the strong degeneracy between the cosmic \hi\ abundance  ($\OmegaHI$) and the linear bias of {\hi} ($\bHI$), we constrain their combination through the amplitude parameter:  $\AHIps = 10^6 \OmegaHI^2(\bHI^2+\fmu)^2$, where $\fmu$ is a factor capturing the angular average of the Kaiser redshift-space distortion term over the wavenumbers we have measured. We find that $\AHIpsConstraintFullBand$ for the full frequency band ($z\sim 1.16$), $\AHIpsConstraintUpperBand$ for the upper sub-band ($z\sim 1.08$), and $\AHIpsConstraintLowerBand$ for the lower sub-band ($z\sim 1.24$). These constraints are dominated by statistical uncertainties from the limited signal-to-noise of our current measurement. 
In \citet{chime-auto-interpretation-paper}, we also compare our measurements with predictions from the IllustrisTNG hydrodynamical simulations.

This detection establishes CHIME's capability for precision \tcm cosmology and opens multiple avenues for future investigation. The CHIME archive now contains nearly 7 years of data.
Including these additional observations will significantly reduce statistical uncertainties and enable more precise constraints on the \hi\ distribution and clustering, and probe its evolution over redshifts. Furthermore, parallel efforts are underway to extend this analysis to the lower-frequency portion of the CHIME band (400–600 MHz), pushing these auto-correlation measurements to higher redshifts ($z\sim1.4$–$2.5$). Our current analysis excludes intracylinder baselines to avoid contamination from diffuse Galactic emission and cross-talk. In addition, our foreground filtering method 
aggressively suppresses large line-of-sight modes (low $k_{\parallel}$). Although these analysis choices enable high SNR detection of the \tcm signal at non-linear scales, they essentially remove the BAO signal. We are continuously working on improving our RFI flagging (both real-time and offline) algorithms, developing full-sky primary beam models, and exploring novel foreground filtering algorithms to recover those linear scales. This will enable future measurements of the BAO scale, providing powerful constraints on dark energy and the universe's expansion history.

More broadly, the novel methodologies developed for this analysis are not specific to CHIME and could be applied to any low-redshift and high-redshift interferometric \tcm surveys, such as CHORD \citep{vanderlinde2019}, Tianlai \citep{li2020-tianlaicyl,wu2021-tianlaidish}, HIRAX \citep{crichton2021}, MeerKat \citep{paul2023-21cmauto},  uGMRT \citep{chakraborty2021},  the Ooty Wide Field Array \citep{subrahmanya2017}, HERA \citep{deboer2017}, MWA \citep{Tingay2013}, LOFAR \citep{van_Harleem_2013} and the upcoming  Square Kilometer Array (SKA). 

The successful detection presented here demonstrates CHIME's capability for  \tcm cosmology and represents a crucial step toward our ultimate goal of measuring BAOs in the auto-power spectrum of \tcm emission to constrain the expansion history of the universe and the nature of dark energy.
As CHIME continues operations and data processing improvements are implemented, we anticipate increasingly powerful constraints on both the distribution of neutral hydrogen and the cosmological parameters that govern our universe's large-scale structure.

\section*{acknowledgment}

We thank the Dominion Radio Astrophysical Observatory, operated by the National Research Council Canada, for gracious hospitality and expertise. The DRAO is situated on the traditional, ancestral, and unceded territory of the syilx Okanagan people.   We are fortunate to live and work on these lands.

CHIME is funded by grants from the Canada Foundation for Innovation (CFI) 2012 Leading Edge Fund (Project 31170), the CFI 2015 Innovation Fund (Project 33213), and by contributions from the provinces of British Columbia, Qu\'ebec, and Ontario. Long-term data storage and computational support for analysis is provided by WestGrid\footnote{\url{https://www.westgrid.ca/}}, SciNet\footnote{\url{https://www.scinethpc.ca/}} and the Digital Research Alliance of Canada \footnote{\url{https://www.alliancecan.ca/}}, and we thank their staff for flexibility and technical expertise that has been essential to this work, particularly Martin Siegert, Lixin Liu, and Lance Couture.

Additional support was provided by the University of British Columbia, McGill University, and the University of Toronto. CHIME also benefits from NSERC Discovery Grants to several researchers,  funding from the Canadian Institute for Advanced Research (CIFAR), and from the Dunlap Institute for Astronomy and Astrophysics at the University of Toronto, which is funded through an endowment established by the David Dunlap family.  This material is partly based on work supported by the NSF through grants (2008031)  (2006911) (2006548) (2510770) (2510771) (2510772) and (2510773) and by the Perimeter Institute for Theoretical Physics, which in turn is supported by the Government of Canada through Industry Canada and by the Province of Ontario through the Ministry of Research and Innovation. M.D. is supported by a CRC Chair, NSERC Discovery Grant, CIFAR, and by the FRQNT Centre de Recherche en Astrophysique du Qu\'ebec (CRAQ). U.P.  is supported by the Natural Sciences and Engineering Research Council of Canada (NSERC) [funding reference number RGPIN-2019-06770, ALLRP 586559-23, RGPIN-2025-06396], Canadian Institute for Advanced Research (CIFAR), Ontario Research Fund – Research Excellence (ORF-RE Fund, 72074697), AMD AI Quantum Astro. AC and AO are supported by the NSERC Brockhouse Prize.
J.M.P. acknowledges the support of an NSERC Discovery Grant (RGPIN-2023-05373). We acknowledge the support of the Natural Sciences and Engineering Research Council of Canada (NSERC) [funding reference number 569654]. We thank the Beus Center for Cosmic Foundations at Arizona State University for supporting a workshop where part of this work was completed. K.W.M holds the Adam J. Burgasser Chair in Astrophysics.

We thank the Sloan Digital Sky Survey and eBOSS collaborations for publicly releasing the galaxy and quasar catalogs and supporting mock catalogs used in this work.  Funding for the Sloan Digital Sky Survey IV has been provided by the Alfred P. Sloan Foundation, the U.S. Department of Energy Office of Science, and the Participating Institutions. SDSS-IV acknowledges support and resources from the Center for High Performance Computing at the University of Utah. The SDSS website is \url{www.sdss.org}.

\software{
   CAMB \citep{lewis1999,camb_zenodo},
   caput \citep{caput},
   ch\_pipeline \citep{ch_pipeline},
   cora \citep{cora},
   Cython \citep{Cython},
   draco \citep{draco},
   driftscan \citep{driftscan},
   emcee \citep{foreman-mackey2013},
   FFTW \citep{FFTW05},
   GetDist \citep{getdist-jcap,getdist_zenodo},
   h5py \citep{h5py},
   hankl \citep{karamanis2021},
   HDF5 \citep{HDF5},
   healpy \citep{healpy},
   Matplotlib \citep{Matplotlib,matplotlib_zenodo},
   mpi4py \citep{mpi4py_2021,mpi4py_2023},
   NumPy \citep{NumPy},
   OpenMPI \citep{OpenMPI},
   SciPy \citep{SciPy},
   zarr \citep{zarr_zenodo},
   Skyfield \citep{Skyfield}.
   }

\pagebreak

\appendix
\twocolumngrid

\section{Algorithms for Radio Frequency Interference Detection and Masking}
\label{app:rfi}

\subsection{Radiometer Test Method}
\label{app:rfi_radiometer}

\begin{figure*}[p]
    \centering
    \includegraphics[width=0.98\linewidth,keepaspectratio]{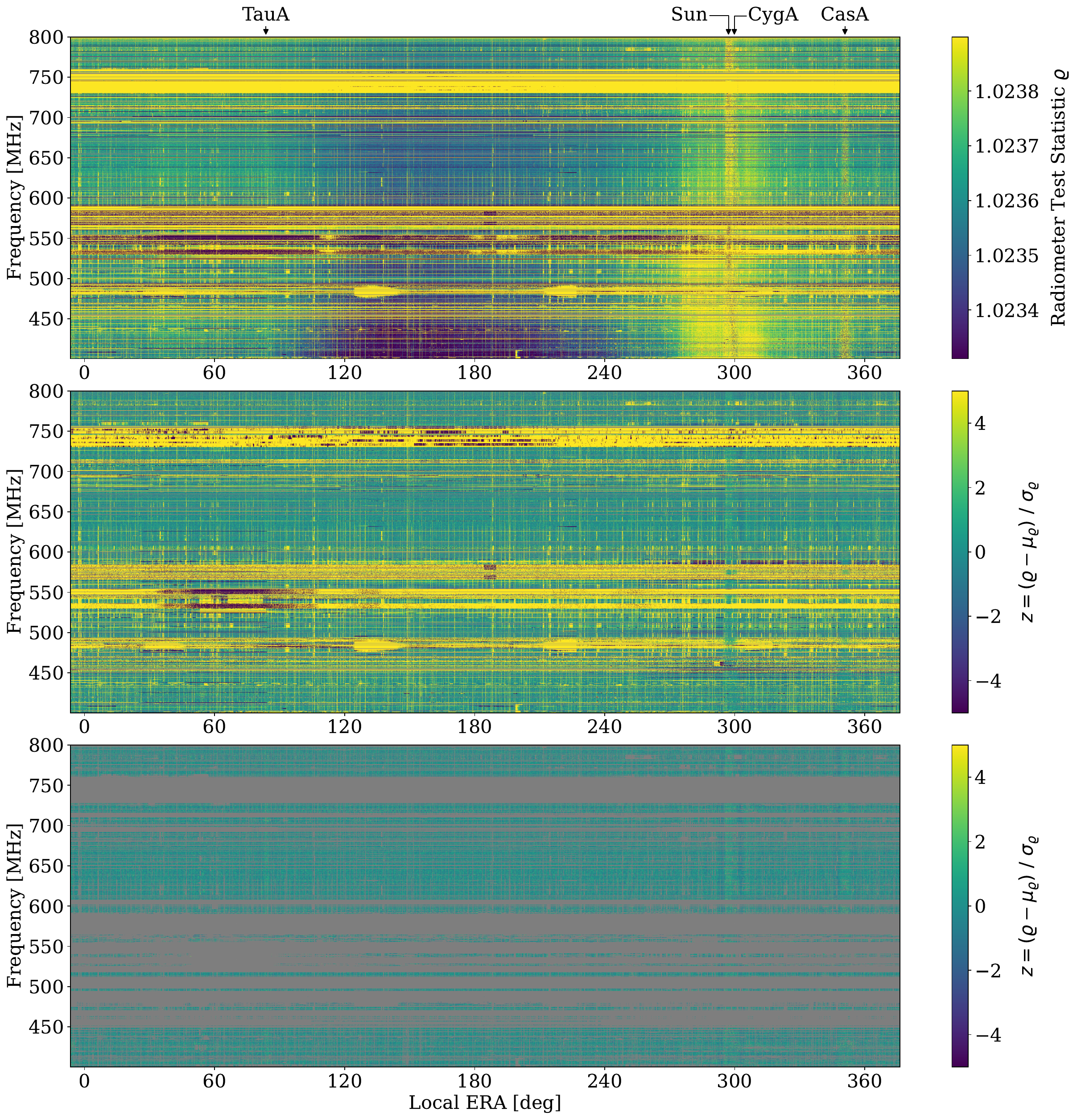}
    \caption{\textit{Top:} Radiometer test statistic $\radiometer$ (\cref{eq:radiometer}) as a function of frequency and local ERA for XY–polarization baselines on 2019-01-15. Frequencies that are missing due to non-operational GPU nodes appear in gray.  The color scale is offset by 1.023, implying the measured noise is at least $2.3\%$ higher than predicted from the autocorrelations. The mild dependence of $\radiometer$ on frequency and local ERA tracks the Galactic plane’s transit through the field of view. During transits of bright sources the scatter in $\radiometer$ increases due to partial noise correlation across baselines.  The displayed dynamic range is narrow ($<0.1\%$), and bright RFI features that clip the scale are often much larger than the maximum value shown. \textit{Middle:} Robust $z$-score map (\cref{eq:norm_radiometer}), after subtracting a slowly varying background $\mu_{\radiometer}$ (\cref{eq:avg_radiometer}) and normalizing by a MAD-based scatter estimate $\sigma_{\radiometer}$ (\cref{eq:sig_radiometer}). \textit{Bottom:} Same quantity as the middle panel, with masked times and frequencies displayed in gray.  The fraction of data missing or masked relative to the full time–frequency grid are: $12\%$ missing, $27\%$ automatically masked by the static persistent-RFI mask, and $25\%$ masked by the algorithm, for a total of $64\%$ missing or masked.
    \label{fig:radiometer_test}}
\end{figure*}

The first step in our RFI excision procedure is to construct a test statistic that measures deviations in the measured variance of the visibilities from the expectation for Gaussian, uncorrelated noise. The correlator begins by applying a four-tap polyphase filterbank (PFB) to the raw voltage timestream from each feed, in which each \SI{2.56}{\micro\second} block of 2048 samples (at \SI{1.25}{\nano\second} spacing) is transformed into 1024 complex frequency channels. These channelized samples, produced at a \SI{2.56}{\micro\second} cadence, form the baseband data. They are spatially cross-correlated to generate visibilities, which are accumulated to a cadence of \SI{30}{\milli\second}. At this stage, two intermediate quantities are accessible internally as data move from GPU to CPU: the unnormalized visibilities $v_{fab\tau}$ and the corresponding number of accumulated baseband samples $n_{f\tau}$. Note that the number of samples need not be constant across time or frequency, as it can be reduced by packet loss in network transport or by real-time RFI excision. When averaging these \SI{30}{\milli\second} visibilities into coarse \dtraw integrations, the correlator also computes variance statistics from the same fine-resolution products, as described below.

Let $M$ denote the number of fine samples per coarse integration ($M\simeq 332$), and define the fine-sample index set for coarse time $t$ as
\[
  \mathcal{I}_t \equiv \{\, \tau \in \mathbb{Z} : Mt \le \tau < M(t+1) \,\}.
\]
Let $\mathcal{E}_t \subset \mathcal{I}_t$ denote the even offsets in this interval,
\[
  \mathcal{E}_t \equiv \{\, \tau \in \mathcal{I}_t : (\tau - Mt) \bmod 2 = 0 \,\},
\]
so that for every $\tau \in \mathcal{E}_t$, the pair $(\tau,\tau+1)$ gives one even–odd pair of fine samples. The correlator accumulates four quantities:
\begin{align}
    v_{fabt} &\equiv \sum_{\tau \in \mathcal{I}_t} v_{fab\tau}, \label{eq:vsum}\\[3pt]
    N_{ft}   &\equiv \sum_{\tau \in \mathcal{I}_t} n_{f\tau}, \label{eq:Nsum}\\[3pt]
    \Delta v^{2}_{fabt} &\equiv \sum_{\tau \in \mathcal{E}_t} \big|\, v_{fab,\tau+1} - v_{fab\tau} \,\big|^{2}, \label{eq:dvsq}\\[3pt]
    \Delta n^{2}_{ft}   &\equiv \sum_{\tau \in \mathcal{E}_t} \big( n_{f,\tau+1} - n_{f\tau} \big)^{2}. \label{eq:dnsq}
\end{align}
From these quantities, it then outputs for each coarse integration an estimate of the normalized visibilities,
\begin{equation}
    V_{fabt} \equiv \frac{v_{fabt}}{N_{ft}}, \label{eq:vis}
\end{equation}
and an estimate of the variance of the visibilities,
\begin{equation}
    \sigma^{2}_{fabt} \equiv \frac{1}{N_{ft}^{2}} \left[\Delta v_{fabt}^{2} - \frac{\Delta n_{ft}^{2}}{N_{ft}^{2}} \left|v_{fabt}\right|^{2} \right], \label{eq:varmeas}
\end{equation}
where the second term in \cref{eq:varmeas} removes the bias that arises when the even and odd accumulations are formed from unequal numbers of baseband samples. The \SI{30}{\milli\second} cadence is short enough that the sky visibility is approximately constant across each fine-sample pair, and therefore cancels in the difference $v_{fab,\tau+1}-v_{fab\tau}$, leaving a combination of thermal noise and RFI.  The expected variance for Gaussian, uncorrelated noise can be obtained directly from the autocorrelations of each feed.  Specifically, for visibility $V_{fabt}$ between feeds $a$ and $b$,
\begin{equation}
    \mathrm{Var}\!\left\{V_{fabt}\right\} = \frac{V_{faat}\,V_{fbbt}}{N_{ft}}, \label{eq:radeq}
\end{equation}
where $V_{faat}$ and $V_{fbbt}$ are the autocorrelations. Any measured excess relative to this prediction indicates a departure from the Gaussian, uncorrelated-noise assumption and is taken as evidence for RFI.

Because the array contains relatively short baselines, RFI tends to appear coherently across them, whereas the receiver noise contributions remain independent. Averaging the variance estimates over all baselines within a given polarization product, therefore, enhances sensitivity to RFI while suppressing uncorrelated noise fluctuations. The radiometer test statistic $\radiometer_{pft}$ is defined as the square root of the ratio of the baseline-averaged measured variance to the baseline-averaged, autocorrelation-based prediction,
\begin{equation}
    \radiometer_{pft} \equiv \left( \frac{\sum_{ab \in \mathcal{B}_{p}} \sigma^{2}_{fabt}}
    {\tfrac{1}{N_{ft}} \sum_{ab \in \mathcal{B}_{p}} V_{faat} V_{fbbt}} \right)^{1/2}, \label{eq:radiometer}
\end{equation}
where $\mathcal{B}_{p}$ denotes the set of all baselines formed from input pairs $(a,b)$ whose polarizations correspond to the polarization product $p \in \{\mathrm{XX},\mathrm{XY},\mathrm{YY}\}$.

\Cref{fig:radiometer_test} (top panel) shows the radiometer test statistic for the XY polarization over a full sidereal day. The measured variance is consistently at least $2.3\%$ larger than the expected variance from the radiometer equation, a behavior observed across days and polarizations whose origin remains unknown. The autocorrelation-based normalization removes nearly all large-scale variations associated with the instrumental bandpass and changing system temperature, but residual variations at the $\sim0.05\%$ level persist and track the transit of the Galactic plane. The scatter in the statistic also increases during the transit of bright sources (Sun, Cyg A, Cas A, Tau A), whose correlated contributions to the system temperature do not average down across baselines.

Under ideal assumptions of Gaussian noise that is uncorrelated in time and between baselines, the mean of $\radiometer$ would be unity and its variance could be computed analytically. In practice, however, correlations between baselines introduced by the sky noise complicate this calculation and would require a detailed sky-dependent model. Rather than attempt to model these effects, we estimate the background and scatter of $\radiometer$ directly from the data using robust median filters. For each polarization $p$, the algorithm then constructs a frequency–time mask by iteratively identifying outliers in $\radiometer$ relative to these empirically determined background and scatter levels.

Before the main iterative procedure, two preliminary masks are applied. First, a static frequency mask excludes channels known to be persistently contaminated in the CHIME environment. Second, for each frequency channel, the 15th percentile of $\radiometer$ over the sidereal day is computed and then smoothed across frequency with a rolling median of width $191$ channels. Channels whose percentile deviates from this baseline by more than 5 times a robust local estimate of scatter (computed using a median absolute deviation (MAD) scaled by 1.4826, estimated with the same 191-channel window) are masked uniformly in time. This step suppresses channels with persistently abnormal behaviour over the sidereal day.

Outlier identification proceeds through a five-iteration process. At the start of each iteration, the mask from the previous iteration is applied so that previously-identified outliers are excluded when calculating statistics. A two-dimensional weighted median filter is used to estimate a smooth background of the radiometer test statistic,
\begin{equation}
    \mu_{\radiometer, pft} \equiv \left[\mathrm{MedFilt}_{w_t,w_f,1}\!\left(\radiometer\right)\right]_{pft}, \label{eq:avg_radiometer}
\end{equation}
where $\mathrm{MedFilt}_{w_t,w_f,1}$ denotes a median filter with time width $w_t=181$ samples, frequency width $w_f=37$ channels, and polarizations treated independently, applied while excluding samples flagged by the current mask $Q_{pft}$. The local scatter is then estimated with a weighted median absolute deviation filter,
\begin{equation}
    \sigma_{\radiometer,pft} \equiv 1.4826 \times \left[ \mathrm{MedFilt}_{w_t',w_f',1}\!\left(\,|\radiometer-\mu_{\radiometer}|\,\right) \right]_{pft}, \label{eq:sig_radiometer}
\end{equation}
with $w_t'=31$ samples and $w_f'=101$ channels, again excluding masked samples. The background-subtracted residual is
\begin{equation}
    \delta \radiometer_{pft} \equiv \radiometer_{pft} - \mu_{\radiometer,pft}, \label{eq:diff_radiometer}
\end{equation}
and the normalized deviation is
\begin{equation}
    z_{pft} \equiv \frac{\delta \radiometer_{pft}}{\sigma_{\radiometer,pft}}. \label{eq:norm_radiometer}
\end{equation}
Thresholds are progressively lowered over the five iterations, so that the brightest excursions are flagged first; this prevents large outliers from biasing the subsequent estimates of $\mu_{\radiometer,pft}$ and $\sigma_{\radiometer,pft}$.

\Cref{fig:radiometer_test} (middle panel) shows the normalized deviation $z$ after the final iteration for the same sidereal day and baseline. The background variations have been removed, making this metric a highly sensitive indicator of RFI. The CHIME site exhibits a striking diversity of RFI morphologies, including persistent contamination from TV channels and cellular bands, frequent broadband impulses, narrowband carriers with varying duty cycles, extended emission from Meridian satellites in Molniya orbits, and brief \SI{6}{\mega\hertz}-wide features produced by distant TV stations reflecting off objects in the sky.

A preliminary MAD mask is first constructed by flagging samples with $|z_{pft}| > \eta^{(k)}$. The threshold begins at $\eta^{(0)}=25.3125$ and is reduced by a factor of $1.5$ at each iteration, ending with $\eta^{(4)}=5$ on the final pass. This preliminary mask also defines the basis for flagging digital TV channels: within each 6\,MHz band beginning at 398\,MHz, the fraction of frequency bins with $|z_{pft}| > \eta_{\rm tv}^{(k)}$ is evaluated. If more than half of the bins in a channel exceed this level, the entire TV channel is masked for that time sample. The threshold $\eta_{\rm tv}^{(k)}$ is significantly lower than the $\eta^{(k)}$ threshold, and is chosen such that the false-positive rate is equal to that given by the $\eta^{(k)}$ threshold for the single sample flagging assuming a Gaussian noise model.  The union of this MAD-based mask and the TV-band mask is then carried forward as the starting point for a variant of the \texttt{SumThreshold} procedure \citep{rfisumthresh}.

The \texttt{SumThreshold} algorithm operates directly on the background-subtracted residuals $\delta \radiometer_{pft}$ (\cref{eq:diff_radiometer}), while propagating the per-sample variance $\sigma_{\radiometer,pft}^2$ (\cref{eq:sig_radiometer}) through convolutions. For each window length $l=1,2,4,\ldots,l_{\max}(=64)$, we form
\begin{align}
    D_{l, pft} &= \left[K_l \circledast \delta \radiometer\right]_{pft}, \\
    \Sigma_{l, pft}^{2} &= \left[K_l^{2} \circledast \sigma_{\radiometer}^{2}\right]_{pft},
\end{align}
where $K_l$ denotes a top-hat kernel of width $l$, and masked samples are set to zero in both convolutions. A detection is declared whenever
\begin{equation}
    |D_{l,pft}| > \eta^{(k)} \, \Sigma_{l, pft}. \label{eq:thresh_radiometer}
\end{equation}
Windows that trigger a detection are symmetrically expanded to cover all $l$ contributing samples. Convolutions are applied first along the time axis and then along the frequency axis, with masks from earlier passes carried forward to later ones. This variance-propagating formulation differs from the canonical \texttt{SumThreshold} algorithm \citep{rfisumthresh}, which lowers the per-sample threshold for longer windows; here the scaling with $l$ arises naturally from variance propagation. The strength of the \texttt{SumThreshold} method lies in its sensitivity to low-level but extended contamination: by summing over contiguous samples, it identifies faint RFI whose aggregate power is inconsistent with Gaussian noise even when no individual pixel exceeds threshold. The use of rectangular windows ensures that the flagged regions naturally match the morphology of broadband or time-extended RFI.

After \texttt{SumThreshold}, the algorithm blends the MAD and \texttt{SumThreshold} masks depending on the local ERA. During transits of bright sources (within one beamwidth of Cas~A, Cyg~A, Tau~A, and B0329+54, and within three beamwidths of the Sun), the MAD mask is preferred to avoid over-flagging astrophysical signals broadened by the primary beam. At all other times, the \texttt{SumThreshold} mask is used. For times near source transits, the mask is expanded with a one-dimensional scale-invariant rank operator \citep{rfisir} applied along the time axis to prevent streak-like RFI from being under-flagged during transits; the default aggressiveness is $\eta=0.2$. Unlike the general scale-invariant rank option, this restricted version is applied only in the time direction and only during transit intervals.

The background and scatter are recomputed after each iteration using the updated mask, and the process is repeated for a total of five iterations. The resulting mask $Q^{\rm rad}_{pft}$ for each polarization is then combined with a logical OR to produce the global frequency–time mask $Q^{\rm rad}_{ft}$. \Cref{fig:radiometer_test} (bottom panel) shows this mask overlaid on the normalized deviation $z$. This final mask incorporates the static exclusions and the 1D outlier test, as well as the MAD, digital TV band, and \texttt{SumThreshold} detections, and is used to excise contaminated data from subsequent analysis.

\subsection{Fringe-rate Filtering Method}
\label{app:rfi_time_filt}

\begin{figure*}[hp]
    \centering \includegraphics[width=0.98\linewidth, keepaspectratio]{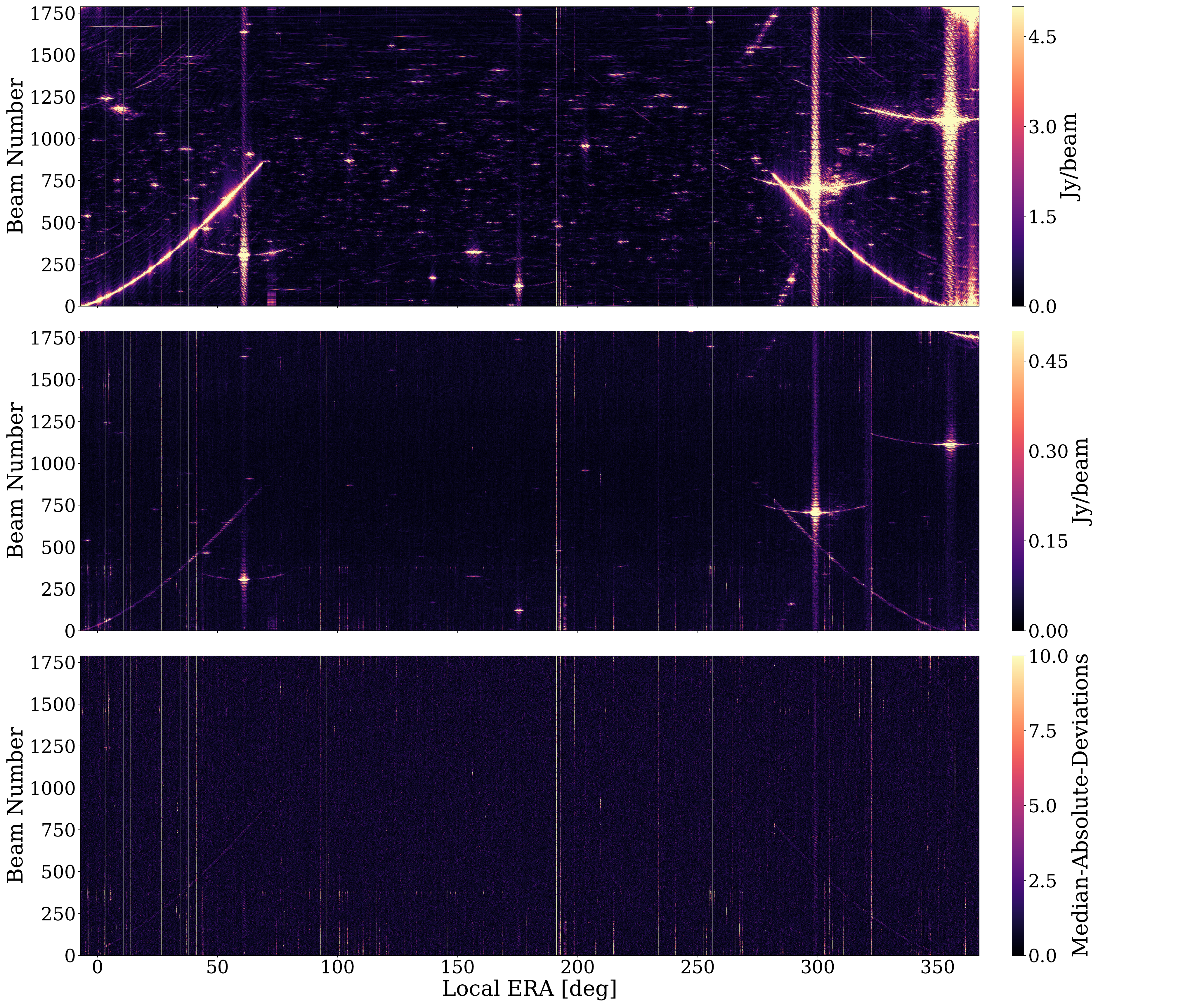}
    \caption{\textit{Top}: Map produced using \cref{eq:fft_map} from the sum of XX- and YY-polarization inter-cylinder baselines on 2021-03-25 at $706.25$ MHz. Time samples that were flagged in real time appear as thin gray lines. RFI features can be seen as bright vertical lines which do not clearly correspond to a bright source, and are most frequently seen near the horizons. The flagging algorithm uses a map that includes intra-cylinder baselines - these are omitted here in order to make RFI and point sources more visible for demonstration purposes. \textit{Middle}: Map produced using all baselines, after subtracting a low-pass filtered background $\bar{V}^{I}_{ft}$ (\cref{eq:fringe_background}). Transient RFI is strongly visible against the residual sky. Additional RFI is visible, originating from intra-cylinder baselines not included in the top panel. \textit{Bottom}: Robust median-absolute-deviations map derived from the middle panel using the method described in equations \eqref{eq:avg_radiometer}-\eqref{eq:norm_radiometer}.
    }
    \label{fig:stokesi_rfi_metric}
\end{figure*}

The second RFI flagging step targets transient RFI based on the assumption that this type of RFI has an impulse-like time-conjugate Fourier response -- that is, power from transient RFI is expected to appear in more rapidly-varying Fourier modes than the background sky, whose contribution should be limited by the fringe pattern of the instrument. Although not all signals with this Fourier behavior are necessarily caused by RFI, we do not have to differentiate for this particular analysis.

In order to reduce the volume of data that we need to filter, we compute and use only the Stokes I polarization parameter for each baseline pair $\left(a, b\right)$:
\begin{equation}
    V^{I}_{fabt} = V^{\rm XX}_{fabt} + V^{\rm YY}_{fabt}
\end{equation}
Empirically, we see no significant differences in the resulting mask if all polarization pairs are included.

An approximate sky map is produced using a fast Fourier transform across the flattened baseline axis, ordered based on two levels of sorting: first, by signed \ns feed pair separation, and second by increasing \ew cylinder separation:
\begin{equation}
    \hat{V}^{I}_{fkt} = \mathrm{FFT}\left\{V^{I}_{f\left(ab\right)t}\right\}.
    \label{eq:fft_map}
\end{equation}
This process is significantly more computationally efficient compared to forming a true ringmap, where each frequency is first beamformed along the \ns telescope axis using a weighted discrete Fourier transform, followed by a weighted FFT in the \ew direction. The omission of baseline weighting and the process of concatenating all \ew separations in \cref{eq:fft_map} produces an incorrect map - the formed beams do not exactly map to the declinations expected from the telescope layout, and the measured intensity of the sky is incorrect. These errors, however, are irrelevant in the context of this process; we simply wish to localize RFI that may not be otherwise visible in the visibilities prior to flagging, and the beam axis is not included in the final mask.

As with the radiometer test method, we first apply a static frequency mask to exclude channels known to be consistently bad. For each frequency $f$ and beam $k$, the background sky is estimated by constructing and applying a low-pass filter with cutoff $\mu_{c}$ set by the fringe rate of the longest east-west CHIME baseline $b_{ew}$,
\begin{equation}
    \mu_{c} = \frac{f}{c}\frac{b_{ew}}{\cos{\Lambda}}
\end{equation}
where $\Lambda$ is the latitude of the telescope. The filter uses a flattop window and order equal to $\mu_{s}/\mu_{c}$, where $\mu_{s} = T^{-1}$ for sampling rate $T\approx10s$. We divide the filtered data by the filtered mask in order to normalize for samples that were masked prior to this stage. The low-pass filtered background is then subtracted from the unfiltered data 
\begin{equation}
    \bar{V}^{I}_{ft} = \frac{M_{ft}\hat{V}^{I}_{ft} \circledast K_{\mu_{c}}}{M_{ft} \circledast K_{\mu_{c}}}
    \label{eq:fringe_background}
\end{equation}
\begin{equation}
    \delta \hat{V}^{I}_{ft} = \hat{V}^{I}_{ft} - \bar{V}^{I}_{ft}
\end{equation}
where $M_{ft}$ is the prior data mask.

\begin{figure*}[hp]
    \centering \includegraphics[width=0.95\linewidth, keepaspectratio, trim = 0 0 0 0]{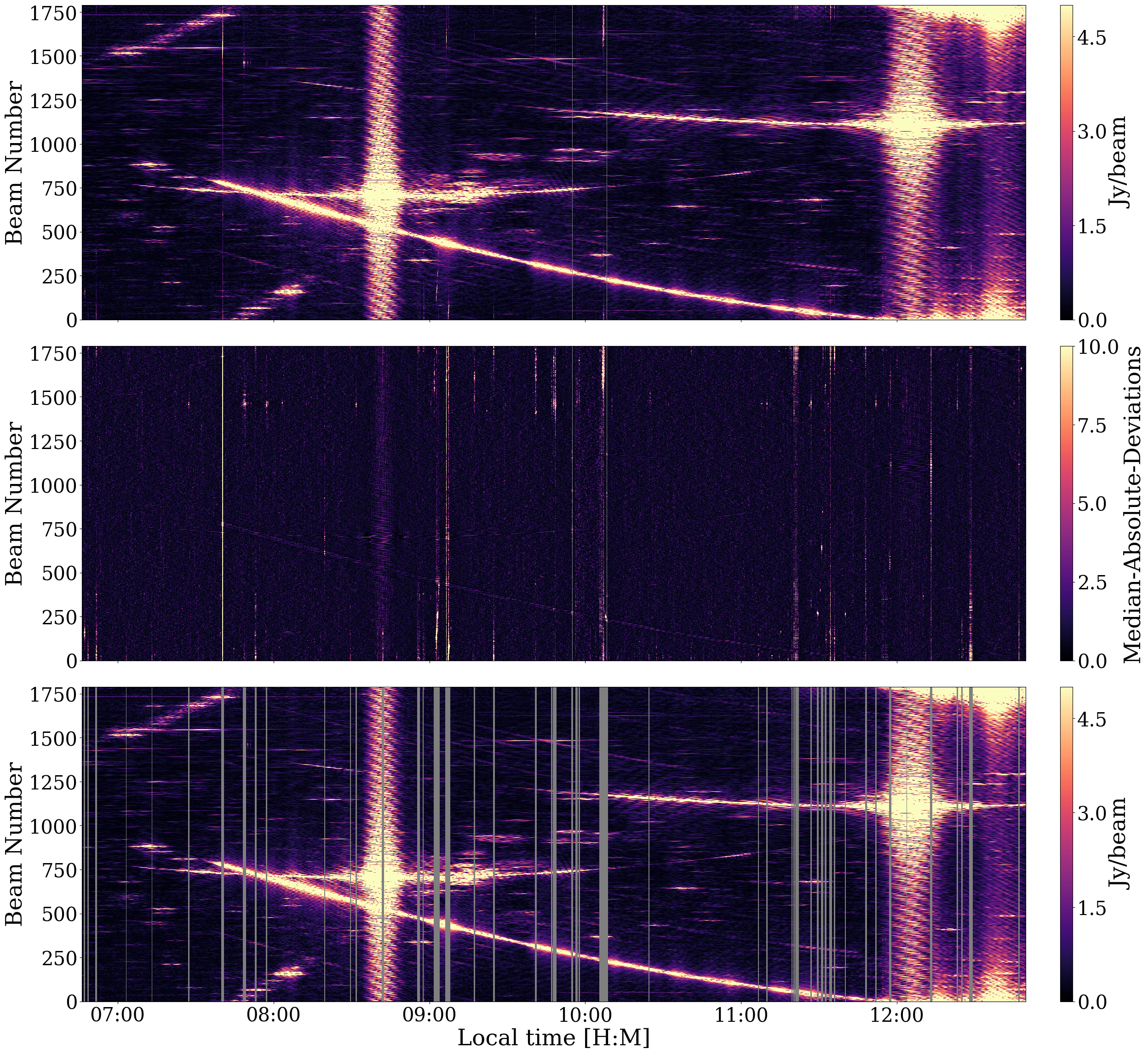}
    \caption{\emph{Top and Middle}: A 6-hour slice of the map and corresponding median-absolute-deviation metric shown in \cref{fig:stokesi_rfi_metric}. A time range during which Cyg~A, Cas~A, and the Sun are chosen to demonstrate the effectiveness of the background subtraction process. \emph{Bottom}: The same map shown in the top panel, after application of the final RFI mask produced using \cref{eq:fringe_rate_final_mask}, which is reduced across beams.
    }
    \label{fig:stokesi_rfi_example}
\end{figure*}

Outlier detection is done independently for each frequency using a robust median-absolute-deviations metric derived from the background-subtracted data using the method described in equations \eqref{eq:avg_radiometer}-\eqref{eq:norm_radiometer}. In this case, 
the one-dimensional windows for background estimation (\cref{eq:avg_radiometer}) and scatter estimation (\cref{eq:sig_radiometer}) have time widths $w_{t}=101$ samples and $w_{t}=51$ samples, respectively.
Rather than using a single threshold, however, flagging is done based on a dual-threshold criterion: a pixel is flagged if it exceeds some high threshold or if it both exceeds a lower threshold and is connected (in an adjacency context) to a pixel which exceeds the higher threshold. This approach proves to be more robust to overflagging than using a single threshold, while consistently flagging RFI events spanning multiple time samples in their entirety.

The final frequency-time mask $Q^{\text{t-filt}}_{ft}$ is obtained by flagging a frequency-time if the fraction of flagged beams for that sample exceeds some threshold $\tau$,
\begin{equation}
    Q^{\text{t-filt}}_{ft} = \Theta \left(\frac{1}{n_{k}}\sum_{k}{Q^{\text{t-filt}}_{fkt}} - \tau \right)
    \label{eq:fringe_rate_final_mask}
\end{equation}
where $n_{k}$ is the total number of beams, $\Theta(\cdot)$ is the Heaviside function, and $Q^{\text{t-filt}}_{fkt}$ is the beam-dependent mask. A fraction $\tau=0.01$ was empirically determined to adequately flag RFI events while minimizing false positives. Similarly to the variance-based mask, this mask is used to excise contaminated data from the analysis.

\subsection{Spectral Filtering Method}
\label{app:rfi_spec_filt}

\begin{figure*}
    \centering \includegraphics[width=\linewidth, keepaspectratio]{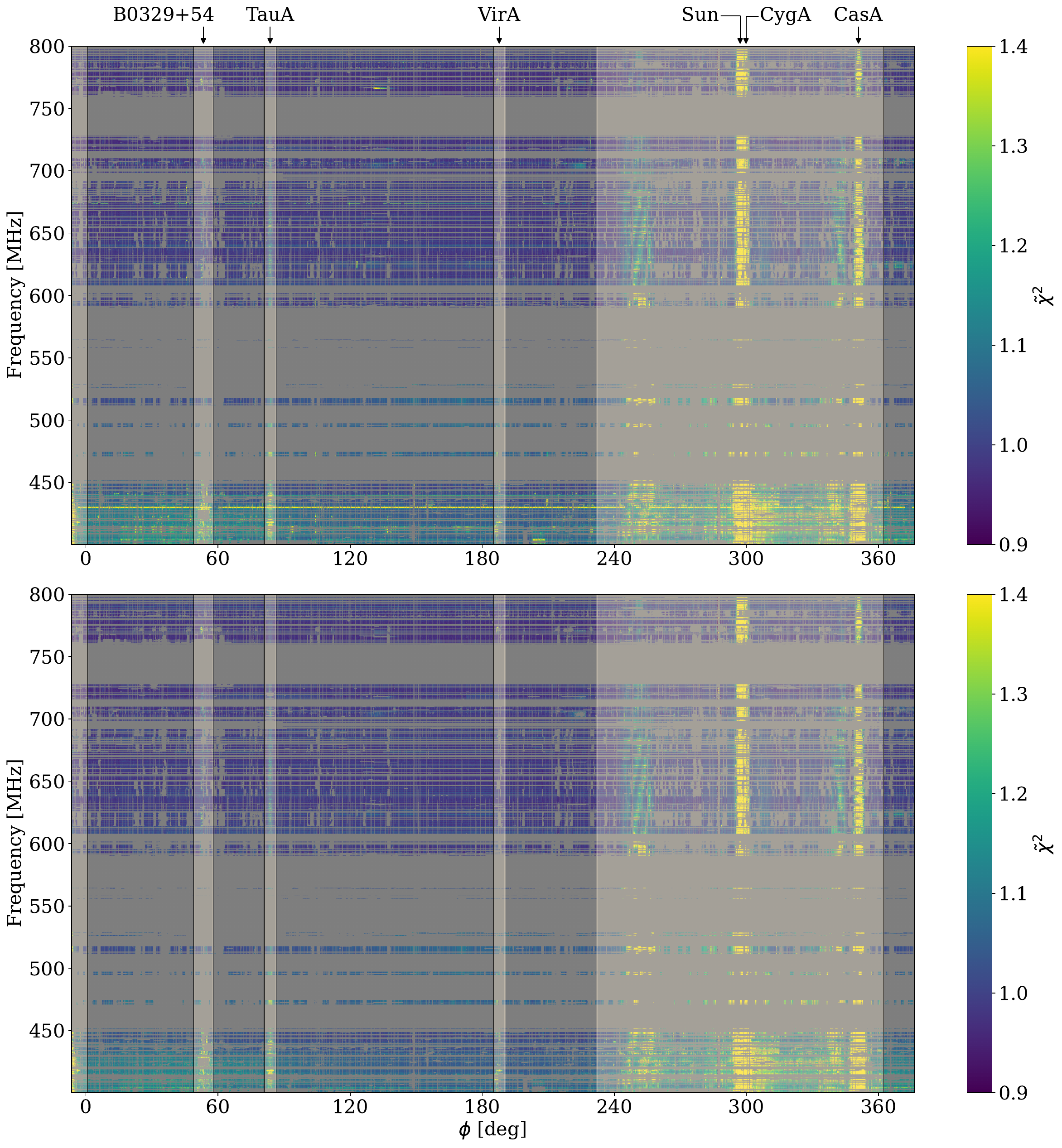}
    \caption{  
    $\chi^{2}$ per degree of freedom test statistic ($\tilde{\chi}^{2}$) as a function of local ERA and frequency on 2019-01-15. Masked samples are shown in gray. Local ERAs when the Sun is above the horizon and when bright sources transit are shaded. 
    \emph{Top:} Combined mask from the radiometer and fringe-rate filters, $Q^{\text{rad}} \lor Q^{\text{t-filt}}$, applied prior to filtering and constructing $\tilde{\chi}^{2}$. 
    \emph{Bottom:} Same as above, but including the additional mask $Q^{\text{f-filt}}$, derived directly from $\tilde{\chi}^{2}$ and applied after filtering. No $Q^{\text{f-filt}}$ mask is constructed within the shaded intervals.
    }
    \label{fig:rfi_chisq}
\end{figure*}

Our final step examines whether, after applying the previously determined masks, the data are consistent with the expected level of thermal noise at high delays. The purpose of this step is to isolate residual RFI and other non-thermal signals by applying an aggressive high-pass filter in delay that removes the contribution from the smooth radio sky. Foregrounds are spectrally smooth, whereas RFI typically exhibits sharp spectral features; applying an aggressive high-pass filter along frequency removes the foreground signal while allowing RFI to pass through and remain detectable.

Let $Q_{ft}\equiv Q^{\rm rad}_{ft}\lor Q^{\text{t-filt}}_{ft}$ denote the combined binary mask from Appendices~\ref{app:rfi_radiometer} and \ref{app:rfi_time_filt}, and define the good-data flag $G_{ft}\equiv1-Q_{ft}$.  The diagonal masking operator is then given by 
\begin{equation}
\left[\mathbf{\mask}_{t}\right]_{ff'} \equiv G_{ft} \ \delta_{ff'} \ .
\end{equation}
The per-time DAYENU high-pass operator \citep{ewall-wice2021} is then
\begin{equation}
\mathbf{H}^{\mathrm{rfi}}_{t}
=\big(\mathbf{\mask}_{t}\,\mathbf{C}\,\mathbf{\mask}_{t}\big)^{+},
\end{equation}
where ${}^{+}$ denotes the Moore–Penrose pseudoinverse, and $\mathbf{C}$ is the model frequency–frequency covariance of smooth foregrounds from \cref{eq:dayenu_cov}, with $\tau_{\mathrm{cut}}=\SI{400}{\nano\second}$ and ridge $\epsilon=10^{-12}$. The value of $\tau_{\mathrm{cut}}$ is set by the geometric horizon delay for the longest baseline in the array ($\approx \SI{100}{\meter}$, corresponding to $\tau_{\rm hor} \approx \SI{330}{\nano\second}$), plus an additional buffer of $\approx \SI{70}{\nano\second}$ to allow for primary-beam leakage. This choice effectively removes smooth sky contributions, even from sources near the horizon, while leaving non-smooth signals, such as RFI, present at high delays.

The filter is applied to both the visibilities,
\begin{equation}
    V^{\mathrm{hpf,rfi}}_{pfent} = \sum_{f'} H^{\mathrm{rfi}}_{ff't} \ V_{pf'ent} ,
\end{equation}
and to the noise variance estimate to obtain the inverse variance of the filtered visibilities,
\begin{equation}
    w^{\mathrm{hpf,rfi}}_{pfent} = \left( \sum_{f'} \left| H^{\mathrm{rfi}}_{ff't} \right|^{2} \ \sigma^{2}_{pf'ent} \right)^{-1} .
\end{equation}
Note that the variance estimates used here already have the radiometer mask $Q^{\rm rad}$ applied, so they closely match the expectation for uncorrelated Gaussian noise derived from the autocorrelations, and they have been median-filtered along the time axis using a \SI{5}{\minute} window to suppress sample variance, following the standard data processing pipeline.  From these two quantities, we can construct a $\chi^{2}$ per degree of freedom test statistic by averaging over baselines:
\begin{equation}
    \tilde{\chi}^{2}_{ft} = \frac{1}{N_{b}} \sum_{pen} w^{\mathrm{hpf,rfi}}_{pfent} \left| V^{\mathrm{hpf,rfi}}_{pfent} \right|^{2} ,
\end{equation}
where the sum over $p$ includes all polarization products $p \in \{\mathrm{XX}, \mathrm{XY}, \mathrm{YY}\}$, the sum over $n$ includes north–south baselines $n \in [-239, 239]$, and the sum over $e$ includes only inter-cylinder baselines $e \in [1, 2, 3]$. The quantity $N_{b}$ denotes the total number of baselines included in the sum. If the data are consistent with thermal noise at high delay, then $\mathbb{E}[\tilde{\chi}^{2}]=1$ and $\mathrm{Var}[\tilde{\chi}^2] = 2 / N_b$.

\Cref{fig:rfi_chisq} (top panel) shows the $\tilde{\chi}^2$ test statistic over a full sidereal day.  Outliers in this statistic are identified through a simple two-step procedure. First, for each frequency we compute the median of $\tilde{\chi}^{2}$ over time, using only unmasked samples, and compare it to a smooth baseline across frequency. The baseline is obtained using the \emph{asymmetric reweighted penalized least squares} (arPLS) method \citep{baek:2015}, which iteratively estimates a smooth trend by minimizing a weighted combination of squared curvature and residuals, with the weights updated to down-weight positive outliers. We use a regularization parameter of $10^{5}$. Channels whose medians deviate from this baseline by more than $5$ MAD are masked uniformly in time, removing channels that exhibit persistent excess power at high delay throughout the sidereal day.

Next, we subtract a moving weighted median in time to remove slow background variations in the test statistic,
\begin{equation}
    \delta \tilde{\chi}^{2}_{ft} \equiv \tilde{\chi}^{2}_{ft} - \left[\mathrm{MedFilt}_{w_{t},1} \left(\tilde{\chi}^{2} \right) \right]_{ft} ,
\end{equation}
where the window size is $w_{t} = 601$ samples (\SI{100}{\minute}). We then flag times and frequencies where $|\delta \tilde{\chi}^{2}_{ft} |>5 \sigma_{\tilde{\chi}^{2}}$, with $\sigma_{\tilde{\chi}^{2}} = \sqrt{2 / N_{b}}$ representing the expected scatter.

Elevated values of $\tilde{\chi}^{2}$ are observed during several specific periods. Around the transits of bright sources (Cyg~A, Cas~A, Tau~A, and Vir~A), excess power at high delay is primarily due to residual bandpass errors that couple foregrounds to high-delay modes. In the main pipeline, these effects are mitigated using the HyFoRes algorithm (\secref{sec:hyfores}) and by masking the brightest sources (\secref{sec:spatial_mask}). Elevated values are also observed when the Sun is in the primary beam during transit, as well as when it is in the far sidelobes near sunrise and sunset. In the latter case, although the beam response is small, the Sun’s extreme brightness combined with the near-horizon geometry of the sidelobes produces significant power at high delays. Finally, the bright pulsar B0329+54 exhibits significant power at high delays, which is thought to arise from its strong diffractive scintillation \citep[e.g.][]{wang:2008}.  For these reasons, we do not construct a mask during daytime, as these data are not used in the cosmological analysis. We also do not generate a mask during the transits of bright sources or B0329+54 within a $1\sigma$ window around transit, where $\sigma$ is the approximate width of a Gaussian fit to the primary beam.

The resulting mask from this procedure is denoted $Q^{\text{f-filt}}$, where samples identified as outliers in the high-delay $\chi^2$ test are flagged. Unlike the radiometer-based method, no iterative background re-estimation is performed.  \Cref{fig:rfi_chisq} (bottom panel) shows the $\tilde{\chi}^2$ test statistic after this mask has been applied.  An additional $1.5\%$ of the data outside of the regions mentioned above was masked on this particular sidereal day.

Finally, we define the overall RFI mask as the logical OR of the three masks introduced in this appendix,
$Q_{ft} \equiv Q_{ft}^{\rm rad} \lor Q_{ft}^{\text{t-filt}} \lor Q_{ft}^{\text{f-filt}}$.  This global mask is used to excise contaminated data from all subsequent analyses.

\section{Details  of the Spatial and Spectral Mask Construction}
\label{app:masking}

In this appendix, we provide the technical details for constructing the masks used to mitigate contamination from bright continuum foreground sources (\secref{sec:spatial_mask}) and narrow-band absorption features (\secref{sec:absorbers}).

For each source $s$, we construct a smoothly tapered mask defined by the source coordinates $(\nu_{s}, \theta_{s}, \alpha_{s})$, where $\nu_s$ is the peak-frequency channel of the narrow-band feature (for absorber candidates), and $\alpha_{s}$ and $\theta_{s}$ are the right ascension and declination of the source in CIRS coordinates at the mean epoch of the observations used to construct the map. For continuum sources, we construct purely spatial masks in the $(y,\phi)$ plane that are applied uniformly across frequency. For absorber candidates, we use a three-dimensional mask that depends on both the spatial coordinates and the observed frequency.

We first define a generalized (normalized) distance $d$ from a map voxel at coordinates $(\nu, y, \phi)$ to the effective location of the source at coordinates $\left(\nu_{s}, y_{s}(\phi_{s}), \phi_{s}\right)$:
\begin{align}
\label{eq:elliptical_dist_3d}
d^{(s)}(\nu, y, \phi)
  = \Bigg[ &
     \left(\frac{\nu - \nu_s}{w_\nu} \right)^{2}
     + \left( \frac{y - y_{s}(\phi_{s})}{w_y} \right)^{2} \notag \\
     & + \left( \frac{(\phi - \phi_{s}) \cos\theta_s}{w_\phi} \right)^{2}
     \Bigg]^{1/2}.
\end{align}
The parameters $w_\nu$, $w_y$, and $w_\phi$ set the extent of the mask in the $\nu$, $y$, and $\phi$ directions, respectively. For continuum sources, we enforce frequency independence by taking $w_\nu \rightarrow \infty$, so that the first term in \cref{eq:elliptical_dist_3d} vanishes and $d^{(s)}$ depends only on $(y,\phi)$. For absorber candidates, we use a finite $w_\nu$ (defined below), so that the taper is genuinely three-dimensional.

The spatial extents $w_y$ and $w_\phi$ are defined differently in the two directions. The extent in $y$ is set by the width of the synthesized beam, whereas the extent in $\phi$ is determined by the width of the primary beam, since we have significant grating lobes along $\phi$. To obtain a single spatial mask that is valid across all frequencies, we conservatively set $w_y$ and $w_\phi$ using the beam widths at the lowest frequency in our analysis band (\SI{608.2}{\mega\hertz}), ensuring that the masked regions are sufficiently large at all frequencies. We adopt $w_y = 0.018$ (corresponding to approximately $1^{\circ}$ at zenith), which is roughly three times the FWHM of the \ns synthesized beam, and $w_\phi = 4.8^{\circ}$, which is roughly twice the FWHM of the primary beam.

We then apply a smooth cosine taper as a function of the generalized distance $d$ to ensure a gradual transition from fully masked to unmasked regions:
\begin{equation}
\label{eq:taper_function}
T(d) = \begin{cases}
     0\ , & d \leq d_{\rm in}, \\
      0.5\left[1 + \cos\left(\frac{\pi (d - d_{\rm in})}{w_{\rm taper}}\right)\right]\ , & d_{\rm in} < d \leq d_{\rm out}, \\
      1\ , & d > d_{\rm out},
\end{cases}
\end{equation}
where $d$ is the normalized distance defined in \cref{eq:elliptical_dist_3d}, $d_{\rm in} = 1.0$ defines the inner radius where the taper begins, $d_{\rm out} = d_{\rm in} + w_{\rm taper}$ defines the outer radius where the taper reaches unity, and $w_{\rm taper} = 1.0$ is the taper width parameter. This yields a smooth transition from fully masked regions ($T = 0$) to unmasked regions ($T = 1$) over a distance equal to the mask window size.

As mentioned in \secref{sec:spatial_mask}, we construct several families of tapered masks, each targeting a specific source of contamination:

\begin{enumerate}
\item \textbf{Transit Mask:}
For all continuum sources in the \emph{specfind v3} catalog \citep{Stein2021} with flux densities greater than \SI{10}{\jansky} at \SI{600}{\mega\hertz}, we construct a spatial mask $T_{\rm transit}^{(s)}(y, \phi)$ for each source $s$ by treating it as a point at
\begin{equation}
    y_{s} \equiv \sin(\theta_{s} - \Lambda)\ , \qquad
    \phi_{s} \equiv \alpha_{s}.
\end{equation}
We evaluate the distance metric $d^{(s)}$ using \cref{eq:elliptical_dist_3d} under the continuum-source convention $w_\nu \rightarrow \infty$, and apply the taper function in \cref{eq:taper_function} to obtain $T_{\rm transit}^{(s)}(y,\phi)$, which is applied uniformly across all frequency channels.

\item \textbf{Track Mask:}
For the brightest continuum sources, with flux densities greater than \SI{60}{\jansky} at \SI{600}{\mega\hertz}, we construct extended track masks $T_{\rm track}^{(s)}(y, \phi)$ that follow the full path of each source $s$ across the sky. The track is parameterized in terms of a local ERA coordinate $\phi_{s}$, which denotes the source location in local ERA, as
\begin{align}
   \label{eqn:track_path}
   y_{s}(\phi_{s}) & \equiv \cos\Lambda \,\sin\theta_s
   - \sin\Lambda \,\cos\theta_s \,\cos(\phi_{s} - \alpha_s).
\end{align}
We sample $\phi_{s}$ on a grid that is upsampled by a factor of 10 relative to the native map resolution in $\phi$, and restrict this grid to the range of local ERA over which the source is above the horizon. For each sampled point $(y_s(\phi_s^{(k)}), \phi_s^{(k)})$, we evaluate \cref{eq:elliptical_dist_3d} with $w_\nu \rightarrow \infty$ to obtain a distance $d_k^{(s)}(y,\phi)$ at each map pixel. The mask value at a given pixel $(y,\phi)$ is then set by the minimum distance to any point on the track,
\[
d_{\min}^{(s)}(y,\phi) = \min_k d_k^{(s)}(y,\phi),
\]
and the taper function in \cref{eq:taper_function} is applied to $d_{\min}^{(s)}$ to define $T_{\rm track}^{(s)}(y,\phi)$.

\item \textbf{Galactic Plane Mask:}
To mask sidelobe pickup from the inner Galactic plane within the NGC field, we construct a Galactic-plane mask $T_{\rm gal}(y, \phi)$ using the same track-mask procedure. We select a set of dimmer known continuum sources in the inner Galactic plane whose tracks approximately coincide with the residual contamination observed in the maps at delays within the transition region of the delay filter. For each such source $g$, we build a track mask $T_{\rm track}^{(g)}(y,\phi)$ as above (again with $w_\nu \rightarrow \infty$), and define
\begin{equation}
T_{\rm gal}(y,\phi) = \prod_{g} T_{\rm track}^{(g)}(y,\phi).
\end{equation}

\item \textbf{Absorber Mask:}
As described in \secref{sec:absorbers}, we identify 40 narrow-band absorption candidates with signal-to-noise ratio $> 7$ within the NGC region. For each candidate system $s$ with sky position $(\theta_s,\alpha_s)$ and observed frequency $\nu_s$ (corresponding to the peak of the feature), we construct a three-dimensional mask $T_{\rm abs}^{(s)}(\nu, y, \phi)$ using the distance definition in \cref{eq:elliptical_dist_3d}. The spectral extent is defined as
\begin{equation}
w_{\nu} = n_{\nu} \,\Delta_{\nu},
\end{equation}
where $\Delta_{\nu} = \dnu$ is the channel width and we adopt $n_{\nu} = 13$ to cover 13 channels on both sides of the peak channel, encompassing the spectral sidelobes (see Section~\ref{sec:absorbers}). We use the same spatial parameters $w_y$ and $w_\phi$ as for the transit masks, and apply the taper function in \cref{eq:taper_function} to the full three-dimensional distance to obtain $T_{\rm abs}^{(s)}(\nu, y, \phi)$.
\end{enumerate}

\begin{figure}
    \centering
    \includegraphics[width=0.98\linewidth,keepaspectratio]{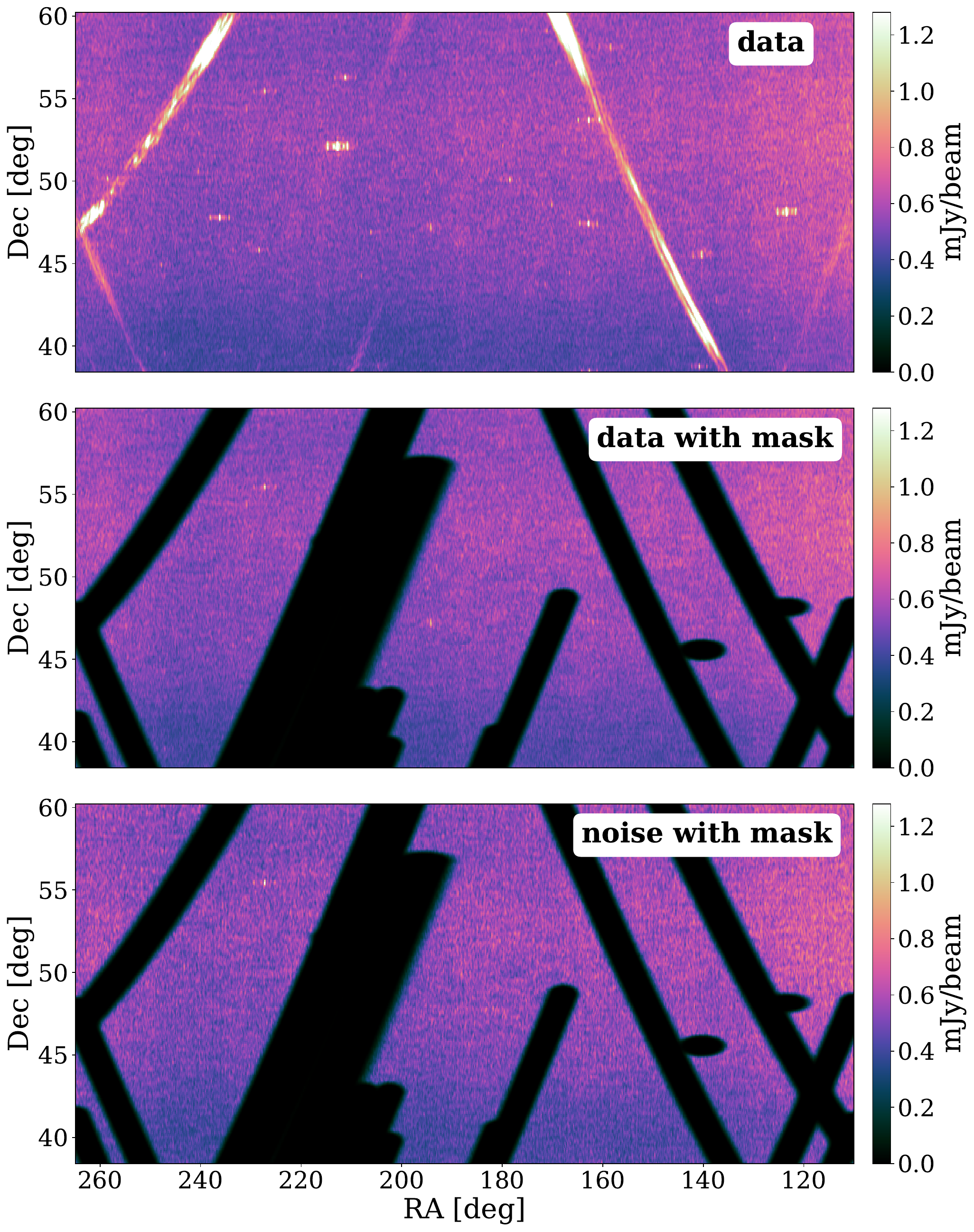}
        \caption{\emph{Top panel:} RMS of the map over the frequency axis, without applying any mask. The ``U''-shaped tracks correspond to bright continuum foreground sources in the far sidelobes of the primary beam, and the point-source-like features are predominantly narrow-band absorbtion features. \emph{Middle panel:} Same data after applying the combined spatial mask to mitigate continuum foreground contamination and the three-dimensional absorber mask to remove narrow-band absorption features, flagging $\sim 33\%$ of the spatial area and $1.29\%$ of voxels, respectively. \emph{Bottom panel:} A single noise realization, scaled to account for sky-temperature variations (see \secref{sec:power_spectrum_estimation:noise_covariance_estimation}), with the identical mask applied. After masking, the data and noise are in excellent agreement. A point-source-like feature is visible in both the masked data and noise maps (near RA $\sim 227^{\circ}$, Dec $\sim 55^{\circ}$). This arises from an unknown variable source that also appears in the even–odd jackknife map used to derive the spatial scaling of the noise simulations, and is therefore imprinted in the noise realization as well. To assess its impact, we generate an additional transit mask to exclude this feature and recompute the power spectrum, finding no significant change in the measured bandpowers. We conclude that this source does not bias our results.}
    \label{fig:rms_over_freq}
\end{figure}

We compute the RMS of the map over the frequency axis and it is shown in Figure~\ref{fig:rms_over_freq}. The top panel shows the unmasked data, clearly revealing systematic contamination from the brightest point sources.  These sources produce characteristic ``U''-shaped tracks in the map as they drift through the far sidelobes of the primary beam.  Additionally, numerous point-source-like features are visible, which are predominantly candidate absorbers. The middle panel shows the same data after applying both the spatial mask (to mitigate continuum foreground source contamination) and the spectral mask (to remove narrow-band absorption features). The bottom panel presents a single realization of the noise simulation, scaled to account for spatial variations in sky temperature (as detailed in \secref{sec:power_spectrum_estimation:noise_covariance_estimation}), with the identical mask applied. The masked data and noise realization exhibit excellent agreement.

However, both the masked data and noise maps contain a single point-source-like feature at the same location. This feature is not associated with any known bright point source in the sky, but exhibits variability across sidereal days. Because it is variable, this source appears in the even-odd jackknife map, which is used to derive the spatial scaling factor for the noise simulations (\secref{sec:power_spectrum_estimation:noise_covariance_estimation}). Consequently, this variability is imprinted in the noise realizations as well. The physical origin of this source requires further investigation. To assess its impact on this analysis, we generate an additional spatial transit mask to exclude this feature and recompute the power spectrum. We find no significant deviation in the measured bandpowers, confirming that this source does not bias our result.


\section{Varying the Signal Prior in the Delay Spectrum Estimation }
\label{app:varying_prior}

In \secref{subsec:delayspectrum} we add a small regularisation matrix
$\mathbf{C}_{\rm reg}$ to the propagated noise covariance $\mathbf{N}$ when
forming inverse–covariance weights along the frequency axis,
$(\mathbf{C}_{\rm reg} + \mathbf{N})^{-1}$. The role of $\mathbf{C}_{\rm reg}$
is to provide a smooth, simulation–informed approximation to the expected
\tcm frequency–frequency covariance while remaining subdominant to the noise,
so that it stabilises the matrix inverse without significantly changing the
weighting of the data. Here we describe how $\mathbf{C}_{\rm reg}$ is constructed.

We start from a simple model for the \tcm signal covariance in delay space,
which we take to be diagonal with constant variance:
\begin{equation}
  [\mathbf{S}]_{aa'} = A_{\rm prior}\,\delta_{aa'} \, .
\end{equation}
The parameter $A_{\rm prior}$ sets the overall variance of the \tcm prior in
delay space. We calibrate $A_{\rm prior}$ using an ensemble of \tcm simulations
passed through our telescope model and analysis pipeline. From the resulting
maps, we form a sample frequency–frequency covariance by averaging over local
ERA and across \tcm realizations, then Fourier transform it along the
frequency axis to obtain an effective delay–space covariance. We find that the
variance decreases approximately exponentially with delay. Rather than imposing
this scale dependence explicitly, we adopt a flat prior in delay space with
constant amplitude $A_{\rm prior} = 5 \times 10^{-6}$, chosen to match the mean
variance observed in simulations over the range of delays used for power
spectrum estimation.

The same simulations also show significant structure in the frequency–frequency
covariance induced by the primary beam. To ensure that $\mathbf{C}_{\rm reg}$
reflects this behaviour, we introduce a diagonal matrix $\mathbf{B}$ whose
entries are the RMS of the EW synthesised beam for the pixel under
consideration,
\begin{equation}
  \bigl[\mathbf{B}\bigr]_{ff'}
  = \delta_{ff'}\,
    \left[
      \frac{1}{2 N_{\phi}}
      \sum_{p} \sum_{r}
        \left| B^{\rm EW}_{\text{synth},pfd}(\phi_{r}) \right|^{2}
    \right]^{1/2} \, ,
\end{equation}
where $B^{\rm EW}_{\text{synth},pfd}(\phi)$ is given by \cref{eq:synth_ew}. We
then construct the regularisation matrix as
\begin{equation}
  \mathbf{C}_{\rm reg}
  = \mathbf{W}\,\mathbf{\mask}\,\mathbf{H}\,
    \bigl(\mathbf{B}\,\mathbf{F}\,\mathbf{S}\,\mathbf{F}^\dagger\,\mathbf{B}\bigr)\,
    \mathbf{H}^\dagger\,\mathbf{\mask}\,\mathbf{W} ,
\end{equation}
where the $\mathbf{W}$, $\mathbf{\mask}$, $\mathbf{H}$, and $\mathbf{F}$ operators implement the sub-band selection, frequency channel mask, foreground filter, and delay-to-frequency transform, respectively, and are defined in \secref{subsec:delayspectrum}.  Compared to the raw frequency–frequency signal covariance measured directly from the end-to-end \tcm simulations, this construction captures the dominant frequency structure introduced by the primary beam while smoothing over sample-variance fluctuations.

\begin{figure}
    \centering
\includegraphics[width=0.98\linewidth,keepaspectratio]{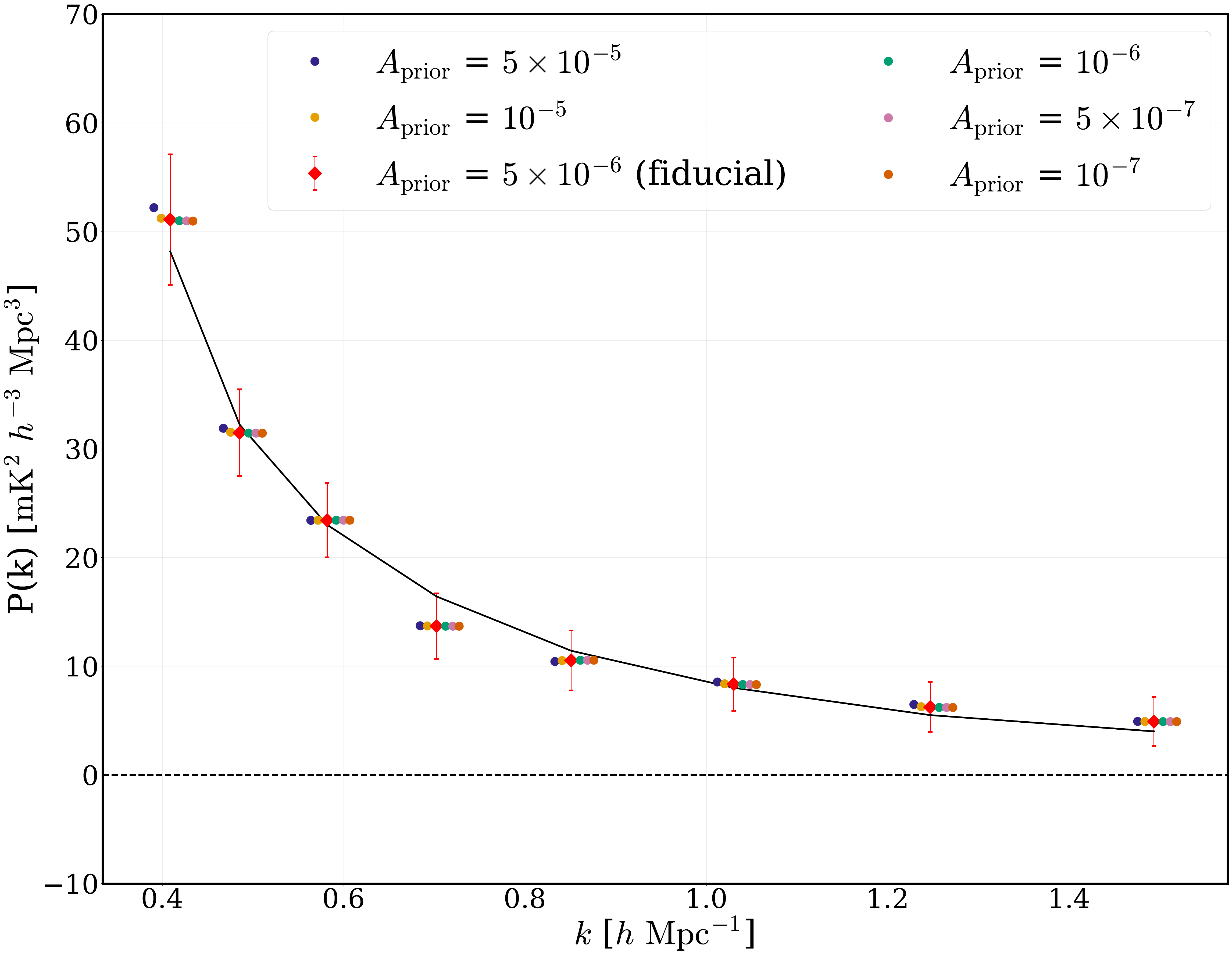}
\caption{Effect of varying the signal-prior variance $A_{\rm prior}$
on the 1D power spectrum. The different-colored points show the measured
$P(k)$ for several choices of $A_{\mathrm{prior}}$ between
$10^{-7}$ and $5 \times 10^{-5}$, while the red open diamond corresponds
to our fiducial choice $A_{\mathrm{prior}} = 5\times10^{-6}$. The solid
black line shows the best-fit model from \secref{sec:results:theoretical_interpretation}.
The curves are nearly indistinguishable, with differences far smaller than
the statistical uncertainties, indicating that the measurement is effectively
insensitive to the choice of $A_{\mathrm{prior}}$ over this range.}
    \label{fig:check_priors}
\end{figure}

It is important to verify that our results do not depend sensitively on the choice of $A_{\rm prior}$. To test this, we estimate power spectra for a wide range of $A_{\rm prior}$ values between $10^{-7}$ and $5 \times 10^{-5}$. The resulting 1D power spectra are shown in Fig.~\ref{fig:check_priors}. As $A_{\rm prior}$ is decreased, the recovered $P(k)$ rapidly converges to a stable curve; our fiducial choice $A_{\rm prior} = 5\times10^{-6}$ lies well within this convergence regime. Differences between the power spectra for different
$A_{\rm prior}$ are small compared to the statistical uncertainties, demonstrating that $A_{\rm prior}$ acts only as a mild regularisation of the inverse in \cref{eq:delay_operator} and does not bias the final power spectrum estimates.


\section{Details of Detection Significance {and Goodness-of-Fit Calculations}}
\label{app:detection_significance}

In this appendix, we provide further details of the detection significance calculations presented in \secref{sec:results:detection_significance} {and goodness-of-fit results quoted in \secref{sec:results:theoretical_interpretation:power_spectrum_amplitude}}.

\subsection{{Detection Significance}}
\label{app:detection_significance:detection_significance}

When computing $\chi^2_{\rm null}$ and $\chi^2_\text{best-fit}$ in \cref{eq:Deltachi2} and \cref{eq:chi2bestfit} for each noise realization, we recompute a ``leave-one-out" covariance $\CNoise$ that excludes that realization $\vec{d}$.
We determine $\chi^2_\text{best-fit}$ by using the BFGS algorithm. To ensure that the algorithm does not get stuck in a local minimum, we run it 1000 times with different initial guesses widely distributed in the 4-dimensional model-parameter space\footnote{For this calculation, we use a parameter set that includes $\OmegaHI^2$ instead of $\OmegaHI$, allowing for $\OmegaHI^2<0$ to allow the best-fit model for a given noise power spectrum to have a negative amplitude. Due to parameter degeneracies, the $\chi^2_\text{best-fit}$ parameter values sometimes lie at extreme parameter values. To ensure that such minima are found, we generate the 1000 initial guesses using Latin hypercube sampling of the 4-dimensional space of the (base-10) logarithm of each parameter, with $\log(\OmegaHI^2)$ bounded by $[-2,6]$ and $\log(\bHI)$, $\log(\alphaNL)$, and $\log(\alphaFoG)$ bounded by $[-2,2]$.}, and retain the parameter values that result in the minimum $\chi^2$ value out of these 1000 runs. We compute $\Delta\chi^2$ for the measured power spectrum in the same manner.

In simple cases, Wilks' theorem states that, under the null hypothesis, $\Delta\chi^2$ for the noise power spectra is drawn from a $\chi^2$ distribution with $n_{\rm dof} = n_{\rm signal}-n_{\rm null}$ degrees of freedom, where $n_{\rm signal}$ and $n_{\rm null}$ are the numbers of parameters in the signal and null models, respectively ($4$ and $0$ in our case). However, parameter degeneracies generally break the conditions required for Wilks' theorem to hold, so it is necessary to characterize the distribution of $\Delta\chi^2$ values via Monte Carlo.

A common prescription is to assume that $\Delta\chi^2$ is still $\chi^2$-distributed, but with an \textit{effective} number of degrees of freedom $n_{\rm dof}<n_{\rm signal}-n_{\rm null}$ that captures the reduction due to parameter degeneracies. However, since we have estimated the noise covariance from a set of noise realizations, we expect the relevant distribution to be $F$ instead of $\chi^2$ \citep{Anderson2003-statistics,sellentin-heavens2016}. In this case, the analog of Wilks' theorem with an effective number of degrees of freedom is
\beq
\label{eq:scaled-Deltachi2-F}
\frac{n-p+1}{n_{\rm dof}n} \Delta\chi^2 \sim F_{n_{\rm dof},\, n-p+1}\ ,
\eeq
where $n=n_{\rm mocks}-1$ and $p=n_{\rm data}$ denotes the length of the data vector.\footnote{\cref{eq:scaled-Deltachi2-F} can be understood from the fact that the $(n-p+1)/n$ prefactor and the ``denominator degrees of freedom" $n-p+1$ of the $F$ distribution arise from the Wishart distribution from which the estimated covariance matrix is drawn, while the $n_{\rm dof}^{-1}$ prefactor and ``numerator degrees of freedom" $n_{\rm dof}$ arise from the vector being tested. \Cref{eq:scaled-Deltachi2-F} is a version of Hotelling's $T^2$ distribution where the vectors being tested are not multivariate Gaussian (in which case $n_{\rm dof}$ would be replaced by $p$), but are instead determined by $n_{\rm dof}$ degrees of freedom, as in Wilks' theorem with a known covariance.} Intuitively, fluctuations in the covariance matrix used to compute each $\chi^2$ value will lead to a slightly heavier tail in the distribution of $\Delta\chi^2$ values, slightly decreasing the detection significance inferred from $\Delta\chi^2$ computed from the data compared to what would be obtained from a $\chi^2$ distribution.

We verify this expectation by fitting \cref{eq:scaled-Deltachi2-F} to the sets of 1000 noise realizations for the full band and each sub-band, determining the best-fit value of $n_{\rm dof}$ in each case, and using the Kolmogorov-Smirnov (KS) and Anderson-Darling (AD) tests to assess how well the associated $F$ distribution describes each set of $\Delta\chi^2$ values.\footnote{Note that standard KS and AD $p$-values are unreliable when the reference distribution is determined from the values being tested, so we calibrate our reported $p$-values via Monte Carlo. For each test, we draw $10000$ sets of $1000$ (appropriately-scaled) $\Delta\chi^2$ values from the best-fit $F$ distribution corresponding to our $1000$ noise realizations. We re-fit $n_{\rm dof}$ to each set and compute the KS or AD test statistic. We then compute a $p$-value by comparing this collection of test-statistic values with that computed from the original $1000$ noise realizations and fitted distribution.}
For both sub-bands, the KS and AD tests both pass (with $p$-values greater than $0.05$), while for the full band, the AD test passes while the KS test fails ($p=0.005$).

To investigate whether this failure is a statistical fluctuation in the original set of $1000$ noise realizations, we generate two other independent sets of $1000$ realizations, and repeat the above tests. For the other two sets, the KS and AD tests both pass (with $p>0.2$). Using two-sample KS and AD tests, we verify that each pair of noise realization sets are consistent with being drawn from the same distribution. We also verify that all three sets are consistent with being drawn from a Gaussian distribution with the associated covariance matrix, by comparing the set of $\chi^2_{\rm null}$ values to a $\chi^2$ distribution with $n_{\rm data}$ degrees of freedom. Finally, we perform a ``leave-one-block-out" comparison: for each set of $1000$ noise realizations, we find the best-fit $F$ distribution for the concatenation of the two other sets, and compute the KS and AD $p$-values comparing this distribution to the first set. In all cases, these tests pass with $p>0.05$.
Therefore, we conclude that the original set of $1000$ full-band noise realizations is a mild outlier, with only the self-fit to an $F$ distribution displaying unexpected behavior. For other parts of the analysis in this paper, we use this original set of noise realizations, and have checked that the results change by a negligible amount if the full set of $3000$ realizations is used instead. 

For each set of noise realizations, we use the best-fit $F$ distribution to compute the $p$-value corresponding to the $\Delta\chi^2$ value computed from the data, and use the inverse CDF of a standard normal distribution to translate this into an ``effective number of sigmas" as an interpretable measure of the signal-to-noise. We quote the mean over the three sets of noise realizations in the first S/N column of Table~\ref{tab:sn}, and use these values as our primary estimate of the S/N of our measurements.
The standard error on these mean values is $0.2\sigma$, $0.1\sigma$, and $0.2\sigma$ for the three bands. 

In several cases, the $\Delta\chi^2$ values for each set of noise realizations are inconsistent with $\chi^2$ distributions (the KS test yields $p<0.05$). Nevertheless, for comparison with the $F$-derived results, we quote the S/N obtained from the best-fit $\chi^2$ distribution in the second S/N column of Table~\ref{tab:sn}. As expected, these values are slightly higher than those from the $F$ distributions, due to the lighter tail of the $\chi^2$ distribution. For each band, the best-fit $n_{\rm dof}$ values for the $F$ and $\chi^2$ distributions are equal to two significant figures, falling between $n_{\rm dof}=3.5$ and $3.7$.

\subsection{{Goodness of Fit}}
\label{app:detection_significance:goodness_of_fit}

To assess the goodness of fit of the best-fit signal models described in \secref{sec:results:theoretical_interpretation:model_fitting}, we add the best-fit model for a given band to each of our base set of 1000 noise realizations. We then re-fit the 4-parameter model to each ``signal+noise" realization, using the L-BFGS-B algorithm to minimize the $\chi^2$ and taking the lowest result across 1000 random initial guesses in the parameter space defined by our priors in \secref{sec:results:theoretical_interpretation:model_fitting}. We compute an empirical $p$-value from the fraction of these $\chi^2$ values that exceed the value from the data.
For additional robustness, we repeat this exercise with the other two sets of 1000 noise realizations from Appendix~\ref{app:detection_significance:detection_significance}, and report the mean $p$-value from the three sets. For the full band, upper sub-band, and lower sub-band, we find $p=0.92$, $0.25$, and $0.31$, with standard deviation $0.01$, $0.03$, and $0.03$ respectively.


\bibliography{paper}{}
\bibliographystyle{aasjournal}

\end{document}